\documentclass{aa}
\usepackage{natbib}
\bibpunct{(}{)}{;}{a}{}{,} 
\usepackage{graphicx}
\usepackage{txfonts}

\begin{document}
\title{Hot-spot model for accretion disc variability as random process}
\author{T.~Pech\'a\v{c}ek,\inst{1} V.~Karas,\inst{1} \and B.~Czerny\inst{2}}
\institute{Astronomical Institute, Academy of Sciences, Bo\v{c}n\'{\i}~II 1401, CZ-14131~Prague, Czech Republic \and Copernicus Astronomical Center, Bartycka 18, P-00716~Warsaw, Poland}

\authorrunning{T.~Pech\'a\v{c}ek, V.~Karas, \& B.~Czerny}
\titlerunning{Hot-spot model for accretion disc variability}
\date{Received 4 March 2008; accepted 19 June 2008}
\abstract{}{Theory of random processes provides an attractive mathematical 
tool to describe the fluctuating signal from accreting sources, such as
active galactic nuclei and Galactic black holes observed in X-rays.
These objects exhibit featureless variability on different timescales,
probably originating from an accretion disc.}{We  study the basic
features of the power spectra in terms of  a general framework, which
permits semi-analytical determination of the power spectral density
(PSD) of the resulting light curve. We consider the expected signal
generated by an ensemble of spots randomly created on the accretion disc
surface. Spot generation is governed by  Poisson or by Hawkes processes.
The latter one represents an avalanche mechanism and seems to be
suggested by the observed form of the power spectrum. We  include
general relativity effects shaping the signal on its propagation  to a
distant observer.}{We analyse the PSD of a spotted disc light curve and
show the accuracy of our semi-analytical approach by comparing the
obtained PSD with the results of Monte Carlo simulations. The asymptotic
slopes of PSD are $0$ at low frequencies and they drop to $-2$ at
high frequencies, usually with a  single frequency break. More complex
two-peak solutions also occur. The amplitude of the peaks and their
frequency difference depend on the inherent timescales of the model,
i.e., the intrinsic lifetime of the spots and the typical duration of
avalanches.}{At intermediate frequencies, the intrinsic PSD is
influenced by the individual light curve profile as well as by the type
of the underlying process. However, even in cases when two Lorentzians
seem to dominate the PSD, it does not necessarily imply that two single
oscillation mechanisms operate simultaneously. Instead, it may well be
the manifestation of the avalanche mechanism. The main advantage of our
approach is an insight in the model functioning and the fast evaluation
of the PSD.}
\maketitle

\section{Introduction}
It is widely accepted that massive black holes with accretion discs
reside in cores of active galactic nuclei, where most activity
originates and X-rays are produced (e.g.,
\citeauthor{1990agn..book.....B} \citeyear{1990agn..book.....B}). The
observed light curves, $f\,\equiv\,f(t)$, show irregular, featureless
fluctuations with a very complex behaviour, practically at every studied
frequency \citep{2006ASPC..360.....G}. Variability has been
traditionally analysed by the Fourier method
\citep{1992scma.book.....F}. Remarkably, a number of similarities
appear between the properties of massive black holes in galactic nuclei
and those in X-ray binaries, suggesting that some kind of universal
rescaling operates according to central masses of  these systems
\citep{1998Natur.392..673M}. This concerns also the X-ray power spectra
(e.g., \citeauthor{2003ApJ...593...96M} \citeyear{2003ApJ...593...96M};
\citeauthor{2006Natur.444..730M} \citeyear{2006Natur.444..730M}).

Light curves can be characterised by an appropriate estimator of the
source variability which, in the mathematical sense, is a functional:
$f\rightarrow\mathcal{S}\left[f\right]$. We accept the idea that the
signal resulting from a spotted accretion disc is intrinsically
stochastic, likely originating from turbulence. Hence, 
$\mathcal{S}\left[f\right]$ is a random value. It can be a number (for
example, the `rms' characteristic), function of a single variable (for
example, the power-spectral density -- PSD) or of many variables (e.g.,
poly-spectra, rms--flux relation, etc). The appropriate choice depends
on the type of information we seek and the quality of data available.
The PSD is a traditional and widely utilised method to examine variable
signals, and the AGN light curves are no exception. A typical signal can be
represented by a broad band PSD with the tendency towards flattening at
low frequency
\citep{1987Natur.325..694L,1987Natur.325..696M,1993ApJ...414L..85L,1993ARA&A..31..717M,2002MNRAS.332..231U}. 

There is an ongoing debate about the overall shape of the PSD and the
occurrence of the break frequency or, possibly, two break frequencies 
at which the slope of the PSD can change 
\citep{1999ApJ...510..874N,2003ApJ...593...96M}. 
In the case of a widely-studied Seyfert galaxy, MCG--6-30-15,
\cite{2005MNRAS.359.1469M} have closely examined the slope of PSD,
namely its bending, with RXTE and XMM-Newton data. It is worth noticing
that the accurate fits to the X-ray sources seem  to exhibit a
multi-Lorentzian structure rather than a simple power law.  The same is
true for the  best studied example, the Seyfert~1 galaxy Akn~564
\citep{2007MNRAS.382..985M}.

It was proposed 
\citep{1991A&A...245..454A,1991A&A...246...21Z,1992AIPC..254..251W} that hot
spots contribute to the variability of the AGN variability,  or that
they could even be the dominant process shaping the variability pattern.
These spots should occur on the disc surface following its intermittent
irradiation by localised coronal flares 
\citep{1979ApJ...229..318G,2001MNRAS.328..958M,2004A&A...420....1C}.
Here, the ``spots'' represent a somewhat generic designation for
non-axisymmetric features evolving on the disc surface in connection
with flares. They share the bulk orbital motion with the underlying
disc. The observed signal is thus modulated by relativistic effects as
photons propagate towards a distant observer. 

Various schemes have been discussed in which the fluctuations of the
disc emissivity at different points of space and time are mutually
interconnected in some way. In particular, the avalanche model
\citep{1999MNRAS.306L..31P,2002MNRAS.333..800Z,2005MNRAS.359..308Z}
seems to be physically substantiated within the framework of
magnetically-triggered flares and spots. It is also a promising model
capable to reproduce, for example, the broken power-law PSD profiles.
Notice, however, that other promising ideas were proposed (e.g.,
\citeauthor{1994PASJ...46...97M}
\citeyear{1994PASJ...46...97M}; \citeauthor{1997MNRAS.292..679L}
\citeyear{1997MNRAS.292..679L}), provoking the question of whether a
common mathematical basis could reflect the entire range of models and
provide us with general constraints, independent of (largely unknown)
model details. 

We add to this model by applying the method of random point processes
\citep{1965cox}. Interestingly enough, a rather formal approach can
provide useful  analytical formulae defining the basic form of the
expected power  spectrum. Apart from this practical aspect, we suggest
that the concept discussed here offers much better insight into various
influences that shape the expected form the power spectrum. These are
very attractive features especially with respect to avalanche models,
which may have different flavours, typically with a vast parameter
space, thus proving very demanding to examine in a systematic manner.

Even more important is that the adopted formalism provides a very
general tool and allows for a broader perspective on different
mechanisms of variability \citep{2007arXiv0711.4772P}. We develop the
idea in a systematic way and give the explicit correspondence between
our approach and some of the above-mentioned and widely-known scenarios
\citep{1991A&A...245..454A,1999MNRAS.306L..31P}. This description
provides semi-analytical solutions, convenient to search through a
broad parameter space. Our results can help to identify how the
intrinsic properties of individual flares and the relativistic effects
influence the overall PSD. In particular, we can identify those
situations in which a doubly-broken power law occurs.

We consider stochastic processes (e.g.,
\citeauthor{1971aitp.book.....F} \citeyear{1971aitp.book.....F};
\citeauthor{1994hsmp.book.....G} \citeyear{1994hsmp.book.....G})
in the framework appropriate for modelling the accretion disc 
variability. In particular, in Sect.~\ref{models} we consider 
a simple model of a spotted accretion disc
constrained by the following three assumptions about the creation and
evolution of spots: (i)~each spot is described by its time and place of
birth ($t_j$, $r_j$ and $\phi_j$) in the plain of the disc; (ii)~every
new occurrence starts instantaneously; afterwards, the emissivity decays
gradually to zero (the total radiated energy is of course finite); and
(iii)~the intrinsic emissivity is fully determined by a finite set of
parameters which form a vector, $\mbox{\boldmath $\xi$}_j$, defining the
light curve profile. Later on, we will consider the modulation of the
intrinsic emission by the orbital motion and relativistic lensing. The
disc itself has a passive role in our considerations; we will treat it
as a geometrically thin, optically thick layer lying in the equatorial
plane. 

Because of the apparently random nature of the variability, we adopt a
stochastic model in which the creation of spots is governed by a random
process. The assumption that spots are mutually statistically
independent seems to be a reasonable (first) approximation, however, we
find that we do need to introduce some kind of relationship between
them. This connection is discussed in Sect.~\ref{relationship}. The
statistical dependence among spots can be introduced in several ways. In
Sect.~\ref{results}, we explore in detail the specific models of
interrelated spots using the formalism of Hawkes-type processes.
Conclusions are summarised in Sect.~\ref{conclusions}. Finally, in the
Appendix we provide some mathematical prerequisites, which the reader
may find useful to understand the general background of the paper, and
we also summarise the mathematical notation.

\section{Models of stochastic variability}
\label{models}
\subsection{Orbiting spots and relativistic effects}
We will apply our investigations to models where the signal is produced
by point-like orbiting spots (circular Keplerian motion along the
azimuthal  $\phi$ direction). The intrinsic emission, produced in the
local co-orbiting frame of the spot, is influenced by the Doppler effect and
gravitational lensing, which cannot be ignored at typical distances
of several units or tens of gravitational radii. Photons emitted at
different moments and positions experience different light-travel time
on the way towards the observer, so the {\em observed} timing properties
should reflect this specific modulation.  We adopt the Schwarzschild
metric  for the gravitational field and employ the method of transfer
function \citep{2005A&A...441..855P,2006AN....327..957P} to describe the
light amplification (or dilution); $\theta_{\rm o}$ is inclination angle
of the observer ($\theta_{\rm o}=90$~deg corresponds the edge-on view of
the disc plane). The periodical modulation of the observed signal is
included in the transfer function $F(t,r,\theta_{\rm o})\equiv
F(\phi(t),r,\theta_{\rm o})$. An implicit relation holds for the phase,
\begin{equation}
\phi(t)=\Omega t-\delta t(\phi(t),r,\theta_{\rm o}),
\end{equation}
where the effect of time delays  $\delta t(\phi,r,\theta_{\rm o})$ is
taken into account.  The modulation by $F$ is superimposed onto that due
to intrinsic timescales  present in the signal from spots. Then, for the
final flux $f(t)$ measured by a distant observer, we write
$f(t)=IF(t,r,\theta_{\rm o})$.

In the case of an infinitesimal surface element with intrinsically
constant and isotropic emissivity $I$, orbiting with Keplerian orbital
frequency $\Omega(r)$, the flux measured by a distant  observer varies
just as $F$ changes along the orbit.

We remind the reader that the mass of central black holes in galactic
nuclei is in the range of $\simeq M\approx10^6$--$10^{9}\,M_\odot.$ Mass
of the accretion disc is at least three orders of magnitude smaller than 
the black hole mass, so we neglect it in our calculations (the accretion
disc self-gravity may be important for its intrinsic structure, but the
direct gravitational effect on light is quite small;
\citeauthor{1995ApJ...440..108K} \citeyear{1995ApJ...440..108K}). Hence,
the gravitational field can indeed be described by a vacuum black-hole
spacetime \citep{1973grav.book.....M}. We use geometrical units with
$c=G=1$. Transformation to physical scales can be achieved when the mass
of the central black hole is specified because Keplerian frequency
scales inversely with the gravitational radius. The gravitational radius
of a massive black hole is  $R_{\rm g}\equiv
c^{-2}GM\approx0.48\times10^{-5}M_8\,$pc, and the corresponding
characteristic time-scale is  $t_{\rm g}\equiv
c^{-3}GM\approx0.49\times10^3M_8\,$sec,  where the mass $M_8\equiv
M/(10^8M_\odot)$. All lengths and times can be made dimensionless by
expressing them in units of $M$, so they can be easily scaled to
different masses. For example, for the Keplerian orbital period we
obtain $T_{\rm{}orb}\approx3.1\times10^3\,r^{3/2}M_8$, where the radius
is expressed in units of  $R_{\rm g}$ and $T_{\rm orb}$ is given in
seconds.

Let us note that the intrinsic timescales of the spot evolution and of
avalanches  (both timescales will be discussed below) need not to be
directly connected with the Keplerian orbital period. This internal
freedom of the model can help to bring the predicted frequency of the
breaks of the PSD  profile in agreement with the data.

\subsection{Model driven by a general point process}
Now we will describe the process of the creation of spots from the
statistics point of view. Let us consider a signal of the form
\begin{equation}
f(t)=\sum\limits_j I(t-{\delta_j},\,\mbox{\boldmath $\xi$}_j)\,F(t-\delta_{{\rm p}j},r_j).
\label{proces}
\end{equation}
The underlying process consists of a sequence of events which, in general, can
be either mutually independent, or there can be some statistical dependence
among them. Naturally, the latter case will be more complicated and interesting.
In Eq.~(\ref{proces}), $I(t,\,\mbox{\boldmath $\xi$})=g(t,\,\mbox{\boldmath
$\xi$})\,\theta(t)$ is the light curve profile of a single event;
$\delta_j=t_j+t_{{\rm d}j}$  and $\delta_{{\rm p}j}=\delta_j+t_{{\rm p}j}$ are
the time offsets (determined by the moment of the ignition of the spot)
and the initial time delay (which is an arbitrary but fixed
value); $\theta(t)$ is  the Heaviside function; and
$g(t,\mbox{\boldmath $\xi$})$  is a non-negative function of $k+1$
variables, $t$ and $\mbox{\boldmath  $\xi$}=(\xi^1,\dots,\xi^k)$.
Hereafter, we omit the explicit dependency on $\theta_{\rm o}$.

\begin{figure}[tbh!]
\begin{center}
\includegraphics[width=0.5\textwidth]{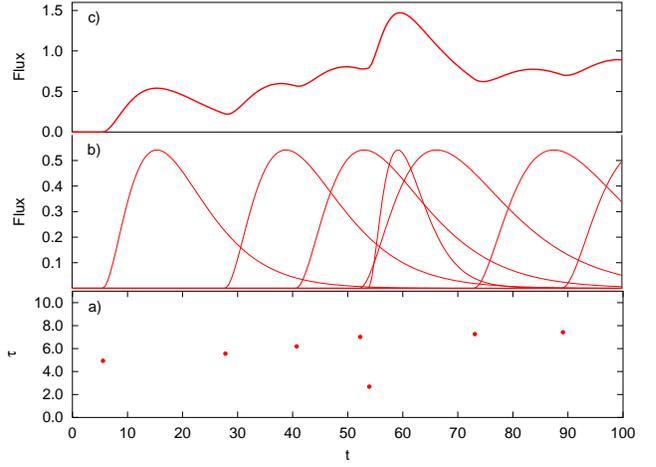}
\caption{Illustrating 
the correspondence between the ignition moments of the elementary
events and the resulting light curve. The model is fully determined by a
set of points in ($t$--$\tau$) plane, representing the pairs of ignition
time $t$ versus the time constant  $\tau$ of each event (panel a), and
the elementary light curves with the profile
$I(t,\tau)=\left(t/\tau\right)^2 \exp\left(-t/\tau\right)\theta(t)$
(panel b). The total light curve (panel c) is represented as a sum
of the individual contributions.}
\label{krivex2}
\end{center}
\end{figure}

Quantities $\mbox{\boldmath $\xi$}_j$, $t_j$, $r_j$, $t_{{\rm p}j}$ and
$t_{{\rm d}j}$ are random values. The vector $\mbox{\boldmath $\xi$}_j$
determines the  duration and shape of each event ($t_j$ is time of
ignition; parameter $t_{{\rm p}j}$ determines the initial phase of  the
periodical modulation of the $j$--th event; and $t_{{\rm d}j}$ is the
corresponding initial time-offset). These assumptions bring the
formulation of the problem close to the framework studied by Br\'emaud
et al.\ (\citeyear{Bremaud:2002:AAP,Bremaud:2005:AAP}). We will
calculate the power spectrum of this process directly from
Eq.~(\ref{powersp}) in the Appendix.

We remark that for the amplitudes of individual events we assume the
identical values (at each given radius). This restriction is imposed
only for the sake of definiteness of our examples; the formalism can
deal with a distribution of amplitudes. Indeed, we do not impose any
serious constraint on the model because the information about the level
of the fluctuating signal can be adjusted by setting the frequency of
the events \citep{1989ESASP.296..499L}. A simple demonstration of this
concept is shown in Fig.~\ref{krivex2}. This plot illustrates how the
model light curve arises from the elementary components. Naturally, we
can approach such decomposition from another  angle, by investigating
how the total light curve can be expressed in terms of some basic
profile. It is important to realise that, for the purposes of  our
present paper, light curves are of secondary importance. Instead, our 
calculations allow us to proceed from the distribution of the onset
times and the characteristics of individual flares directly to the power 
spectral density, which stands as the primary characteristic of the
source  signal.

Equation~(\ref{proces}) represents a very general class  of random
processes. By applying the Fourier transform, we find
\begin{eqnarray}
\mathcal{F}_T\big[f(t)\big](\omega)&=& \frac{2\sin(T\omega)}{\omega}\nonumber
\;\star\;\sum\limits_j
\mathcal{F}\big[I(t-{\delta_j},\,\mbox{\boldmath $\xi$}_j)\\
&&\times\;F(t-\delta_{{\rm p}j},r_j)\big](\omega),
\end{eqnarray}
where $\star$ denotes the convolution operation. In Eq.~(\ref{proces}),
we sum together a set of all events (this infinite sum could in general
pose  problems with convergence, however, as we will see later, we can
restrict the sum to a finite set of events without any loss of
generality). The Fourier transform of a single event,
$I(t-{\delta_j},\,\mbox{\boldmath $\xi$}_j)F(t-\delta_{{\rm
p}j},\,r_j)$, is then
\begin{eqnarray}
&&\mathcal{F}\big[I(t-{\delta_j},\,\mbox{\boldmath $\xi$}_j)
F(t-\delta_{{\rm p}j},r_j)\big](\omega)\nonumber \\
&&\quad=e^{-i\omega\delta_j}\mathcal{F}\big[I(t,\mbox{\boldmath $\xi$}_j)\big]\star\mathcal{F}\big[F(t+t_{{\rm p}j},r_j)\big](\omega).
\end{eqnarray}
Periodical function $F(t,r)$ can be now expanded in a series,
$F(t,r)=(2\pi)^{-1}\sum\limits c_k(r) \exp[{ik\Omega(r) t}]$, and the expanded form
substituted in the incomplete Fourier transform of $f(t)$.

Knowing the incomplete Fourier transform of $f(t)$, we can calculate 
its squared absolute value and perform the averaging over all 
realisations of the process. This can be simplified by assuming
that every single event quickly decays. In principle, between $-T$ and $T$
the process is influenced by all events ignited during the whole
interval $\langle-\infty,T\rangle$, however, because of
the fast decay of  the events, the interval can be restricted to
$\langle-(T+C),T\rangle$, where $C$ is a sufficiently large positive
constant. In other words, every realisation of the process $f(t)$ on
$\langle-T,T\rangle$ can be described by a set of points in
$(k+4)$-dimensional space 
$(t_j,t_{{\rm d}j},t_{{\rm p}j},r_j,\mbox{\boldmath $\xi$}_j)$ 
with $-(T+C)\leq t_j \leq T$.

Values of the initial time delay and phase are functions of initial
position of each spot, i.e.\
\begin{equation}
t_{{\rm d}j}=\delta t(r_j,\phi_j),\qquad
t_{{\rm p}j}=\frac{\phi_j}{\Omega(r_j)}+t_{{\rm d}j}.
\end{equation}
Fourier transform of the resulting signal is then
\begin{equation}
\mathcal{F}\big[I(t-t_{{\rm d}j},\,\mbox{\boldmath $\xi$}_j)\,
F(t-t_{{\rm d}j}+t_{{\rm p}j},r_j)\big](\omega)
=\sum\limits_{k=-\infty}^\infty c_k(r)\,e^{ik\phi}\,\mathcal{F}_k,
\end{equation}
where $\mathcal{F}_k\equiv\mathcal{F}[I(t-\delta
t(r,\phi),\mbox{\boldmath $\xi$})](\omega-k\Omega(r))$ is the Fourier
transform of the event light curve, corrected for the time delay. Every
realisation of this process is completely determined by the set of
points $(t_j,\phi_j,r_j,\mbox{\boldmath $\xi$}_j)$.

Defining the function 
\begin{eqnarray}
s(t,\phi,r,\mbox{\boldmath $\xi$};\omega)&=&
\frac{2\sin(T\omega)}{\omega}\star
\Big(e^{-i\omega t}\!\sum\limits_{k=-\infty}^\infty \!c_k(r) e^{ik\phi}\nonumber\\
&&\times\mathcal{F}\left[I(t-\delta t,\,\mbox{\boldmath $\xi$})\right](\omega-k\Omega(r))\Big)
\label{sfunkce1}
\end{eqnarray}
we can write
\begin{equation}
\big|\mathcal{F}_T[f(t)](\omega)\big|^2
=\Big|\sum\limits_j s(t_j,\,\phi_j,\,r_j, \,\mbox{\boldmath $\xi$}_j;\,\omega)\Big|^2.
\end{equation}
Due to Campbell's theorem (\ref{stred2}),
\begin{eqnarray}
&&\!\!\!{\rm E}\left[\left|\mathcal{F}_T[f(t)](\omega)\right|^2\right]\nonumber
=\int\limits_{A\times A'} s(t,\phi,r,\mbox{\boldmath $\xi$};\omega)\\
&&\!\times\,
s^*(t',\phi',r',\mbox{\boldmath $\xi$}';\omega)\,
m_2(t,\phi,r,\mbox{\boldmath $\xi$},t',\phi',r',\mbox{\boldmath $\xi$}')\,{\rm d}A\,{\rm d}A',
\label{Eft2}
\end{eqnarray}
where $m_2$ is density of the second-order moment measure $M_2(.)$
corresponding to the random point process of 
$(t_j,\phi_j,r_j,\mbox{\boldmath $\xi$}_j)$, $A$ is a Cartesian product
of four sets representing the domains of definitions, i.e.
\begin{equation}
A=\langle-(T+C),T\rangle \times \langle 0,2\pi\rangle\times
\langle r_{\rm min},r_{\rm max}  \rangle\times\Xi.
\end{equation}

Now we can perform the limit of $T\rightarrow\infty$, as given by
Eq.~(\ref{powersp}) in the Appendix. It can be shown that the result is
independent of the value of $C$. Therefore, to obtain an explicit
formula for the PSD we need only to specify the form of $m_2(.)$ in
Eq.~(\ref{Eft2}). Hereafter, we will show how to proceed with this task.

\subsection{Model driven by the Poisson process}
To start with a simple example, we assume mutually independent events
with uniformly distributed ignition times. In other words, in this
subsection we assume that there is no relationship among different spots
-- neither in their position nor in the time of birth (spots are
statistically-independent). The intensity and the second-order measure
are
\begin{eqnarray}
M_{{\rm g}1}({\rm d}t)&=&\lambda\,{\rm d}t,\label{PoissLambda}\\
M_{{\rm g}2}({\rm d}t\,{\rm d}t')&=&\left[\lambda^2 +\lambda\delta(t-t')\right]{\rm d}t\,{\rm d}t',\label{PoissM}
\end{eqnarray}
where $\lambda$ is the mean rate of events. Other parameters are treated as
independent marks with common distribution $G({\rm d}\phi\,{\rm
d}r\,{\rm d}\mbox{\boldmath $\xi$})$. The second-order measure is
\begin{eqnarray}
&&\!\!\!M_2({\rm d}t\,{\rm d}\phi\,{\rm d}r\,{\rm d}\mbox{\boldmath $\xi$}
\,{\rm d}t'\,{\rm d}\phi'\,{\rm d}\mbox{\boldmath $\xi$}')
=\big[\lambda^2\,G({\rm d}\phi\,{\rm d}r\,{\rm d}\mbox{\boldmath 
$\xi$})\nonumber\\
&&\times\,G({\rm d}\phi'\,{\rm d}r'\,{\rm d}\mbox{\boldmath $\xi$}')
+\lambda G({\rm d}\phi\,{\rm d}r\,{\rm d}\mbox{\boldmath $\xi$}) \nonumber\\
&&\times\,\delta(t-t')\,\delta(\phi-\phi')\,\delta(r-r')\,
\delta(\mbox{\boldmath $\xi$}-\mbox{\boldmath $\xi$}')\big]\,{\rm d}t\,{\rm d}t'.
\label{somd}
\end{eqnarray}

The result of integration (\ref{Eft2}) can be simplified for
$T\rightarrow\infty$. The task reduces to evaluation of two limits (see
\citeauthor{2008Pechacek} \citeyear{2008Pechacek}, for details).  After
somewhat lengthy calculations we obtain a general formula for the power
spectrum:
\begin{equation}
S(\omega)=4\pi^2n\!\!\sum\limits_{k,l=-\infty}^\infty 
\int\limits_{\mathcal{K}}c_k(r)\,c^*_l(r)\,e^{i(l-k)\phi}\,\mathcal{F}_k\,\mathcal{F}_l^*\,
G({\rm d}\phi\,{\rm d}r\,{\rm d}\mbox{\boldmath $\xi$}).
\label{PoissPSD}
\end{equation}

Here, we remark that the general relativity effects are included in
the Fourier coefficients $c_k(r)$. Knowing them in advance (i.e.,
pre-calculating the sufficient number of the coefficients that are 
needed to achieve the desired accuracy) helps us to produce the PSD
efficiently. But we will start by neglecting these relativistic
effects,  so that we can clearly reveal the impact of the intrinsic
timescales of individual spots and the form of the avalanche process.

\subsubsection{Example 1: Markov chain}
Let us consider box-shaped events with exponentially-distributed
life-times, i.e.,
\begin{equation}
I(t,\tau)=I_0\,\chi_{\langle 0,\tau\rangle}(t),\quad \zeta(\tau)=\mu\,e^{-\mu\tau},\\
\end{equation}
where $\chi$ is the characteristic function ($\chi=1$ for $0\leq
t<\tau$, $\chi=0$ otherwise); $\zeta(\tau)$ is the probability density;
and $I_0$ is a constant (we will set $I_0=1$ for simplicity). This again
represents a process of the type (\ref{proces}), but  at the same time
it can be considered as a continuous-time Markov chain with  discrete
states \citep{1965cox}. The process PSD is then given by
Eq.~(\ref{PoissPSD}) with
\begin{equation}
G({\rm d}\tau\, {\rm d}r)=\frac{\mu}{r_{\rm max}-r_{\rm min}}\,e^{-\mu\tau}\,{\rm d}\tau\,{\rm d}r,\quad F(t,r)=1.\label{markFF}
\end{equation}
In this example, by setting the transfer function $F(t,r)=1$ we
completely ``switch-off''  the periodical modulation and the
relativistic effects. 

Coefficients $c_k(r)$ are given by the relation
\begin{equation}
c_k(r)=\int\limits_0^{2\pi/\Omega(r)}F(t,r)\,e^{-ik\Omega(r)t}\,{\rm d}t,
\end{equation}
which for the constant function $F(t,\,r)$ leads to
$c_0(r)=(2\pi)^{-1}\doteq0.159$, $c_k(r)=0$ $(k>0)$.
Fourier transform of the profile function is
\begin{eqnarray}
\mathcal{F}\big[I(t,\tau)\big](\omega)&=&\frac{1-e^{-i\omega\tau}}{i\omega},\\
\Big|\mathcal{F}\big[I(t,\tau)\big](\omega)\Big|^2&=&\frac{4\sin^2(\omega\tau/2)}{\omega^2}.
\end{eqnarray}
Substituting into the general formula (\ref{PoissPSD}) we find
\begin{eqnarray}
S(\omega)&=& 4\pi \lambda\int\limits_{r_{\rm min}}^{r_{\rm max}}\big|c_0(r)\big|^2\int\limits_0^\infty
\Big|\mathcal{F}\big[I(t,\tau)\big](\omega)\Big|^2\nonumber\\
&&\times\,\frac{\mu}{r_{\rm max}-r_{\rm min}}\, e^{-\mu\tau}\,{\rm d}\tau\,{\rm d}r
=\frac{2\lambda}{\omega^2+\mu^2}.
\end{eqnarray}
Notice that the theory of Markov chains is usually formulated in another way,
independent of our previous calculations (see \citeauthor{1965cox} \citeyear{1965cox}). 
One can thus obtain the PSD of the Markov process by two different
approaches and verify the results.  Such comparison gives the same
result, as expected.

\section{Introducing a relationship among spots}
\label{relationship}
The assumption that spots are statistically independent seems to be a
reasonable (first) approximation. However, the actual ignition times and
spot parameters should probably depend on the history of a real system.
The statistical dependence among spots can be introduced in several
ways. In this section we discuss different models where an existing
spot gives, with a certain probability, birth to new spots. In this way
a single spot at the beginning can trigger a whole avalanche of its
offsprings. This avalanche can be in principle of arbitrary length,
although, to obtain an infinitely long avalanche with non-diverging rate
of new spots one would have to fine-tune the parameters. In order to
avoid the unlikely fine-tuning and to obtain a stationary process, we
will assume the occurrence of many spontaneous spots distributed by the
Poisson process that keeps triggering new avalanches of finite duration.

The first example can be called ``Chinese process''. By definition, an
existing spot gives birth to exactly one new spot with probability
$0\leq\psi<1$. In other words, every event produces at most one
offspring. The spot of the $k$th generation is always ignited later than
the spot of the $(k-1)$th generation. As mentioned above, spontaneous 
spots arise randomly, according to Poissonian process. In the simplest 
version of this model, delays between the parent spot and its lineal 
descendant are random values obeying the probability density $p(t)$.

\begin{figure}[tbh!]
\includegraphics[width=0.5\textwidth]{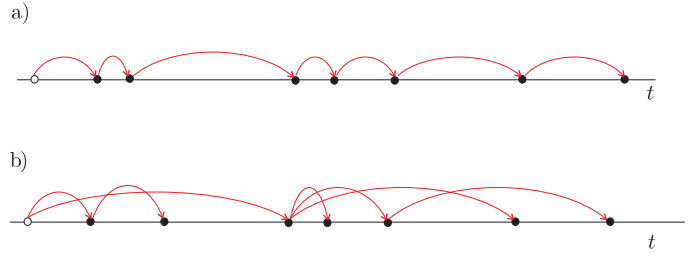}
\caption{Distinguishing between the Chinese process (a) and the Hawkes 
processes (b). In both panels, points represent  ignition times of the
spots. Each sequence starts with a spontaneously generated parent spot
(open circles) and it continues with subsequent secondary ones (filled
circles). Arrows symbolise the parent--daughter relation. The difference
between the two scenarios is described in Sects.~\ref{sec:chinese} and
\ref{sec:hawkes}, respectively. Within the schematic level
of this graph, the pulse avalanche process (see
Sect.~\ref{sec:avalanche}) belongs also to case (b).}
\label{czer-haw}
\end{figure}

More generally, every spot can deliver $n$ new spots, where
$n$ is a random value with Poisson distribution and the mean $\nu$. We
describe this situation in terms of (i)~standard Hawkes' process, and
(ii)~the pulse avalanche model. The temporal distribution of new spots
is now governed either by the function $\mu(t)$ of the Hawkes' process
\citep{Hawkes:1971}, or $\mu(t,\tau\,|\,t_0,\tau_0)$ in the case of avalanches
\citep{1999MNRAS.306L..31P}. Spots of different generations can appear
at the same time. 

The difference between the Chinese process and the latter two processes
is schematically sketched in Fig.~\ref{czer-haw}. Mathematically, all
three examples belong to the class of cluster processes.

\subsection{The cluster processes}
Point processes are characterised by the generating functional,
$\mathcal{G}[.]$, which is defined by its action \citep{Daley:2003}
\begin{equation}
\mathcal{G}\big[h(x)\big]={\rm E}\Big[\!\prod\limits_{x_i\in {\rm supp}\{N\}}\!\!\!h(x_i)\Big],
\label{gen_funkcional} 
\end{equation}
where $h(x):\mathcal{X}\rightarrow \mathbb{C}$ is a function satisfying
the condition $\left|h(x)\right|\leq 1$.

The functional $\mathcal{G}[.]$ satisfies various relations which can be derived
in close analogy with the theory of generating functions of random variables.
For our purposes it will be useful to expand $\mathcal{G}[.]$ into a series in terms 
of factorial measures,
\begin{equation}
\mathcal{G}\left[1+\eta\right]=1+\sum\limits^\infty_{k=1}\!\!\int\limits_{~\mathcal{X}^k}
\!\eta(x_1)\dots\eta(x_k)\,M_{[k]}({\rm d}x_1\dots{\rm d}x_k).
\label{GExpanze}
\end{equation}

The cluster processes consist of two point processes. The first one is
connected with the counting measure $N_{\rm c}(A)$ and defines the central
points $y$ of the clusters. The second process spreads new points around
the central point according to the random measure $N(B\,|\,y)$. The
complete counting measure is then $N(A)=\int_{\mathcal{X}}N(A\,|\,y)\,N_{\rm
c}({\rm d}y)$.

Let $\mathcal{G}\left[h(x)\,|\,y\right]$ be the generating functional of a cluster
with the center at $y$,
\begin{equation}
\mathcal{G}\big[h(x)\big]={\rm E}\Big[\int\limits_{\mathcal{X}} 
\mathcal{G}\big[h(x)\,|\,y\big]\,N_{\rm c}({\rm d}y)\Big].
\end{equation}
If the cluster center process is Poissonian, the latter formula
simplifies  to
\begin{equation}
\mathcal{G}\big[h(x)\big]=\exp\Big\{-\int\limits_{\mathcal{X}} 
\big(1-\mathcal{G}\left[h(x)\,|\,y\right]\big)\,\Lambda_{\rm c}({\rm d}y)\Big\},
\label{GPoisson}
\end{equation}
where $\Lambda_{\rm c}({\rm d}y)$ is the intensity measure of the center
process. One can expand the functional (\ref{GPoisson}) in
terms of the factorial measure of the cluster,
\begin{equation}
\mathcal{G}\big[1+\eta\,|\,y\big]=1+\!\sum\limits^\infty_{k=1}\int\limits_{\mathcal{X}^k}
\eta(x_1)\dots\eta(x_k)M_{[k]}({\rm d}x_1\dots{\rm d}x_k\,|\,y).
\label{GClusExpanze}
\end{equation}
Putting Eqs.~(\ref{GExpanze}) and (\ref{GClusExpanze}) into
Eq.~(\ref{GPoisson}) and collecting the terms with the same order of
$\eta(x)$, we find
\begin{eqnarray}
M_{[1]}(A)\!&=&\!\!\int\limits_{\mathcal{X}}\!M_{[1]}(A\,|\,y)\,\Lambda_{\rm c}\,({\rm d}y),\\
M_{[2]}(A\times B)\!&=&\!\!\int\limits_{\mathcal{X}}\!M_{[2]}(A\times B\,|\,y)\,\Lambda_{\rm c}\,({\rm d}y)+M_{[1]}(A)\,M_{[1]}(B).\label{PoissClusQuSym}
\end{eqnarray}

For a stationary process, the intensity measure stays constant,
$\Lambda_{\rm c}({\rm d}x)=\lambda_{\rm c}{\rm d}x$, and gives
the density of the centers. All factorial
moments are shift-invariant in the sense
\begin{equation}
m_{[k]}(x_1,\dots,\,x_k\,|\,y)=m_{[k]}(x_1-y,\dots,\,x_k-y\,|\,0),
\label{ShiftInvar}
\end{equation}
where $m_{[k]}(.\,|\,y)$ is density of $M_{[k]}(.\,|\,y)$. As a result of
the shift invariance, density of the first-order moment $m_1$ must be
also constant. Moreover, we can always choose $y$ in Eq.~(\ref{ShiftInvar})
equal to one of $x_i$, so the function $m_{[k]}$ depends on only $k-1$
variables. We define the reduced factorial moments,
\begin{equation}
\breve{m}_{[k]}(u_1,\dots,\,u_{k-1})=m_{[k]}(0,\,u_1,\dots,\,u_{k-1}),
\end{equation}
and from Eq.~(\ref{PoissClusQuSym}) it follows that
\begin{equation}
\breve{m}_{[2]}(u)=\lambda\int\limits_\mathcal{X}m_{[2]}(y,y+u\,|\,0)\,{\rm d}y + m^2_1.
\end{equation}
Stationarity of the process implies that the second-order measure
density depends only on the difference of its arguments,
\begin{equation}
m_{[2]}(t,t')=\breve{m}_{[2]}(t-t')=c\,(t-t')+m_1^2,
\label{hawksm}
\end{equation}
where $c(t)$ is an even function.

The second-order measure of a marked cluster process is
\begin{eqnarray}
&&\!\!\!M_2({\rm d}t\,{\rm d}\phi\,{\rm d}r\,{\rm d}\mbox{\boldmath $\xi$}
\,{\rm d}t'\,{\rm d}\phi'\,{\rm d}\mbox{\boldmath $\xi$}')
=\big[\big(m_1^2+c(t-t')\big)\,G({\rm d}\phi\,{\rm d}r\,{\rm d}\mbox{\boldmath 
$\xi$})\nonumber\\
&&\times\, G({\rm d}\phi'\,{\rm d}r'\,{\rm d}\mbox{\boldmath $\xi$}')
+m_1\, G({\rm d}\phi\,{\rm d}r\,{\rm d}\mbox{\boldmath $\xi$}) \nonumber\\
&&\times\,\delta(t-t')\,\delta(\phi-\phi')\,\delta(r-r')
\,\delta(\mbox{\boldmath $\xi$}-\mbox{\boldmath $\xi$}')\big]\,{\rm d}t\,{\rm d}t',
\label{somd2}
\end{eqnarray}
almost identical with that of the Poissonian process. There is only one
additional term associated with the function $c(t)$. The resulting 
spectrum is given by a somewhat lengthy, but still explicit formula. We
find stationary cluster processes to be particularly promising as a
general scheme, encompassing a broad range of models as special cases.
The PSD is
\begin{eqnarray}
&&\!\!S(\omega)=S_1(\omega)+4\pi^2m_1 \nonumber\\
&&\times\sum\limits_{k,l=-\infty}^\infty 
\int\limits_{\mathcal{K}}e^{i(l-k)\phi}c_k(r)\,c^*_l(r)\,\mathcal{F}_k(\omega)\,
\mathcal{F}_l^*(\omega) 
\,G({\rm d}\phi\,{\rm d}r\,{\rm d}\mbox{\boldmath $\xi$})
\label{HawkPSD}
\end{eqnarray}
with $S_1(\omega)\equiv4\pi^2 S_{\rm\! P}(\omega)\,Q_{{_\mathcal{K}}}(\omega)\,{Q_{{_\mathcal{K}}'}}^{\!\!\!*}(\omega)$ 
and
\begin{equation}
Q_{{_\mathcal{K}}}(\omega)\equiv\!\sum\limits_{k=-\infty}^\infty\! \int\limits_{\mathcal{K}}\!e^{-ik\phi}c_k(r)\,
\mathcal{F}_k(\omega)\,
G({\rm d}\phi\,{\rm d}r\,{\rm d}\mbox{\boldmath $\xi$}).
\label{HawkPSD2}
\end{equation}

The reduced quadratic factorial moment appears in the formulae for 
power spectra of cluster processes expressed in terms of $S_{\rm\!
P}(\omega)\equiv\mathcal{F}\left[c(t)\right](\omega)=\mathcal{F}\left[\breve{m}_{[2]}(t)-m^2_1\right](\omega)$.
This is directly related to the two-dimensional Fourier transform of the
quadratic factorial measure as
\begin{equation}
S_{\rm\! P}(\omega)=\lambda\,\tilde{m}_{[2]}(\omega,-\omega\,|\,0),
\end{equation}
where
\begin{equation}
\tilde{m}_{[2]}(\omega,\omega'\,|\,0)=\int\limits_{\mathbb{R}^2}
e^{i(x_1\omega+x_2\omega')}m_{[2]}(x_1,\ x_2\,|\,0)\,{\rm d}x_1\,{\rm d}x_2.
\end{equation}
We will discuss the form of $S_{\rm\! P}(\omega)$, the expression for 
$m_1$, and the resulting PSD in different situations. But before that,
we still need to show how the model properties can be detailed
in term of marks.

\subsection{Marks as a way to specify the model properties}
\label{marks}
Until now the variability patterns have been restricted only by very
general properties of the assumed process. This means that the model is
kept in a very general form. However, formulae (\ref{PoissPSD}) and
(\ref{HawkPSD}) are  too general for any practical use. Their main value
is after defining special  cases. Then these formulae can be readily
applied to derive the analytical form of the PSD. Such special cases are
conveniently defined by means  of marks. We discuss possible choices of
the mark  distribution, $G({\rm d} r\,{\rm d}\phi\,{\rm
d}\mbox{\boldmath $\xi$})$.

We can simplify the situation by assuming axial symmetry. Therefore, all
statistical properties  should depend only on the radius (the azimuthal
part of $G$ is constant). The distribution of marks has now the form
\begin{equation}
G({\rm d} r\,{\rm d}\phi\,{\rm d}\mbox{\boldmath $\xi$})
=\frac{1}{2\pi}\,\zeta_{\rm R}(\mbox{\boldmath $\xi$}\,|\,r)\,\rho(r)
\,{\rm d} r\,{\rm d}\phi\,{\rm d}\mbox{\boldmath $\xi$},
\end{equation}
where $\zeta_{\rm R}(\mbox{\boldmath $\xi$}\,|\,r)$ is the reduced probability
density of the remaining parameters. 
To illustrate this case more clearly, we will now examine the
phenomenological model of \citet{1991A&A...245..454A} and
\citet{1991A&A...246...21Z}, which can be considered as representation
of the spotted accretion disc.

\subsection{Example 2: a spotted disc}
Let $n(r)$ be an average number of spots at radius $r$, each of them 
shining with the average intensity $I_{\rm M}(r)$ for 
average duration $\tau_{\rm M}(r)$. Let us further assume that 
these characteristics scale with the radius as power laws:
\begin{equation}
n(r)=A_n\,r^{\alpha_n},\quad
\tau_{\rm M}(r)=A_\tau\,r^{\alpha_\tau},\label{AbrMargTau}\quad
I_{\rm M}(r)=A_I\,r^{\alpha_I}
\end{equation} 
(here, $\alpha$s and $A$s are constants). This setup falls within
the category of models described by Eq.~(\ref{proces}), with the profile function
$I(t,r,\tau)$ and the mark distribution $G({\rm d} r\,{\rm d}\phi\,{\rm
d}\tau)$. The first two relations (\ref{AbrMargTau}) prescribe the
conditional  marginals of $G$, i.e.,
\begin{equation}
n(r)=\lambda\,\rho(r),\quad
\tau_{\rm M}(r)=\int\limits_{\tau_{\rm min}(r)}^{\tau_{\rm max}(r)}
\tau\,\zeta_{\rm R}(\tau\,|\,r)\,{\rm d}\tau,\label{MargTau1}
\end{equation} 
where the integration limits can depend explicitly on $r$. 
The third equation determines the amplitude of the profile function,
\begin{equation}
I(t,r,\tau)=A_I\,r^{\alpha_I}\,g(t,\tau)\,\theta(t).
\end{equation}
The dependency on $\tau$ is not determined uniquely. In order to obtain
an explicit form of $G$ we have to go beyond the model of
\cite{1991A&A...245..454A} by assuming the distribution of $\tau$,
\begin{equation}
\zeta_{\rm R}(\tau\,|\,r)=K(r)\,\tau^{-p}.
\label{zeta}
\end{equation}
The normalisation
constant then equals 
\begin{equation}
K(r)=\frac{1-p}{\tau_{\rm max}^{1-p}(r)-\tau_{\rm min}^{1-p}(r)},
\label{normconst}
\end{equation}
and for $\tau_{\rm M}(r)$ we find
\begin{equation}
\tau_{\rm M}(r)=K(r)\,\int\limits_{\tau_{\rm min}(r)}^{\tau_{\rm max}(r)}\tau^{1-p}{\rm d}\tau
=\frac{1-p}{2-p}
\frac{\tau_{\rm max}^{2-p}(r)-\tau_{\rm min}^{2-p}(r)}{\tau_{\rm max}^{1-p}(r)-\tau_{\rm min}^{1-p}(r)}.
\label{MargTau}
\end{equation}
The mean, $\tau_{\rm M}(r)$, must satisfy Eqs.~(\ref{AbrMargTau}) 
and (\ref{MargTau}). The choice of
\begin{equation}
\tau_{\rm min}(r)=C_{\rm min}r^{\alpha_\tau},\quad \tau_{\rm max}(r)=C_{\rm max}r^{\alpha_\tau}
\end{equation} 
is consistent with both equations. Constants $C_{\rm min}$, $C_{\rm max}$ and $A_\tau$ are
bound by the relation
\begin{equation}
A_\tau(2-p)(C_{\rm max}^{1-p}-C_{\rm min}^{1-p})=(1-p)(C_{\rm max}^{2-p}-C_{\rm min}^{2-p})
\end{equation}
Because $C_{\rm min}$ and $C_{\rm max}$ are positive, we can write
\begin{equation}
C_{\rm min}=C, \quad C_{\rm max}=\gamma C,
\end{equation}
\begin{eqnarray}
C&=&A_\tau \frac{2-p}{1-p}\frac{\gamma^{1-p}-1}{\gamma^{2-p}-1},\\
K(r)&=&\frac{1-p}{\gamma^{1-p}-1}\left(Cr^{\alpha_\tau}\right)^{p-1}.
\end{eqnarray}
Therefore, by substituting back to Eq.~(\ref{zeta}), we obtain
\begin{eqnarray}
\zeta_{\rm R}(\tau\,|\,r)\,\rho(r)&=&\frac{\alpha_n+1}
{\left(r_{\rm max}^{\alpha_n+1}-r_{\rm min}^{\alpha_n+1}\right)}
\frac{1-p}{\gamma^{1-p}-1}\nonumber\\&\times&
\left(\frac{2-p}{1-p}\frac{\gamma^{1-p}-1}{\gamma^{2-p}-1}A_{\tau}\right)^{p-1}
\,\tau^{-p}\,r^{(p-1)\alpha_{\tau}+\alpha_n},
\end{eqnarray}
where the definition domain is a set
\begin{equation}
\Xi=\left\{(r,\xi(r))\,|\,r\in\langle r_{\rm min},r_{\rm max}\rangle,\xi(r)\in 
C r^{\alpha_\tau}\langle 0,\gamma\rangle\right\}.
\end{equation}

We can set $p=1$ as a specific example. This value of the power-law
index is special in the sense that neither short nor long timescales
dominate the PSD, as we see from Eq.~(\ref{normconst}):
\begin{equation}
C=A_\tau \frac{\ln\gamma}{\gamma-1},\quad K(r)=\frac{1}{\ln\gamma}.
\end{equation}
By substituting back to Eq.~(\ref{zeta}) we obtain
\begin{equation}
\zeta_{\rm R}(\tau\,|\,r)\,\rho(r)=\frac{\alpha_n+1}{\ln\gamma
\left(r_{\rm max}^{\alpha_n+1}-r_{\rm min}^{\alpha_n+1}\right)}\,r^{\alpha_n}\,\tau^{-1}.
\end{equation}

Knowing the concrete form of the distribution of marks, we perform the
integration over $\mbox{\boldmath $\xi$}$. As $I(t,r)$ does not
explicitly depend on  azimuthal angle, the integration is simplified.
Denoting
\begin{equation}
d_n(r)=\frac{1}{2\pi}\int\limits_0^{2\pi}
e^{in[\phi+\Omega(r)\delta t(r,\,\phi)]}\,
{\rm d}\phi,
\end{equation}
we rewrite the PSD formula (\ref{PoissPSD}) in the final form
\begin{eqnarray}
S(\omega)&=&4\pi^2n\!\!\!\sum\limits_{k,l=-\infty}^\infty\,
\int\limits_{r_{\rm min}}^{r_{\rm max}}\!c_k(r)\,c^*_l(r)\,d_{k-l}(r)\nonumber\\
&&\times\int\limits_\Xi\! 
\mathcal{F}_k(\omega)\,
\mathcal{F}_l^*(\omega)
\,\zeta_{\rm R}(\mbox{\boldmath $\xi$}\,|\,r)\,{\rm d}\mbox{\boldmath $\xi$}\,\rho(r)\,{\rm d}r.
\label{PoissIndPSD}
\end{eqnarray}
We find the coefficients by direct evaluation,
\begin{equation}
d_n(r)=\frac{1}{2\pi}\int\limits_0^{2\pi}
e^{iny}\left[1+\Omega(r)\,\frac{\partial\delta t(\phi(y),r)}{\partial{\phi}}\right]^{-1}
{\rm d}\phi
\label{dn}
\end{equation}
with $y=\phi+\Omega(r)\,\delta t(\phi,r)$. We note that the term
(\ref{dn}) corresponds to the effect of delay amplification in
terminology of \cite{2008MNRAS.384..361D}. It influences the observed
signal from a source moving (i.e., orbiting) near a black hole.

\section{Results for the model PSD}
\label{results}
\subsection{Model driven by the Chinese process}
\label{sec:chinese}
Let us denote $\psi$ the probability that an existing spot generates a
new one, and $q_k$ the probability that a family of spots consists of
exactly $k$ members. The value $q_k$ obeys the geometrical distribution,
$q_k=\psi^k(1-\psi)$. 

We interpret probability density $p(t)$ of the delay between successive
spots  as a mean number of first-generation spots that occur at the
ignition time $t>0$, where $t=0$ is a moment of ignition of the parent
spot. Analogically, $p(t)\star p(t)$ is the mean number of
second-generation spots. For a sequence of $k$ spots, we obtain the
intensity measure
\begin{equation}
m_{1}(t\,|\,0,k)=\sum\limits_{j=0}^k p^{\star j}(t),
\end{equation}
where $p^{\star k}(t)$ is the $k$th convolutionary power of $p(t)$.
We can write $m_{1}$ in the form
\begin{equation}
m_{1}(t\,|\,0)=\sum\limits_{k=0}^\infty q_k\, m_{1}(t\,|\,0,k).
\end{equation}
Convolution of two functions can be calculated via the Fourier image.
Defining $\tilde{p}(\omega)=\mathcal{F}\left[p(t)\right](\omega)$ we get
\begin{equation}
\tilde{m}_{1}(\omega\,|\,0)=(1-\psi)\sum\limits_{k=0}^\infty \psi^k\sum\limits_{j=0}^k \tilde{p}^j(\omega)
=\frac{1}{1-\psi \tilde{p}(\omega)}.
\end{equation}
From this we find
\begin{eqnarray}
\int\limits_0^\infty m_{1}(t\,|\,0)\,{\rm d}t&=&\tilde{m}_{1}(\omega\,|\,0)_{|\omega=0}=\frac{1}{1-\psi},
\label{m1t}\\
\int\limits_0^\infty t\, m_{1}(t\,|\,0)\,{\rm d}t&=&-i\,\frac{\rm d}{{\rm d}\omega}\tilde{m}_{1}
(\omega\,|\,0)_{|\omega=0}=\frac{\psi}{\left(1-\psi\right)^2}\,{\rm E}[t].
\label{et}
\end{eqnarray}
The meaning of integral (\ref{m1t}) is the average number of spot offsprings in 
the whole avalanche. The meaning of the last integral is the average duration of the 
avalanche.

Calculation of the quadratic measure is a less intuitive procedure. We
start from the generating functional (\ref{gen_funkcional}) of the
process,
\begin{equation}
\mathcal{G}\left[h(t)\,|\,0\right]=\sum\limits_{l=0}^\infty q_l\, h(0)\,\mathcal{G}_l[h(t)],
\end{equation}
where $\mathcal{G}_l[.]$ denotes a generating functional of finite renewal process 
with $l$ points (see chapt.~5 in \citeauthor{Daley:2003} \citeyear{Daley:2003}).

Substituting $h(t)=1-\eta(t)$ in the expansion (\ref{GExpanze}), we 
obtain the Fourier image
\begin{equation}
\tilde{m}_{[2]}(\omega,\omega'\,|\,0)\!
=\!\frac{\psi\big[\tilde{p}(\omega) +\tilde{p}(\omega')-2\psi\tilde{p}(\omega)\tilde{p}(\omega')\big]}{
\big[1-\psi\tilde{p}(\omega)\big]\big[1-\psi\tilde{p}(\omega')\big]\big[1-\psi\tilde{p}(\omega+\omega')\big]}.
\end{equation}
This equation allows us to find the second-order measure. The PSD is then
given by Eqs.\ (\ref{HawkPSD})--(\ref{HawkPSD2}) with
\begin{eqnarray}
m_1&=&\frac{\lambda}{1-\psi},\\
S_{\rm\! P}(\omega)&=&\frac{2\lambda\psi\,\big[\Re e\,\big\{\tilde{p}(\omega)\big\}+\psi\,|\,\tilde{p}(\omega)\,|\,^2\big]}{(1-\psi)\big[1-2\psi\,\Re e\,\big\{\tilde{p}(\omega)\big\}+\psi^2\,|\,\tilde{p}(\omega)\,|\,^2\big]}.
\label{m1sp}
\end{eqnarray}
Equation~(\ref{m1sp}) can be cast in the form
\begin{equation}
S_{\rm\! P}(\omega)=\lambda\,\frac{|\tilde{m}_{[1]}(\omega\,|\,0)|^2\,\big(1-\psi^2\,|\tilde{p}(\omega)|^2\big)-1}{1-\psi}.
\label{mlsp1}
\end{equation}

\begin{figure*}[tbh]
\begin{center}
\hfill~\includegraphics[width=0.4\textwidth]{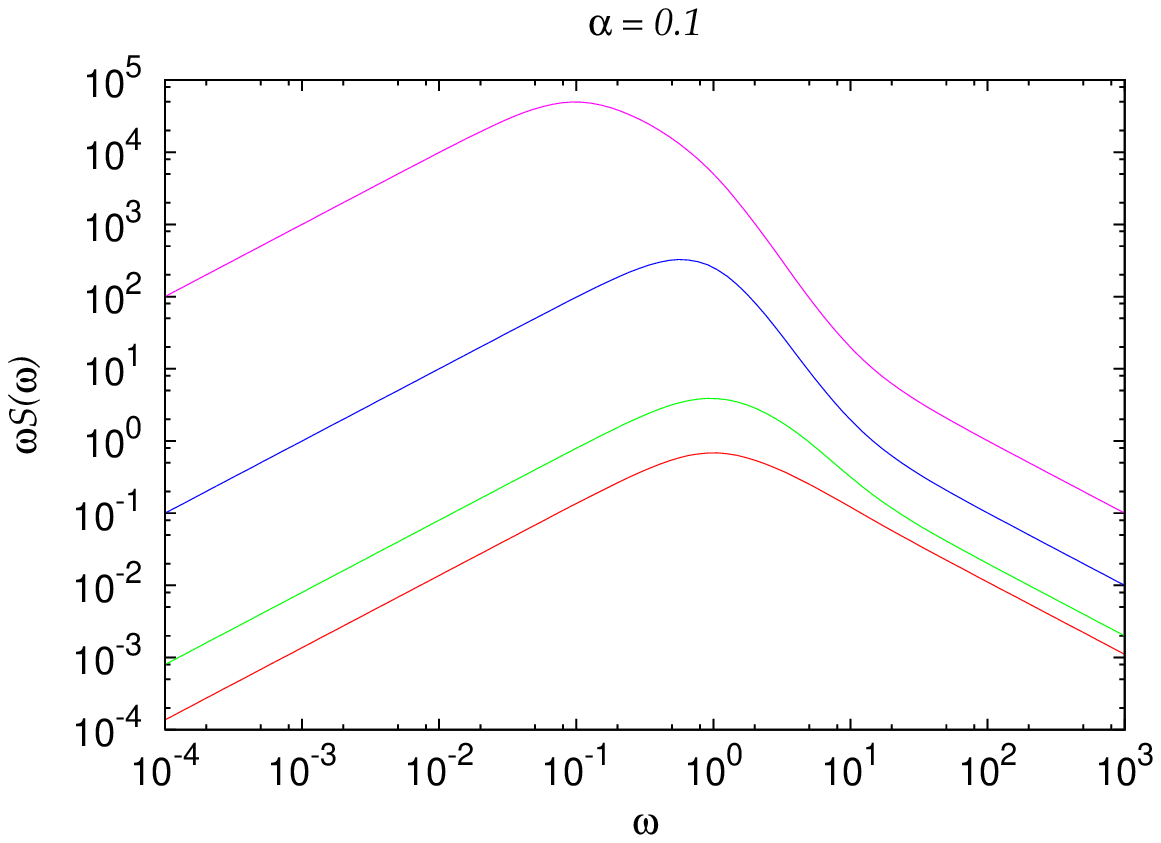}
\hfill~\includegraphics[width=0.4\textwidth]{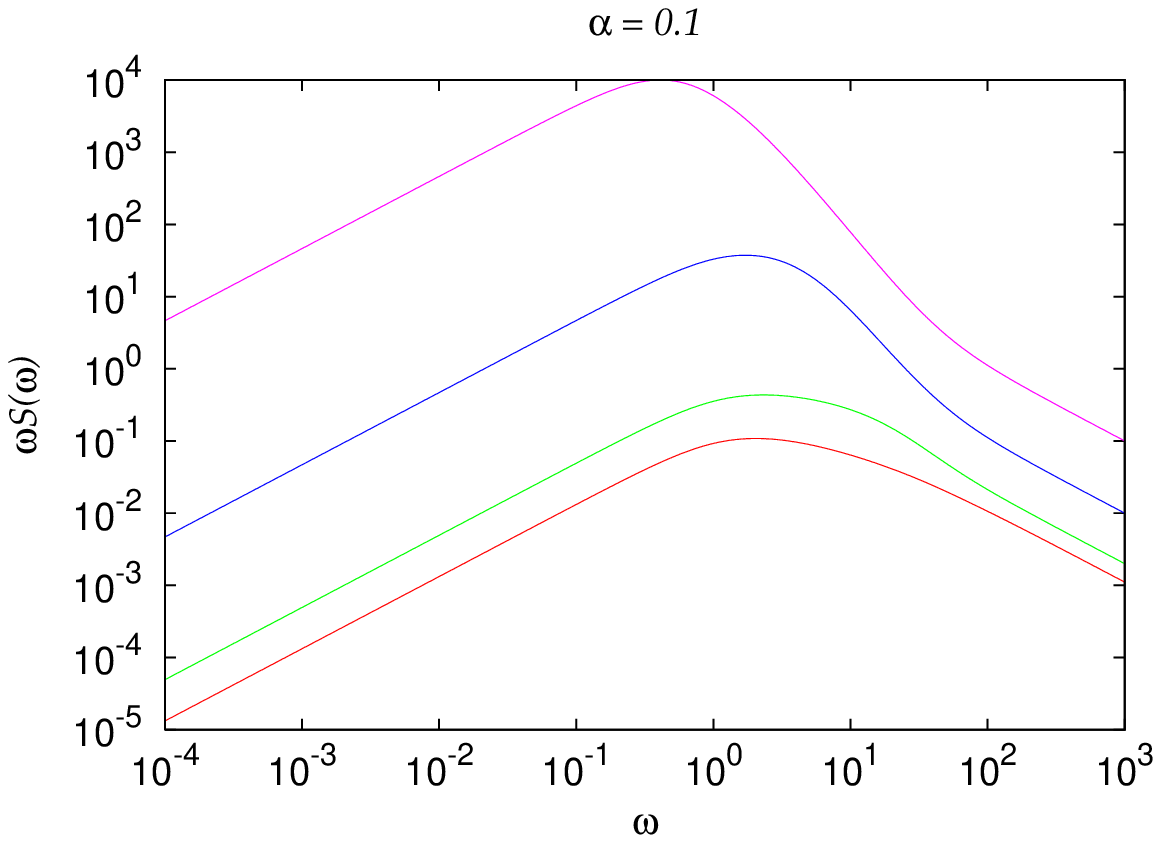}
\hfill~\\
\hfill~\includegraphics[width=0.4\textwidth]{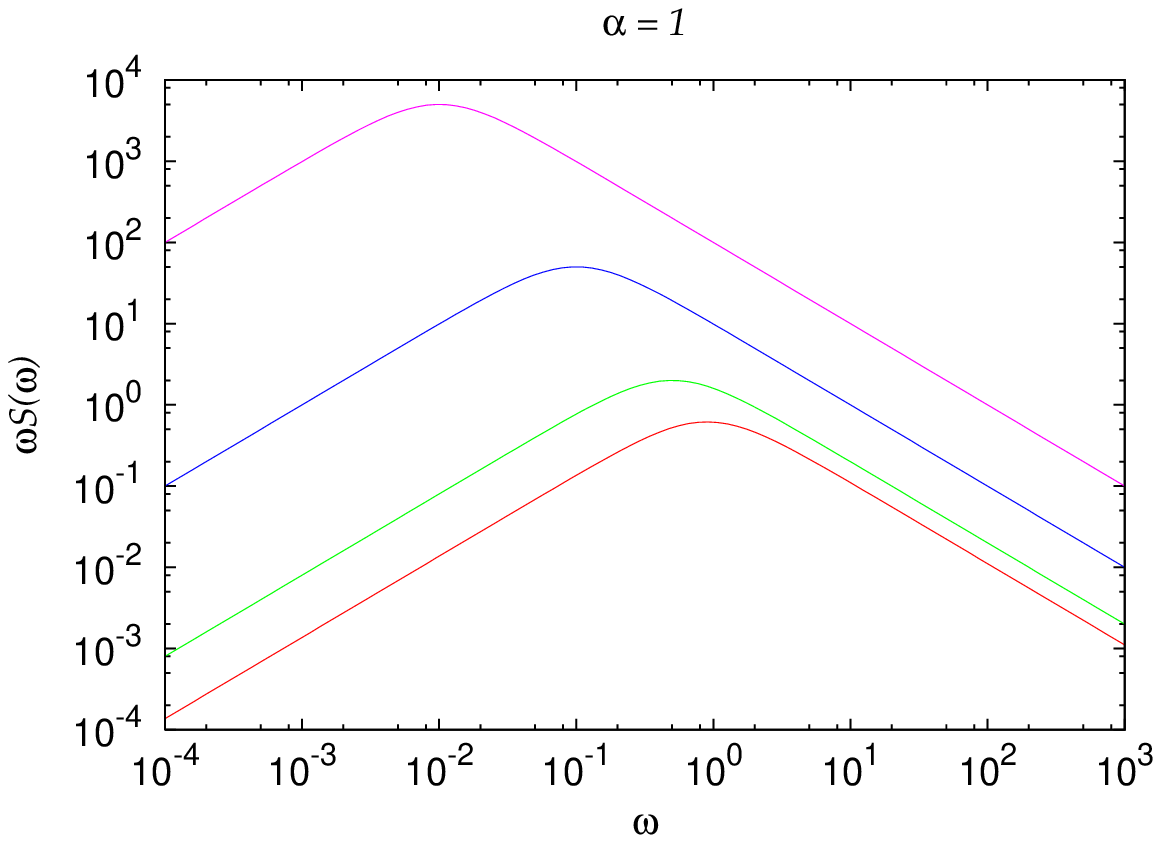}
\hfill~\includegraphics[width=0.4\textwidth]{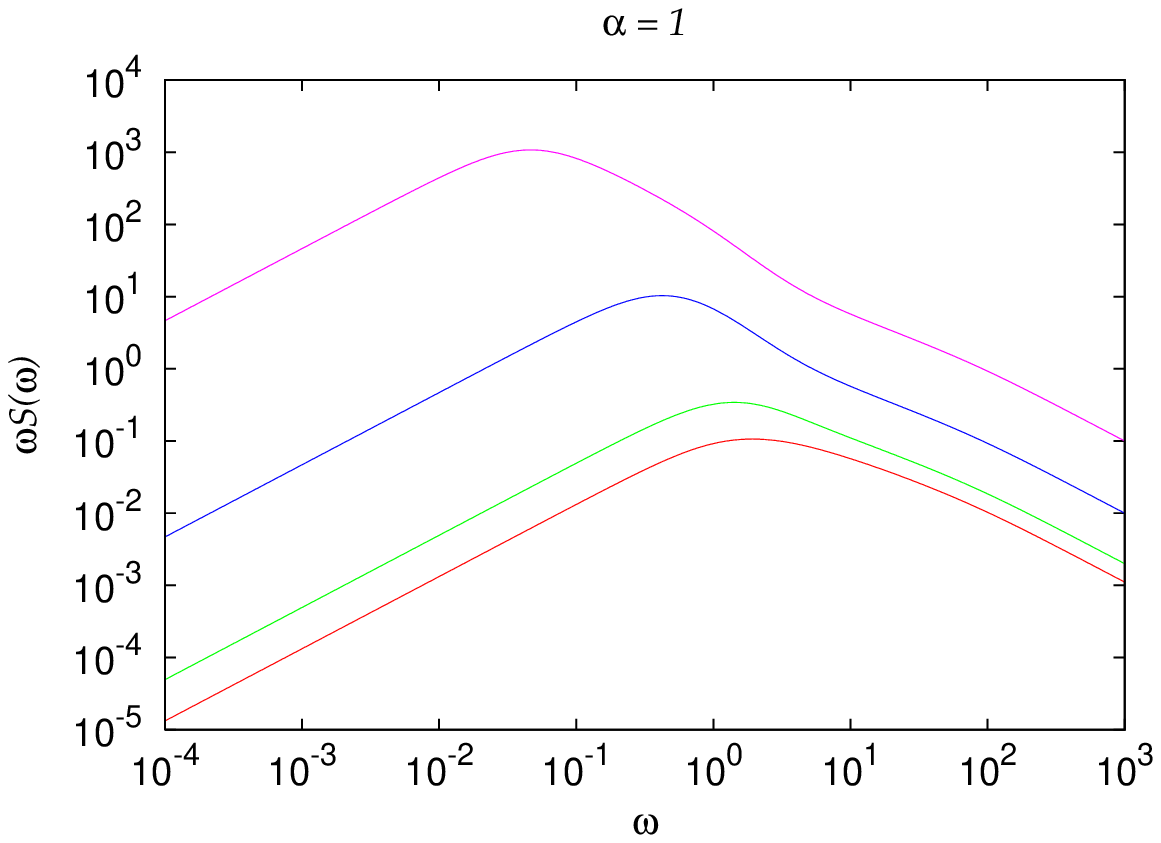}
\hfill~\\
\hfill~\includegraphics[width=0.4\textwidth]{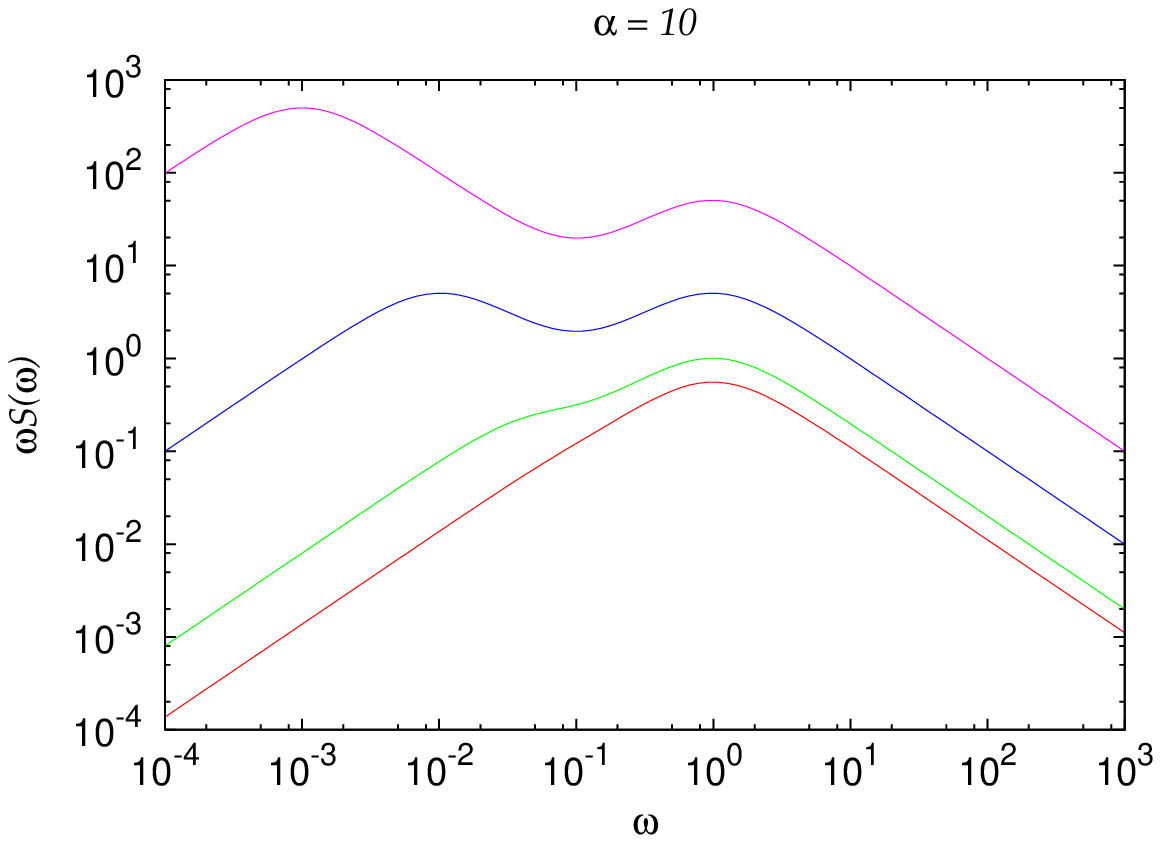}
\hfill~\includegraphics[width=0.4\textwidth]{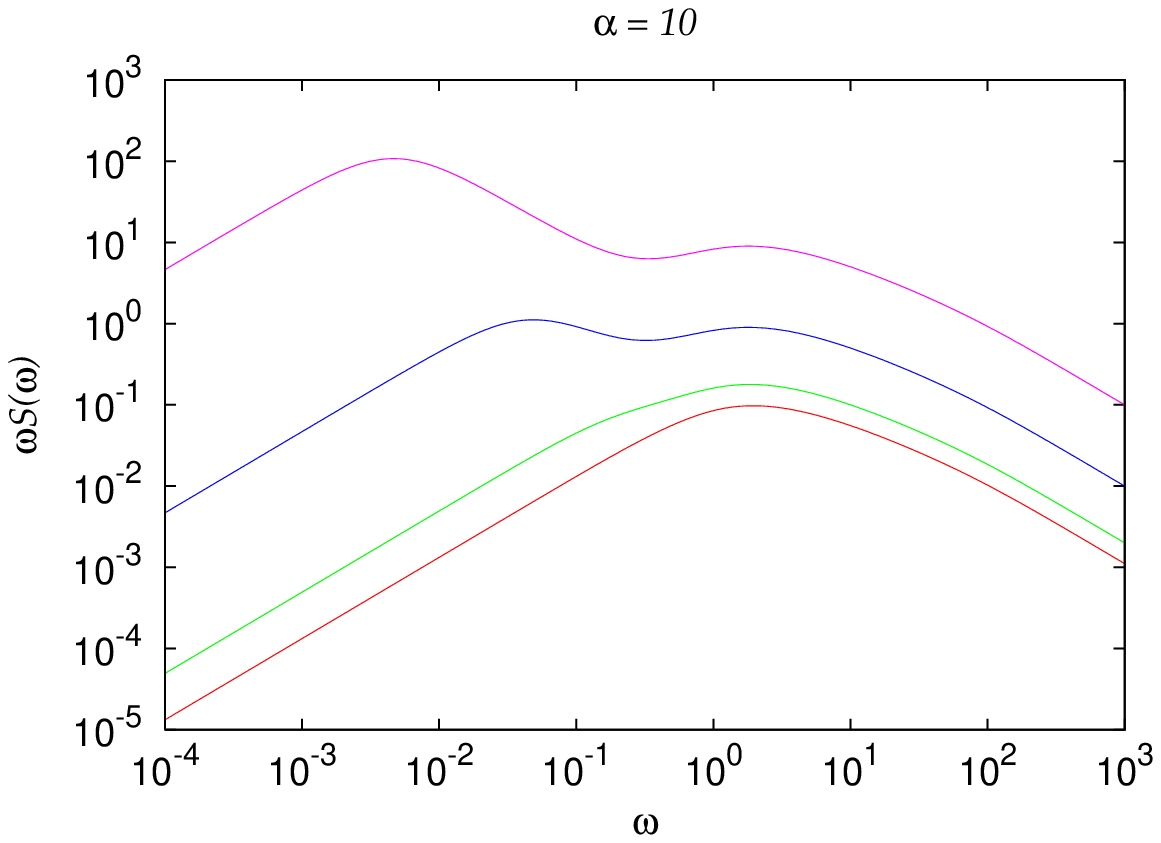}
\hfill~
\end{center}
\caption{Power spectra from the spot model in which the birth
and duration of  spots are governed by the market Hawkes process with
the exponential infectivity (\ref{expinf}). Values of $\alpha$ are given
on top of the plots. Frequency is given in  geometrical units (see the
text for its scaling to physical units). In each frame, the four curves
correspond to different values of $\nu$; from bottom  to top $\nu=0.1$,
$0.5$, $0.9$, and $0.99$. The mean number of spontaneous spots has been 
set to $\lambda=(2\pi)^{-1}$. The relative normalisation of these curves
scales as the mean number of all spots, i.e.\ proportionally to
$1/(1-\nu)$. Left: the case when all profiles $I(t)$ are identical;
$\tau=1$, i.e.\ $\zeta(\tau)=\delta(\tau-1)$. Right: the case when the
life-times of spots are  distributed uniformly, i.e.\ according to
$\zeta(\tau)=1/(\tau_{\rm max}-\tau_{\rm min})$,  between $\tau_{\rm
min}=0.01$ and $\tau_{\rm max}=1$.}
\label{haw1}
\end{figure*}

\begin{figure*}[tbh]
\begin{center}
\hfill~\includegraphics[width=0.4\textwidth]{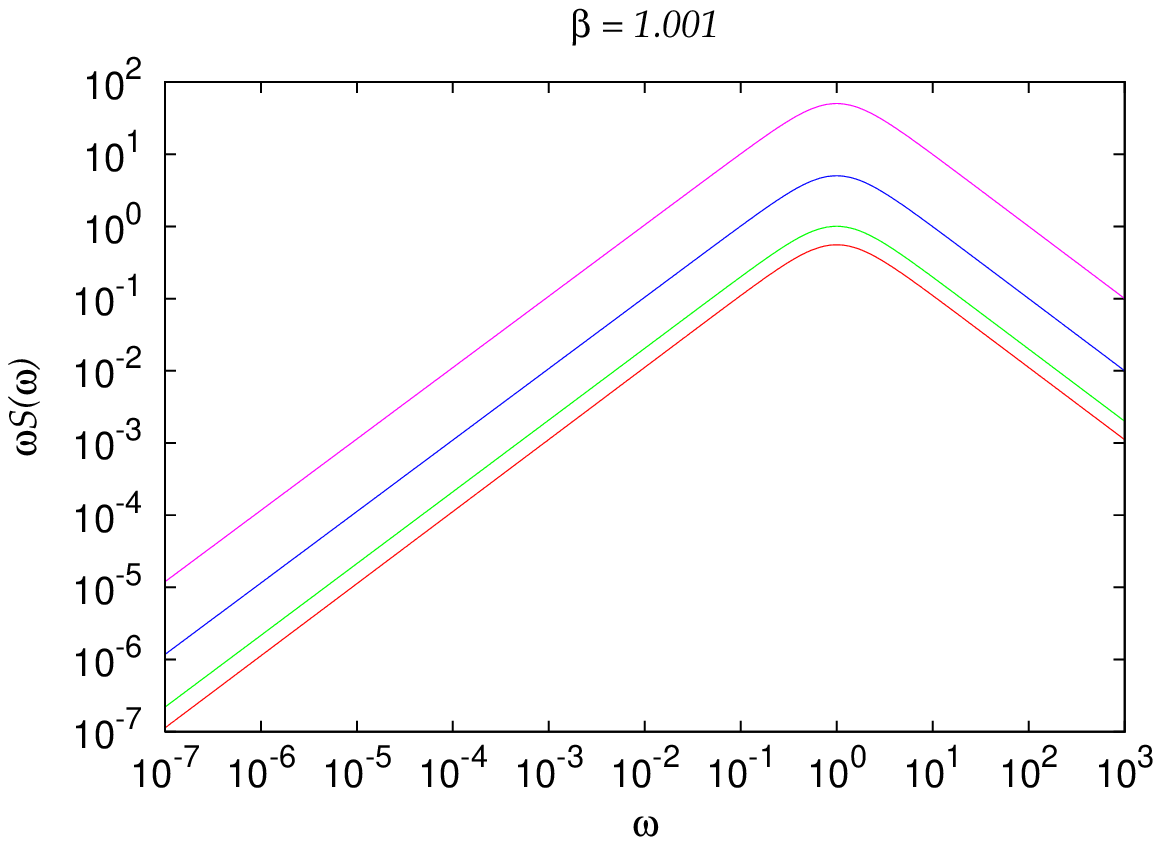}
\hfill~\includegraphics[width=0.4\textwidth]{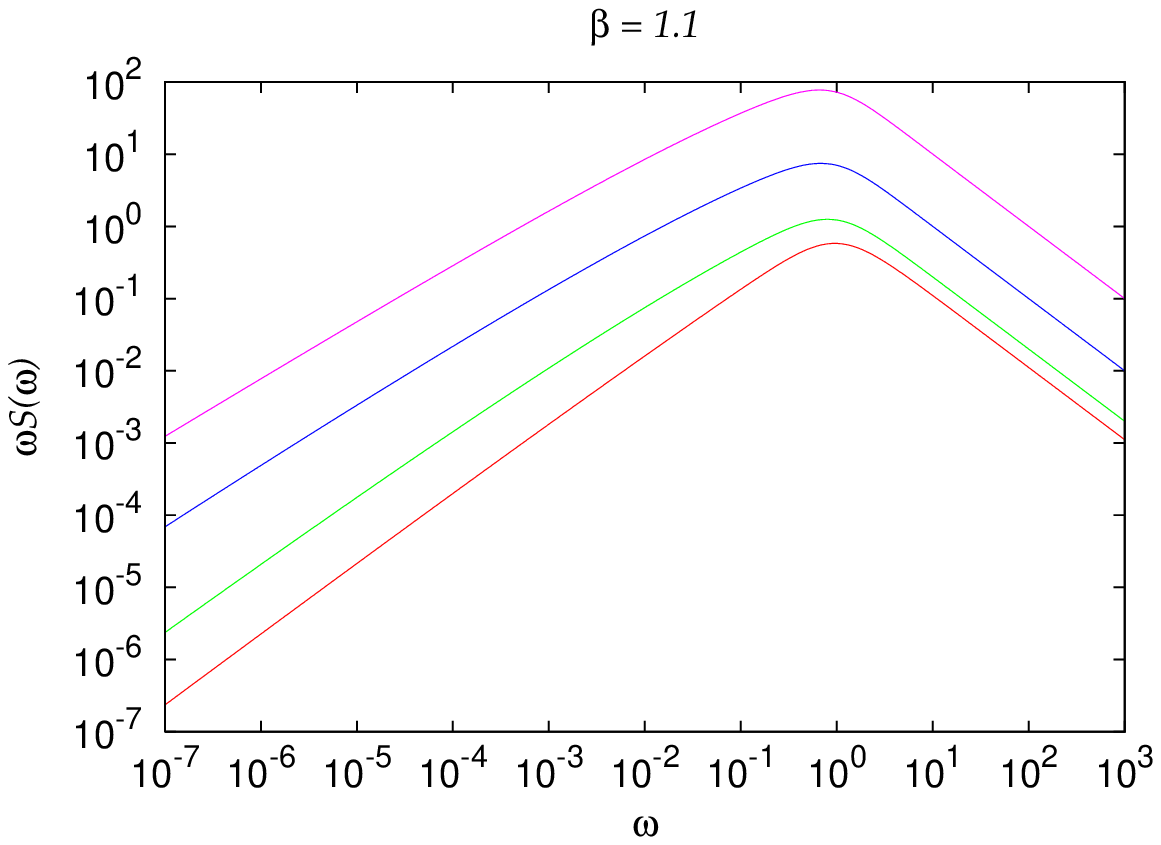}
\hfill~\\
\hfill~\includegraphics[width=0.4\textwidth]{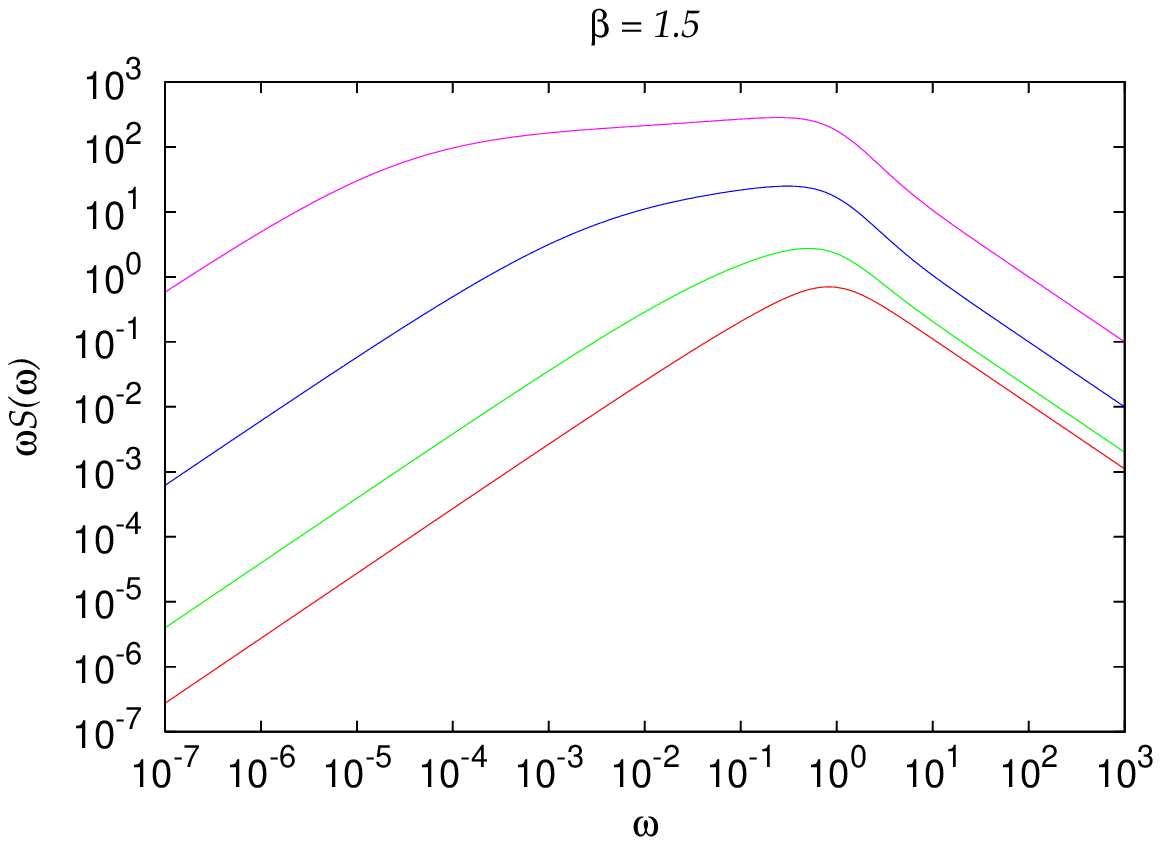}
\hfill~\includegraphics[width=0.4\textwidth]{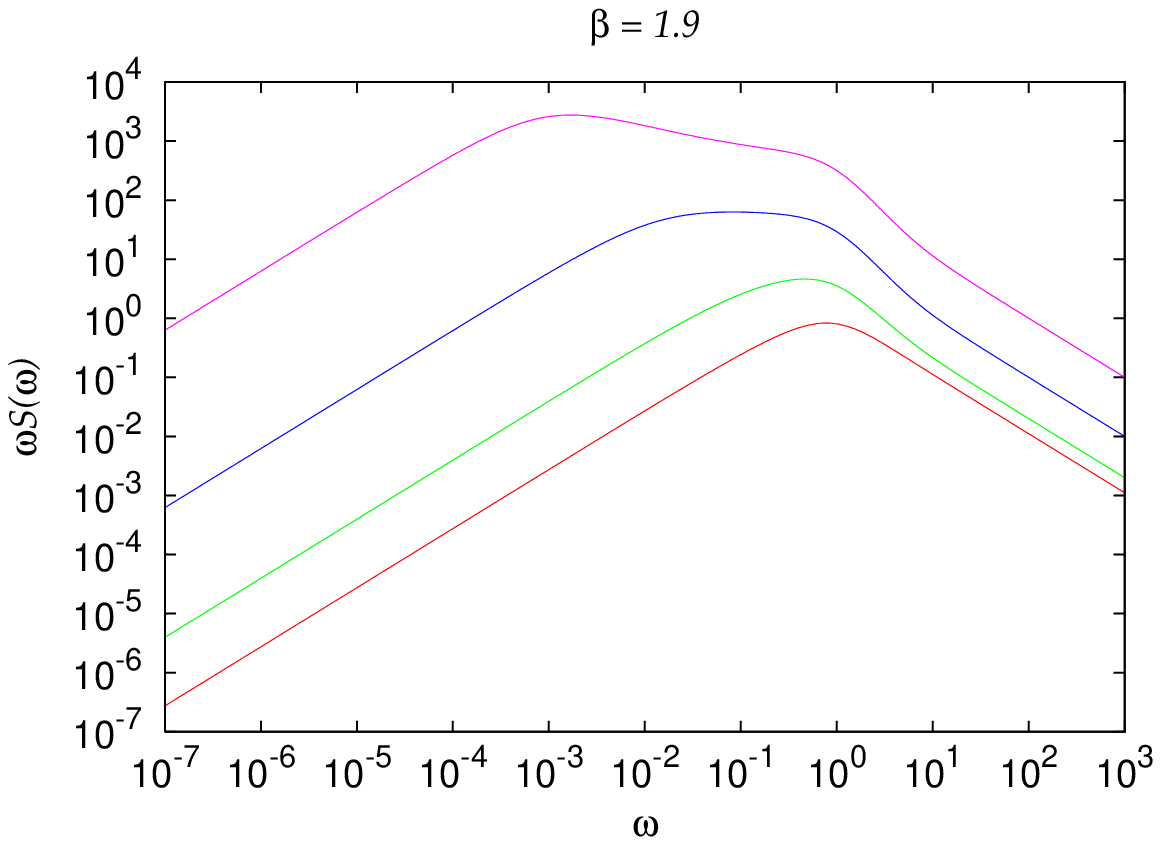}
\hfill~\\
\hfill~\includegraphics[width=0.4\textwidth]{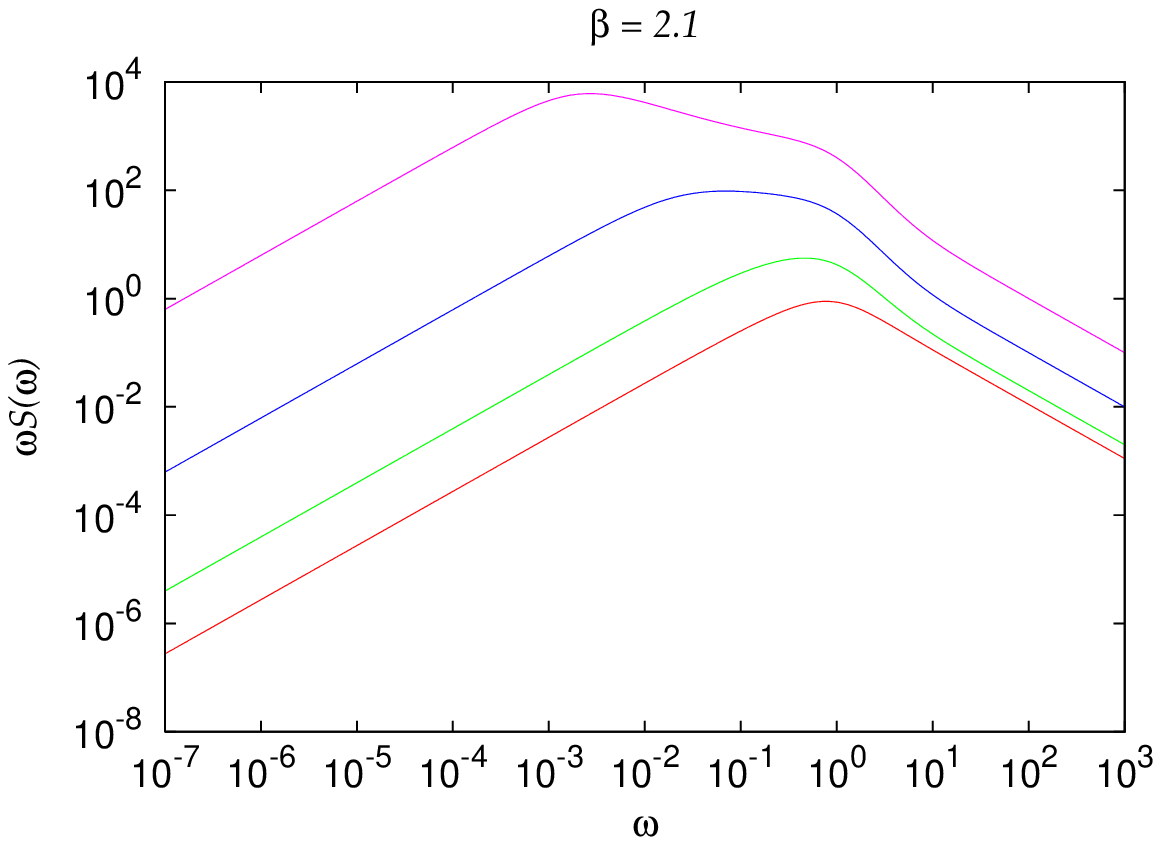}
\hfill~\includegraphics[width=0.4\textwidth]{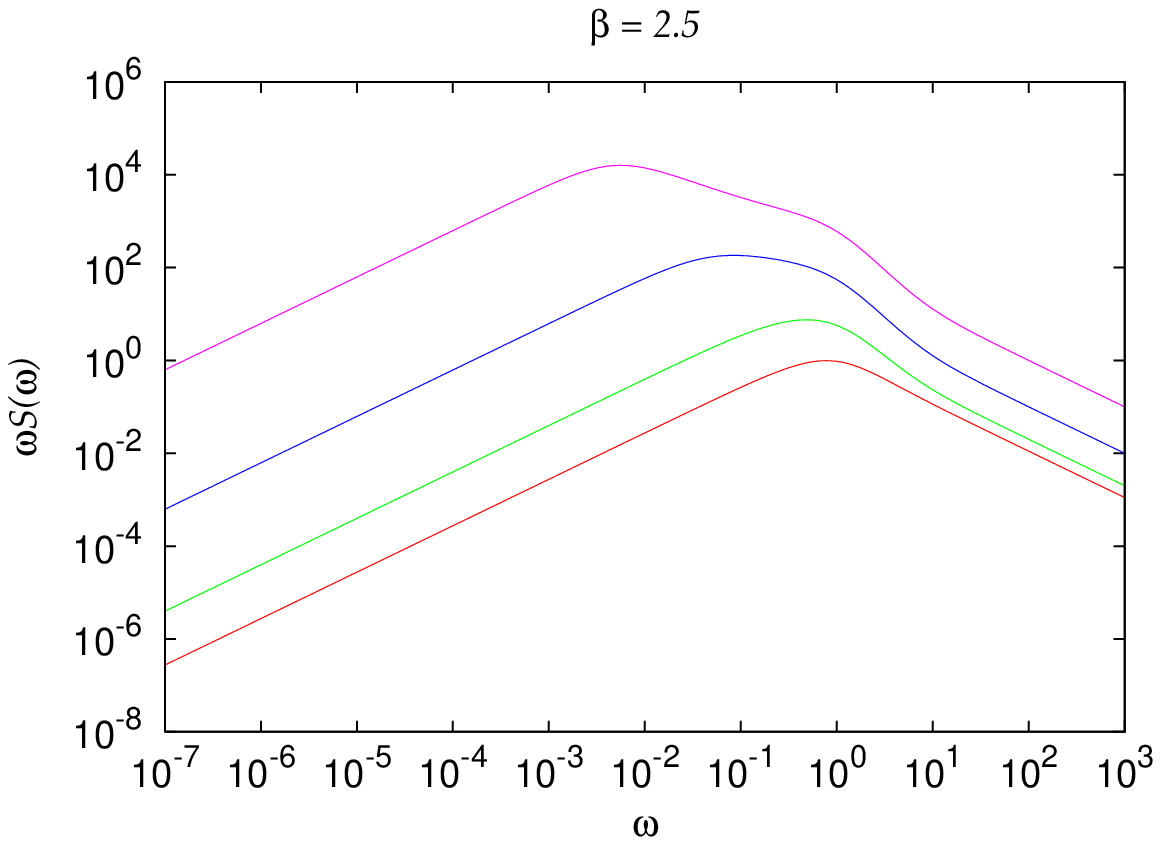}
\hfill~
\end{center}
\caption{Graphs of the model PSD as in the previous figure, but for 
the case of Hawkes process with the power law infectivity (\ref{plinf}). 
Parameter values $\beta$ are given on top of each panel; values of 
$\nu$ are as in Fig~\ref{haw1}. All spots have the same (exponential) 
profile, as in the previous figure. For $\beta\gtrsim1.5$ we notice the
new profile resembles a power law that develops in between the two
peaks. However, this situation is rather  rare within the parameter
space of our models. By parameter tuning -- e.g., by setting
$\nu\rightarrow1$, which means enhancing the contribution of avalanches
while suppressing the importance of  spontaneous parent spots -- the
extent of the flat part of the PSD profile (like the one seen for
$\beta=1.5$) can be stretched farther towards small frequencies.}
\label{haw2}
\end{figure*}

\subsection{Model driven by the Hawkes process}
\label{sec:hawkes}
Hawkes' process \citep{Hawkes:1971,Bremaud:2002:AAP} is more complicated
than the previous example because it consists of two types of events.
First, new spots are generated by the Poisson process operating with
intensity $\lambda$ (let us denote $t_0$ the moment of ignition). 
Second, an existing spot can give birth to new one at a later time, $t$,
according to the Poisson process with varying intensity 
$\mu(t-t_0)$.\footnote{Mathematically, a very similar  model has been
employed to describe the propagation of diseases through a population
\citep{Daley:2003}. In this  context, the function $\mu(t)$ is called
``infectivity'', for obvious reasons. We will adopt the same name for
it, although the medical connotation is irrelevant here.}

The mean number of events is
\begin{equation}
m(t)=\lambda+\sum\limits_{i, t_i<t}\mu(t-t_i)=\lambda+\int\limits_{-\infty}^{\infty} \mu(t-x)\,N({\rm d}x).
\label{mt}
\end{equation}

In analogy with Eq.~(\ref{et}) we define the characteristic time of
avalanche $t_{\rm a}$. It can be proven that $t_{\rm a}$ is related to
the characteristic time of the infectivity $t_{\rm i}$:
\begin{equation}
t_{\rm a}=\frac{\int\limits_0^\infty t\,m_1(t\,|\,0)\,{\rm d}t}
{\int\limits_0^\infty m_1(t\,|\,0)\,{\rm d}t}
=\frac{\nu}{1-\nu}\frac{\int\limits_0^\infty t\,\mu(t)\,{\rm d}t}
{\int\limits_0^\infty \mu(t)\,{\rm d}t}=\frac{\nu}{1-\nu}\,t_{\rm i}.
\label{ta}
\end{equation}

For a stationary process, the first-order moment density is constant. Averaging
the relation (\ref{mt}) we find
\begin{equation}
m_1=\frac{\lambda}{1-\nu},\quad \nu=\int\limits_{0}^\infty\mu(t)\,{\rm d}t.
\label{mt2}
\end{equation}
The meaning of $\nu$ is the mean number of the offsprings. Clearly, it satisfies
the normalisation $\int\mu(x)\,{\rm d}x\equiv \nu \leq 1$. 

The generating functional of the cluster of the Hawkes process fulfils
the integral equation,
\begin{equation}
\mathcal{G}\big[h(x)\,|\,0\big]
=h(0)\exp\Big\{\!-\!\int\limits_{-\infty}^{\infty} 
\left(1-\mathcal{G}\left[h(x)\,|\,y\right]\right)\,\mu(y)\,{\rm d}y\Big\}.
\end{equation}
Substituting $h(x)=1+\eta(x)$ and
expanding both sides into the series (\ref{GClusExpanze}) we find
\begin{eqnarray}
m_{[1]}(t\,|\,0)&=&\int\limits_{-\infty}^{\infty} m_{[1]}(t\,|\,y)\,\mu(y)\,{\rm d}y+\delta(t),\label{HawkClusInt}\\
m_{[2]}(t,t'\,|\,0)&=&\int\limits_{-\infty}^{\infty}  m_{[2]}(t,t'\,|\,y)\,\mu(y)\,{\rm d}y\nonumber\\
&&+m_{[1]}(t\,|\,0)\,m_{[1]}(t'\,|\,0)-\delta(t)\,\delta(t').\label{HawQ}
\end{eqnarray}

\begin{figure*}[tbh]
\begin{center}
\includegraphics[width=0.49\textwidth]{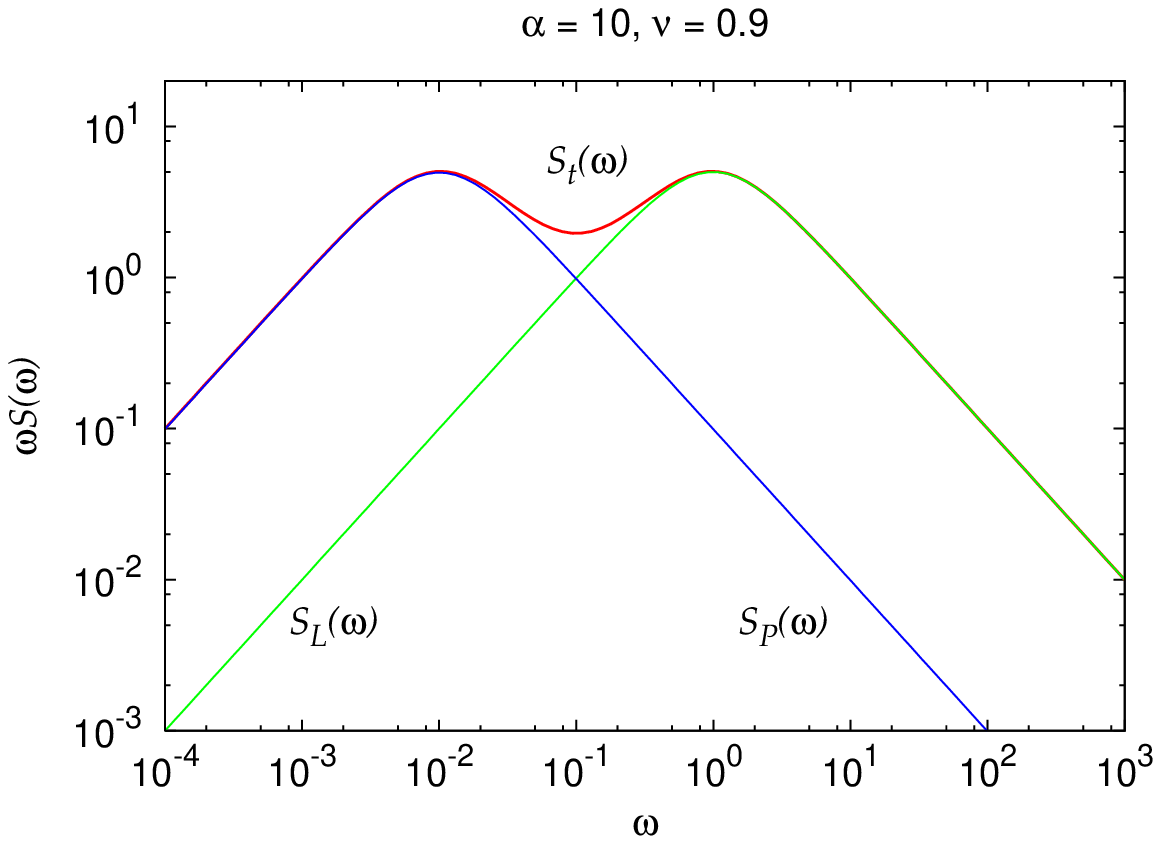}
\includegraphics[width=0.49\textwidth]{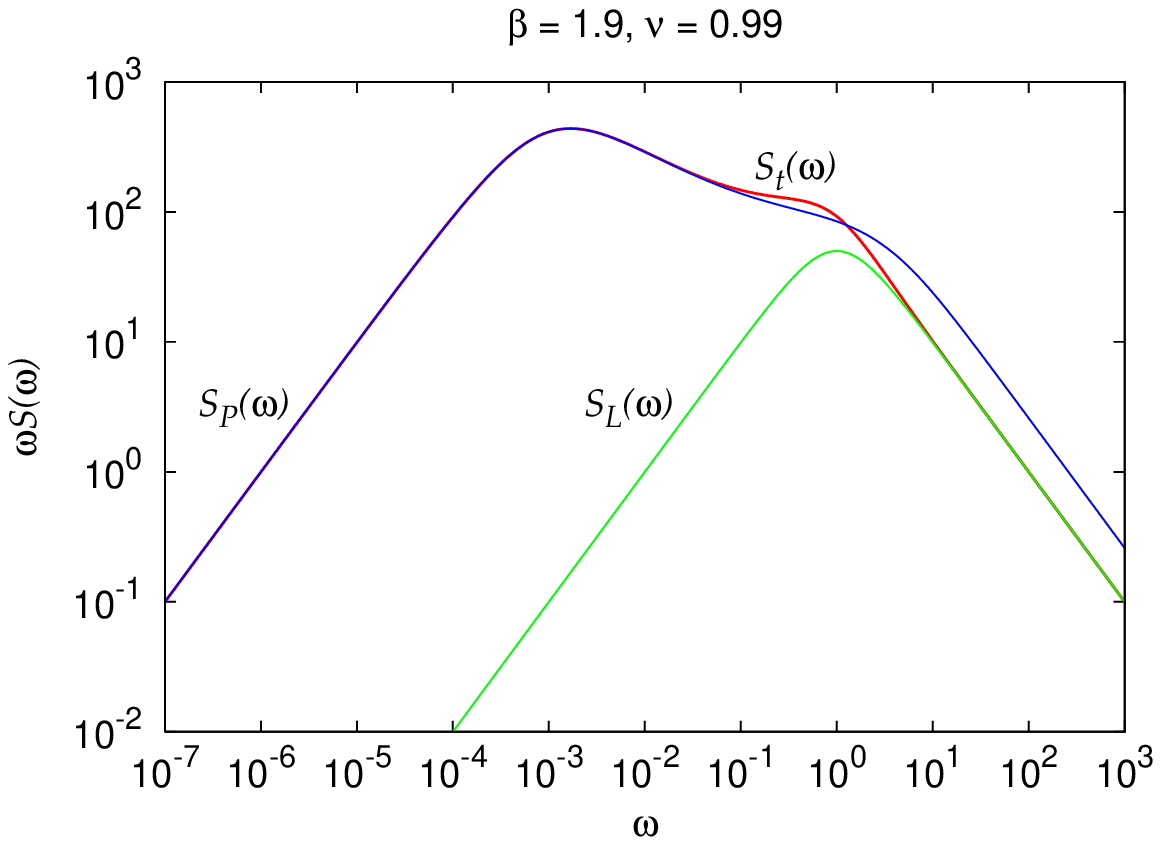}
\end{center}
\caption{Decomposition 
of the total PSD curve, $S_{\!\rm t}(\omega)$, into a  product of two
terms which are responsible for the two peaks in the final profile (see
Eq.~(\ref{HawkPSD4})). Left: this case corresponds to $\alpha=10$,
$\nu=0.5$ curve  shown in the bottom--left panel of Fig.~\ref{haw1}.
Right: this case corresponds to $\beta=1.9$ and $\nu=0.99$ curve  in the
middle--right panel of Fig.~\ref{haw2}. Similar behaviour of the PSD
profile is typical in our model, although in some cases only one peak
dominates the spectrum whereas the other one is suppressed.}
\label{peakshift}
\end{figure*}

To complete the calculation we solve the integral equation
(\ref{HawkClusInt}) for $m_{[1]}(x\,|\,0)$. Because this is a  linear
convolutional integral equation, it can be solved  efficiently by using
the Fourier transform:
\begin{eqnarray}
\tilde{\mu}(\omega)&=&\int\limits_{-\infty}^{\infty} e^{-i\omega t}\mu(t)\,{\rm d}t,\\
\tilde{m}_{[1]}(\omega\,|\,0)&=&\int\limits_{-\infty}^{\infty} e^{-i\omega t}m_{[1]}(t\,|\,0)\,{\rm d}t
=\frac{1}{1-\tilde{\mu}(\omega)}.
\end{eqnarray}
For the Fourier transform of the quadratic factorial measure we find
\begin{equation}
\tilde{m}_{[2]}(\omega,\omega'\,|\,0)
=\frac{\tilde{m}_{[1]}(\omega\,|\,0)\tilde{m}_{[1]}(\omega'\,|\,0)-1}
{1-\mu(\omega+\omega')}.
\end{equation}
Again, the PSD is given by Eqs.~(\ref{HawkPSD}), (\ref{HawkPSD2}), 
and (\ref{mt2}) with
\begin{equation}
S_{\rm\! P}(\omega)=\lambda\,\frac{|\tilde{m}_{[1]}(\omega\,|\,0)|^2-1}{1-\nu}.
\end{equation}
Comparing this equation describing the Hawkes process PSD with the
corresponding  Eq.~(\ref{mlsp1}) for the case of the Chinese process, we
reveal a subtle difference between the two mechanisms. It turns out that
the high-frequency limit is identical for both of them, however, the
difference grows  as one proceeds towards the low-frequency end of the
PSD domain.

\begin{figure*}[tbh]
\begin{center}
\hfill~\includegraphics[width=0.4\textwidth]{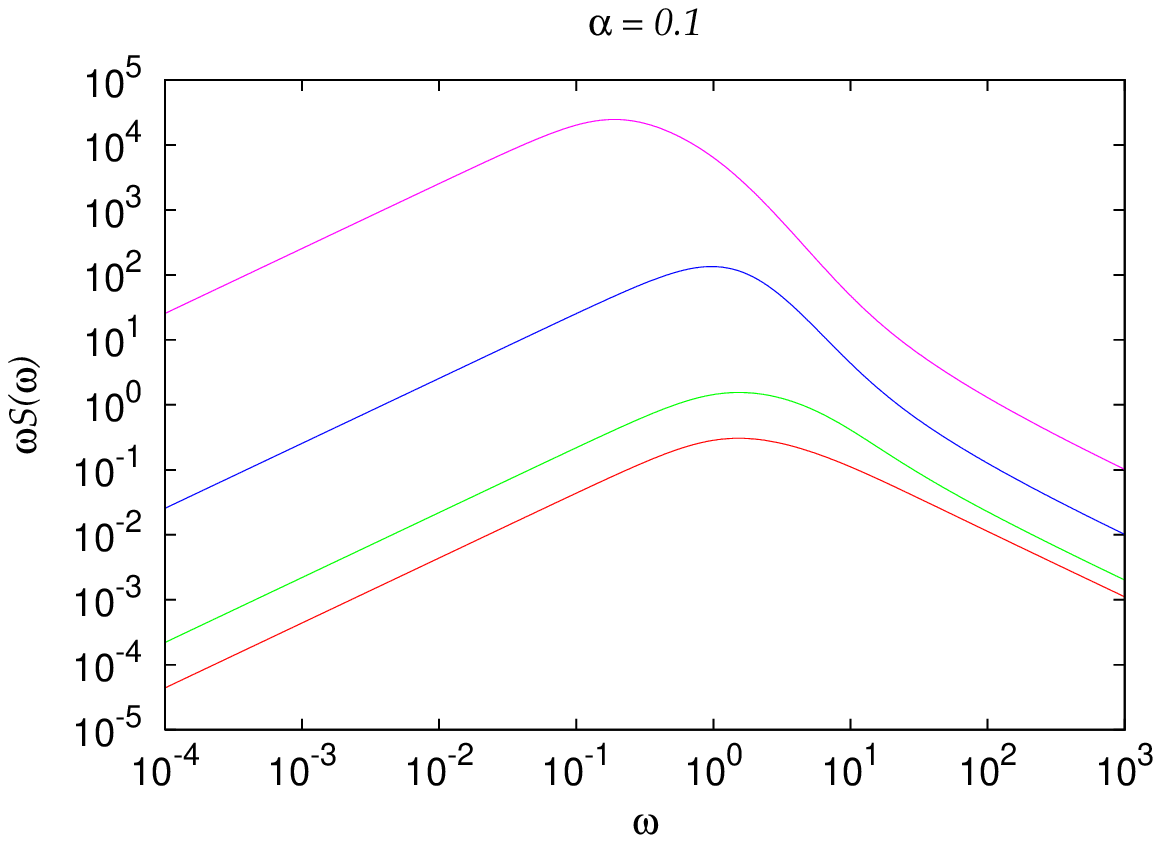}
\hfill~\includegraphics[width=0.4\textwidth]{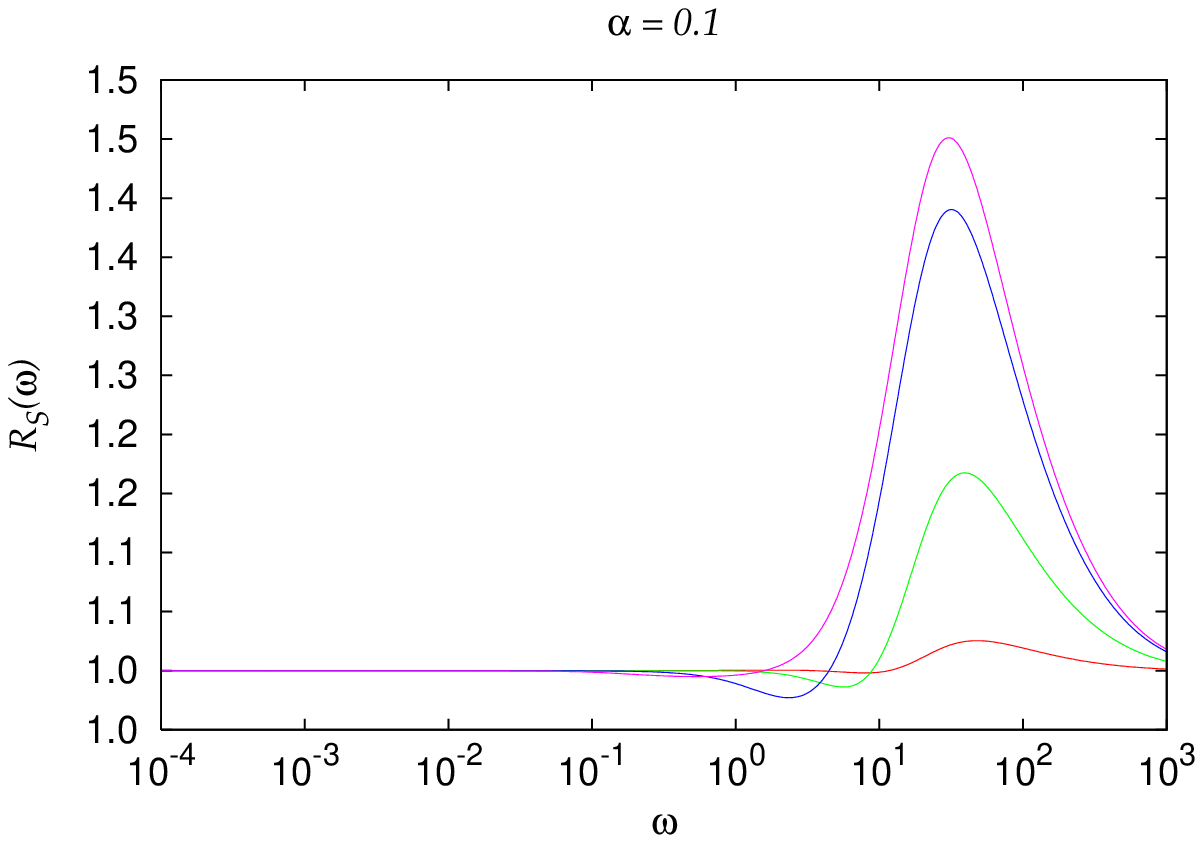}
\hfill~\\
\hfill~\includegraphics[width=0.4\textwidth]{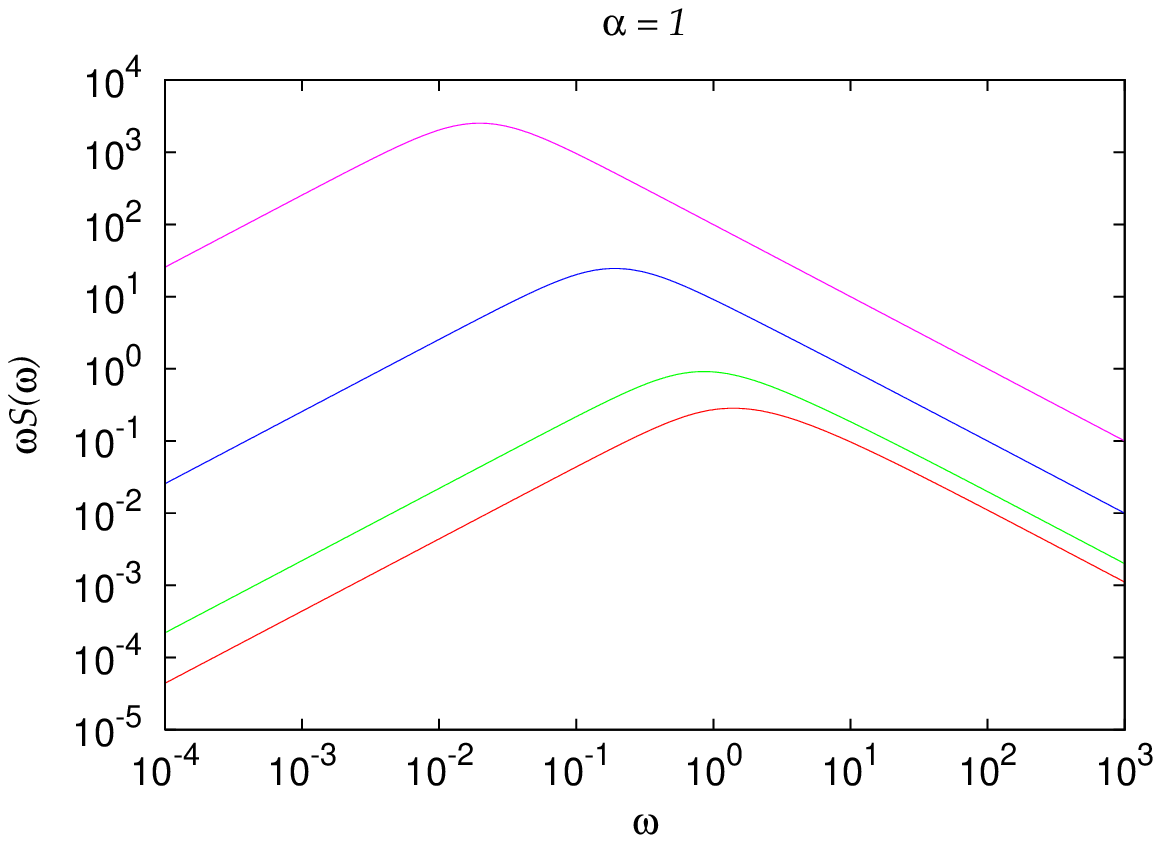}
\hfill~\includegraphics[width=0.4\textwidth]{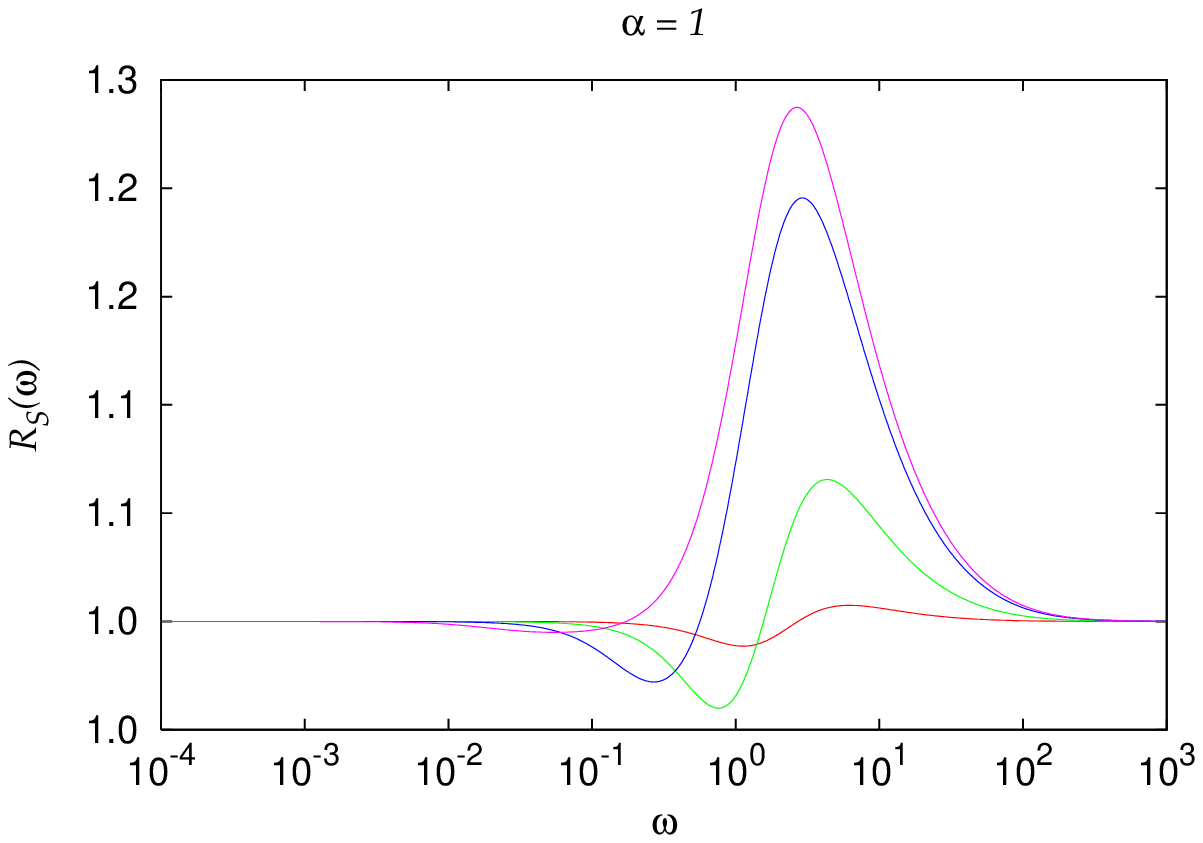}
\hfill~\\
\hfill~\includegraphics[width=0.4\textwidth]{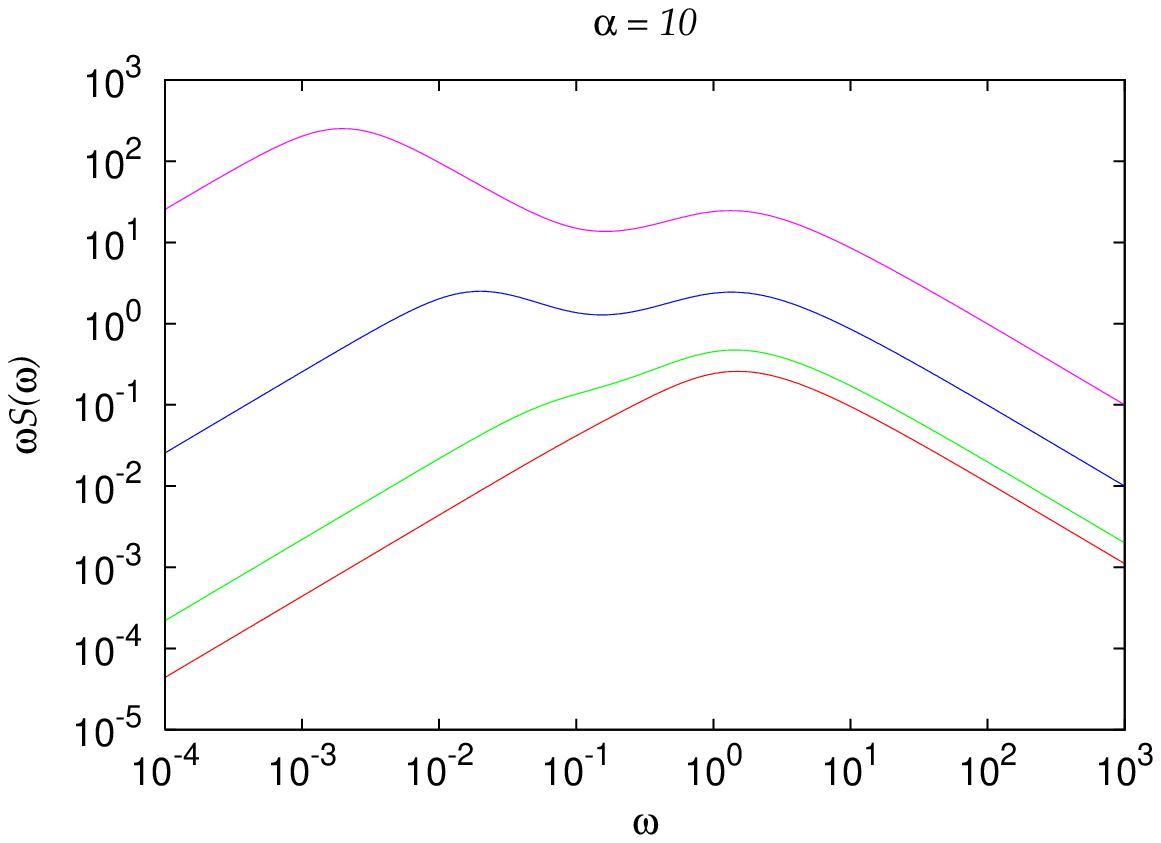}
\hfill~\includegraphics[width=0.4\textwidth]{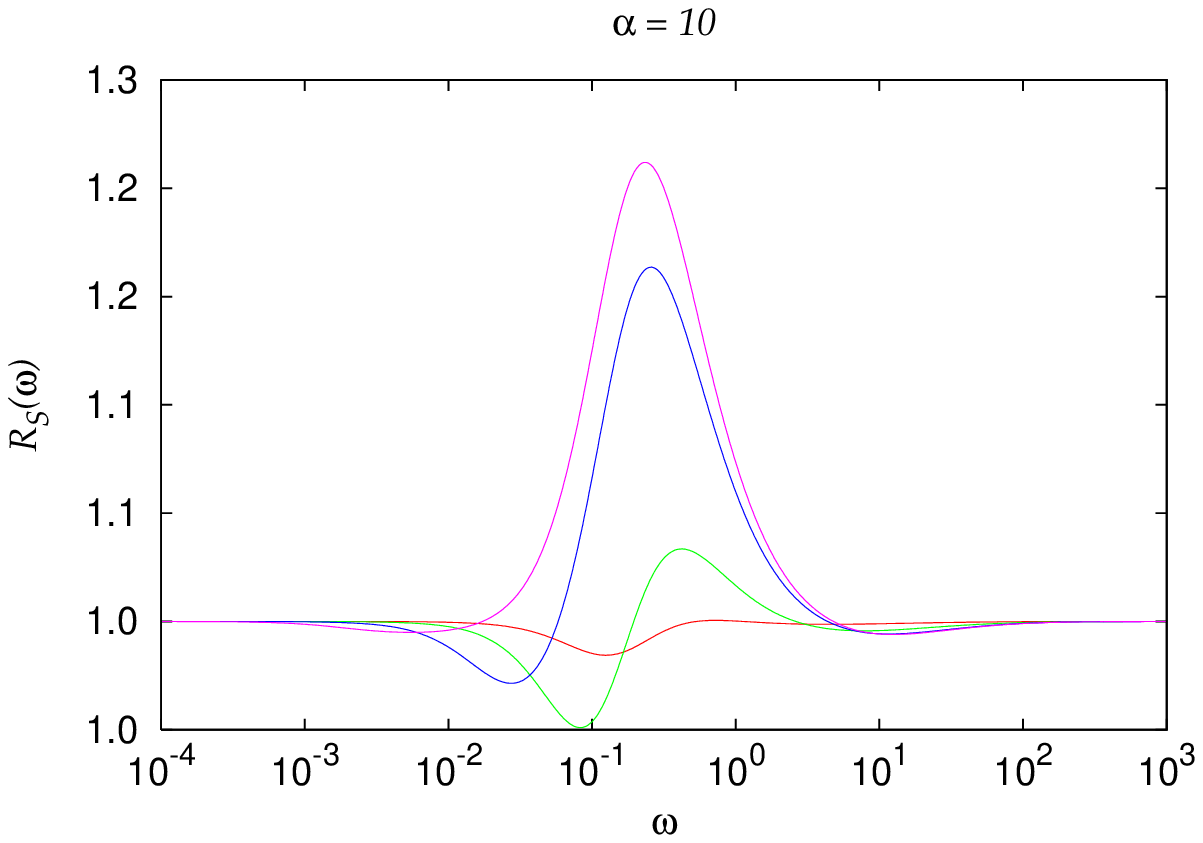}
\hfill~
\end{center}
\caption{The pulse avalanche model for the uniform probability density
$\zeta(\tau)$, constrained by limits $\tau_{\rm min}=0.01$ and
$\tau_{\rm max}=1$. Left: the PSD calculated from the formulae
(\ref{HawkPSD}) and (\ref{paPSD}). Right: the ratio between the PSD of
the pulse avalanche process to the Hawkes process; in the latter the
exponential infectivity $\mu(t)$ is related to the corresponding
infectivity of  the pulse avalanche process by 
$\mu(t)=\mu(t\,|\,\bar{\tau})$.}
\label{haw4}
\end{figure*}

For the exponential form of infectivity measure,
\begin{equation}
\mu(t)=\nu\sigma\,e^{-\sigma t}\,\theta(t),
\label{expinf}
\end{equation}
we obtain the explicit expression of
\begin{eqnarray}
\tilde{m}_{[1]}(\omega\,|\,0)&=&1+\frac{\nu\sigma}{\sigma(1-\nu)+i\omega},\\
S_{\rm\! P}(\omega)&=&\frac{\nu\,(2-\nu)\,\sigma^2}{(1-\nu)^2\,\sigma^2+\omega^2}\,\frac{\lambda}{1-\nu},
\label{sp1}
\end{eqnarray}
where $\sigma>0$ is a constant. 

Figure~\ref{haw1} shows the resulting PSD of this model in a logarithmic
plot of $\omega S(\omega)$ versus $\omega$. Here we can study the occurrence
of break frequency where the PSD slope changes depending on the model
parameters. 
Light curve profiles of individual spots were chosen as exponentials, 
$I(t)=I_0\exp(-t/\tau)\,\theta(t)$, where
$\tau$ is random value with probability density $\zeta(\tau)$ and the
mean $\bar{\tau}$. The
characteristic time of infectivity, $t_i= \sigma^{-1}$, is set to be
$t_i=\alpha\,\bar{\tau}$. In general, we can identify two
characteristic frequencies in the PSD. The first one corresponds to the
characteristic frequency of the profile $I(t)$ (in our case this
frequency is given  by $1/\bar{\tau}$), the second one is given by
the characteristic frequency of the  avalanches ($1/t_{\rm a}$).

We remind the reader that this plot (as well as the subsequent figures
\ref{haw2}--\ref{haw4}) does {\em not} include general relativity
effects; they will be recovered later in the paper. This is merely to
simplify calculations -- the relativistic effects  complicate the
derivation of the analytical formula for the PSD and it is somewhat
difficult to distinguish them from the intrinsic properties of the
signal. 

The typical form of the model PSD resembles a Lorentzian or
doubly-Lorentzian profile at central frequencies. 
In most cases, there is only one
break in the spectra corresponding to the characteristic frequency of
the profiles $I(t)$.
When two peaks are
visible, as we notice from Fig.~\ref{haw1} and Eq.~(\ref{sp1}), two
characteristic frequencies occur. The upper frequency is scaled to unity
in our figures; this peak corresponds to the e-folding time of the
exponentially decaying spots, i.e.\ 
$I(t)\propto\exp(-t/\tau)$.\footnote{A single exponentially-decaying
pulse is described by function $I(t,\xi)=I_0\,e^{-t/\tau}\theta(t)$.
Notice that in this simple case the only parameter, $\tau\equiv\xi$, has
a meaning of the decay timescale.} The lower frequency peak is then set
by a combination of two timescales,
\begin{equation}
\frac{\bar{\tau}}{t_{\rm a}}=\frac{1-\nu}{\alpha},
\label{combination}
\end{equation} 
where $\bar{\tau}\equiv{\rm E}[\tau]$. Notice that the model is
sufficiently complex, and so one cannot easily disentangle the two
timescales in any straightforward way directly from the PSD.

Another natural choice of infectivity is a power-law function of the form
\begin{equation}
\mu(t)=\frac{K}{(t+L)^\beta}\,\theta(t),
\label{plinf}
\end{equation}
where $\beta$, $K$, and $L$ are positive constants satisfying
\begin{equation}
\beta>1, \quad K=\nu\,(\beta-1)\,L^{\beta-1}.
\label{plinf2}
\end{equation}
One can thus ask how the PSD form depends on the assumptions about the
form of the infectivity function $\mu$. For every $b>0$ the function
(\ref{plinf}) satisfies $\left.\mu(t)\right|_{L=b} 
=b^{-1}\left.\mu(t/b)\right|_{L=1}$. Therefore, we can set $L=1$ without
any loss  of generality (the value of $L$ can be  recovered by a simple
rescaling of the result). 

We note that the characteristic times $t_{\rm a}$
and $t_{\rm i}$, as defined in Eq.\ (\ref{ta}), diverge for the
infectivity (\ref{plinf}) and  $1<\beta\leq2$. In this case, $t_{\rm a}$
does not exist. Again, we derive an analytical form of the PSD for the
adopted infectivity function. The procedure is entirely analogical as
above, but we do not give the explicit  form of $S_{\rm\! P}(\omega)$
because the final formula is rather complicated.
The resulting PSD curves with the power-law infectivity 
are plotted in Fig.~\ref{haw2}. 

It is also interesting to note at this point, how the peaks of the PSD
move when the model parameters are shifted. For example, see the panel
$\alpha=1$ of Fig.~\ref{haw1}. Although in this case the two timescales
are equal to each other, changing the other parameter, $\nu$, from
$0.1$ to $0.99$ brings the peak over two orders of magnitude. In other
words, the maximum of the PSD can appear at a frequency lower than the
inverse of the spot decay time. The frequency shift of the peaks
is again given by factor $t_{\rm a}/\bar{\tau}$, as explained in 
Eq.~(\ref{combination}).

In order to understand better the behaviour of the PSD peaks, we rewrite 
Eq.~(\ref{HawkPSD}) in the form
\begin{equation}
S(\omega)=m_1\,{\rm E}\big[\big|\mathcal{F}[I](\omega)\big|^2\big]+
S_{\rm P}(\omega)\;\big|{\rm E}\big[\mathcal{F}[I](\omega)\big]\big|^2
\label{HawkPSD3}
\end{equation}
(i.e., relativistic effects neglected). Equation~(\ref{HawkPSD3}) can be 
simplified by assuming that all spots have the identical, exponentially
decaying light curves, $I(t,\tau)=I_0\,e^{-t/\tau}\,\theta(t)$. We obtain
\begin{equation}
S(\omega)=S_{\rm L}(\omega)\,\big(m_1+ S_{\rm P}(\omega)\big),
\label{HawkPSD4}
\end{equation}
where $S_{\rm L}(\omega)=\tau^2/(1+\omega^2\tau^2)$ is the Lorentzian
PSD, and $m_1$ is a constant (see the discussion following
Eq.~(\ref{ShiftInvar})). In this simple case the two terms, $S_{\rm
L}(\omega)$ and $S_{\rm P}(\omega)$, give rise to two comparably
significant peaks, although their amplitudes  are generally different. 
The situation is shown in Fig.~\ref{peakshift}. The left panel is 
particularly transparent because it shows the Hawkes process with
exponential infectivity, for which $S_{\rm P}(\omega)$ adopts the
Lorentzian shape, dominating the spectrum at frequencies where $S_{\rm
L}(\omega)\approx1$. Therefore, Eq.~(\ref{HawkPSD4}) is effectively a
sum of two identical Lorentzians. However, this example does not apply
to a general case when the amplitudes of the PSD peaks can be very
different.  In the right panel, we note that $S_{\rm{}P}(\omega)$ itself
dominates the whole spectrum and has a complicated (non-Lorentzian)
shape.

\begin{figure*}[tbh]
\begin{center}
\includegraphics[width=0.32\textwidth]{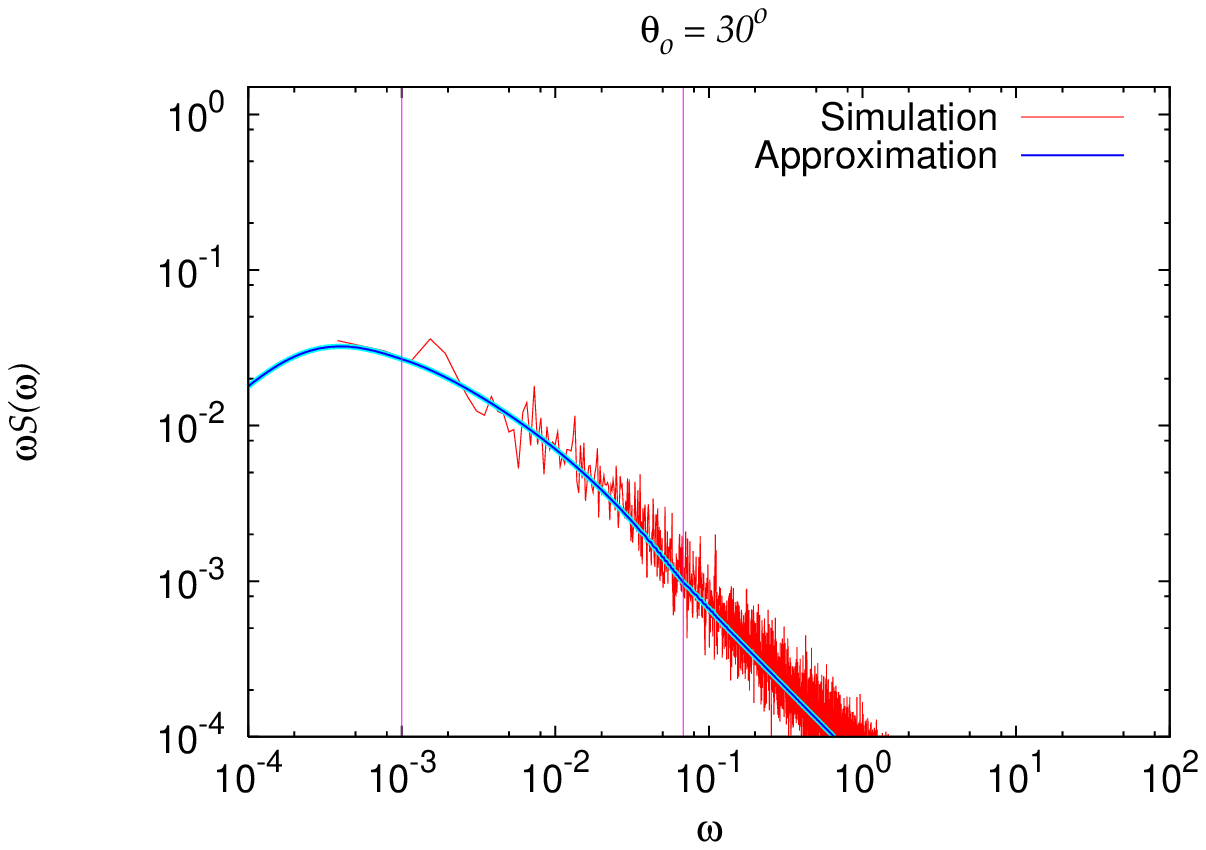}
\includegraphics[width=0.32\textwidth]{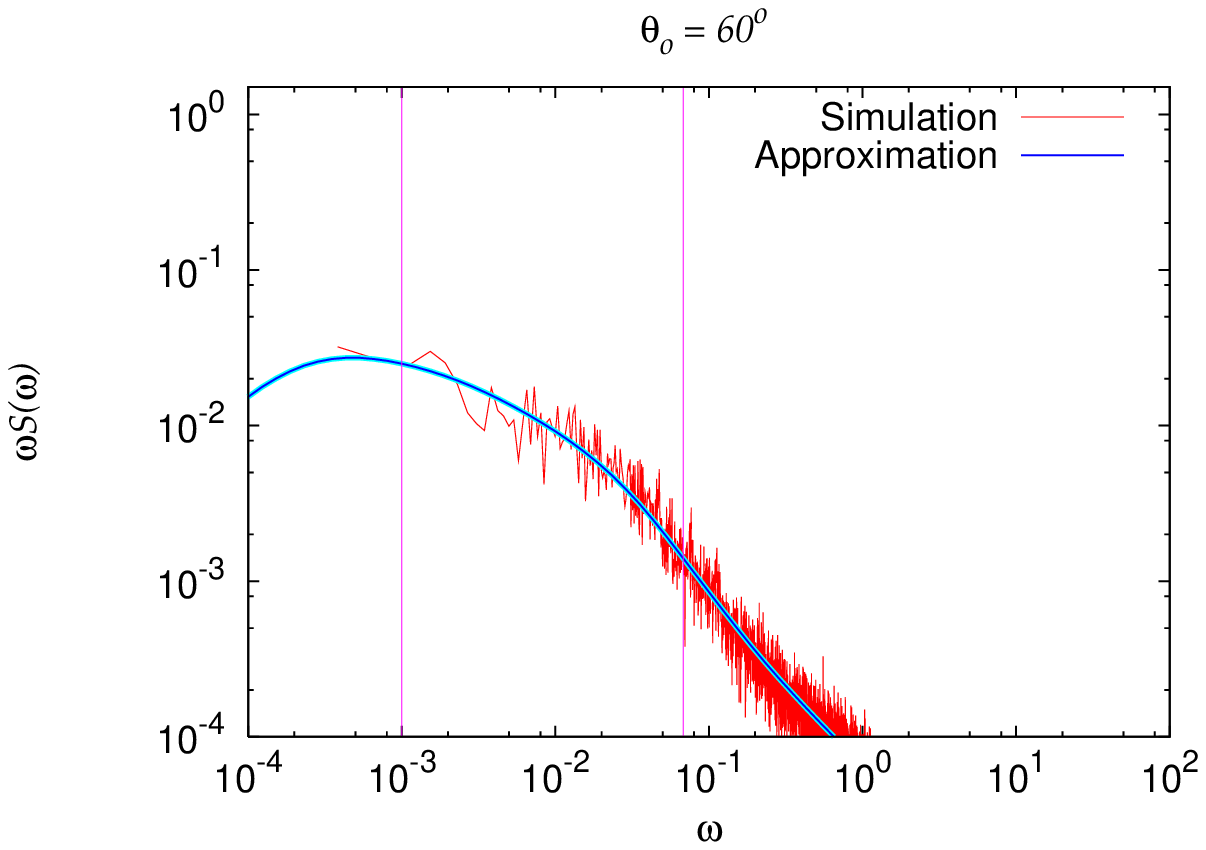}
\includegraphics[width=0.32\textwidth]{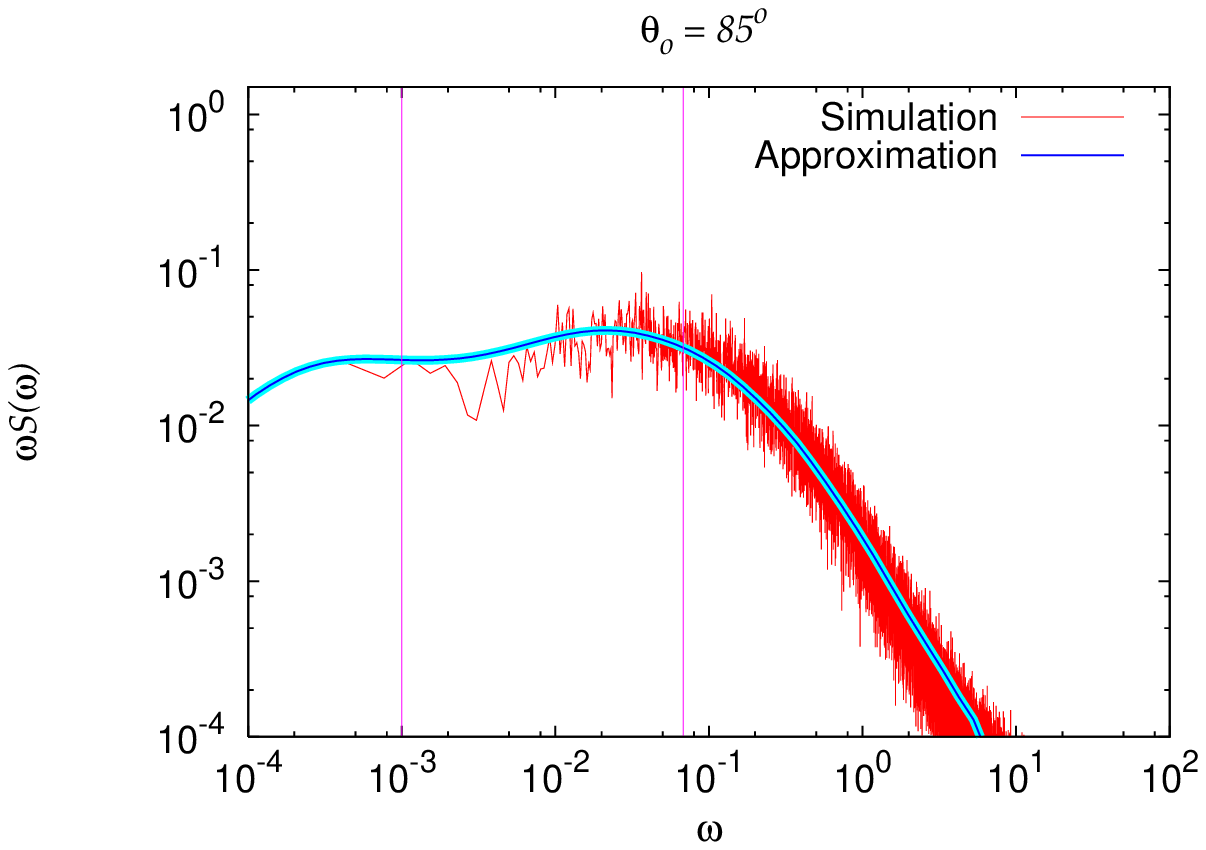}
\includegraphics[width=0.32\textwidth]{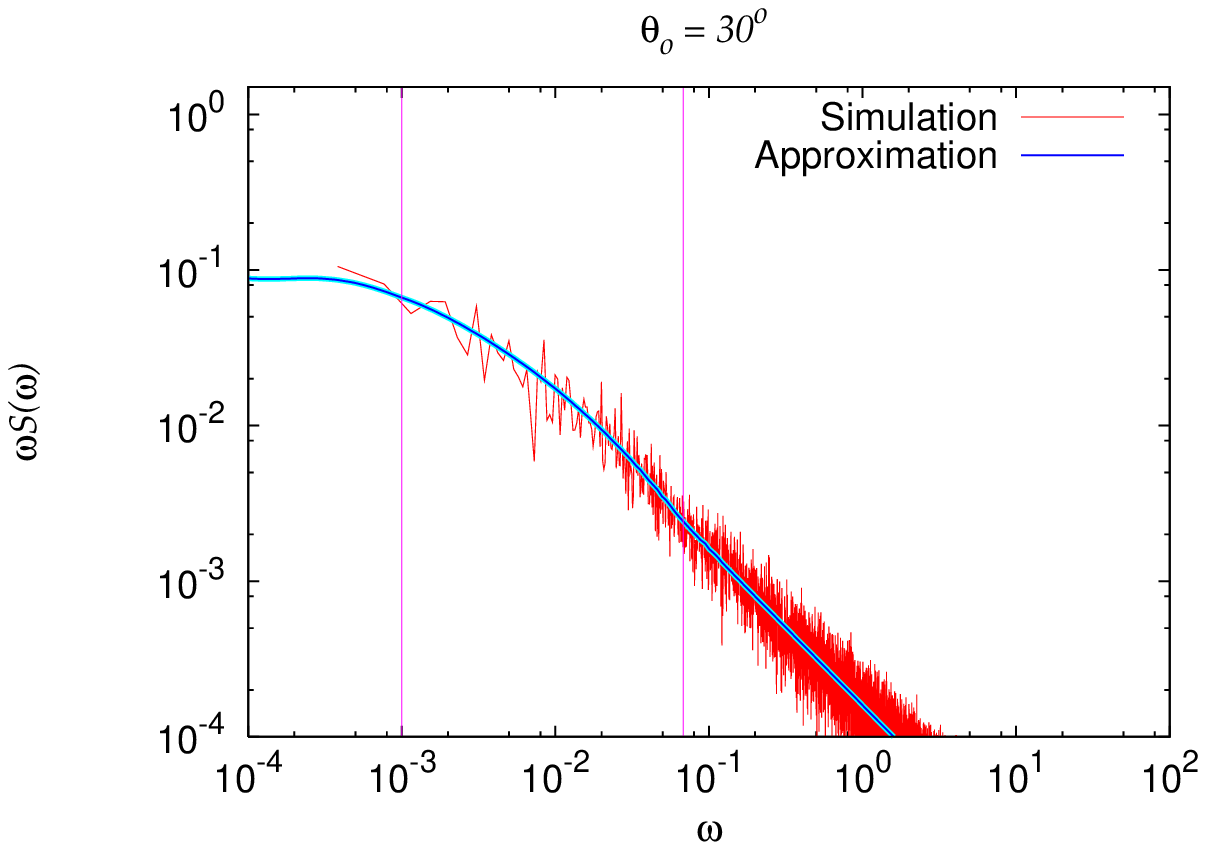}
\includegraphics[width=0.32\textwidth]{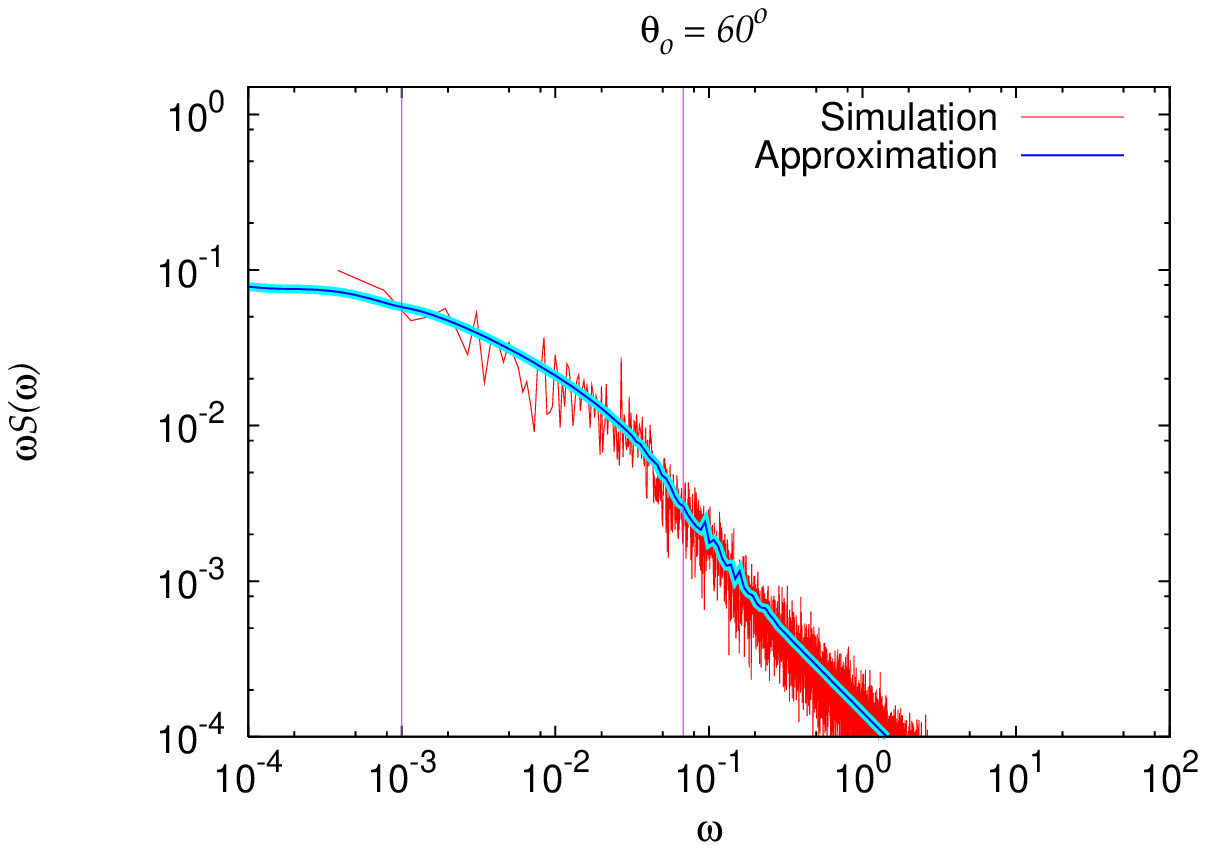}
\includegraphics[width=0.32\textwidth]{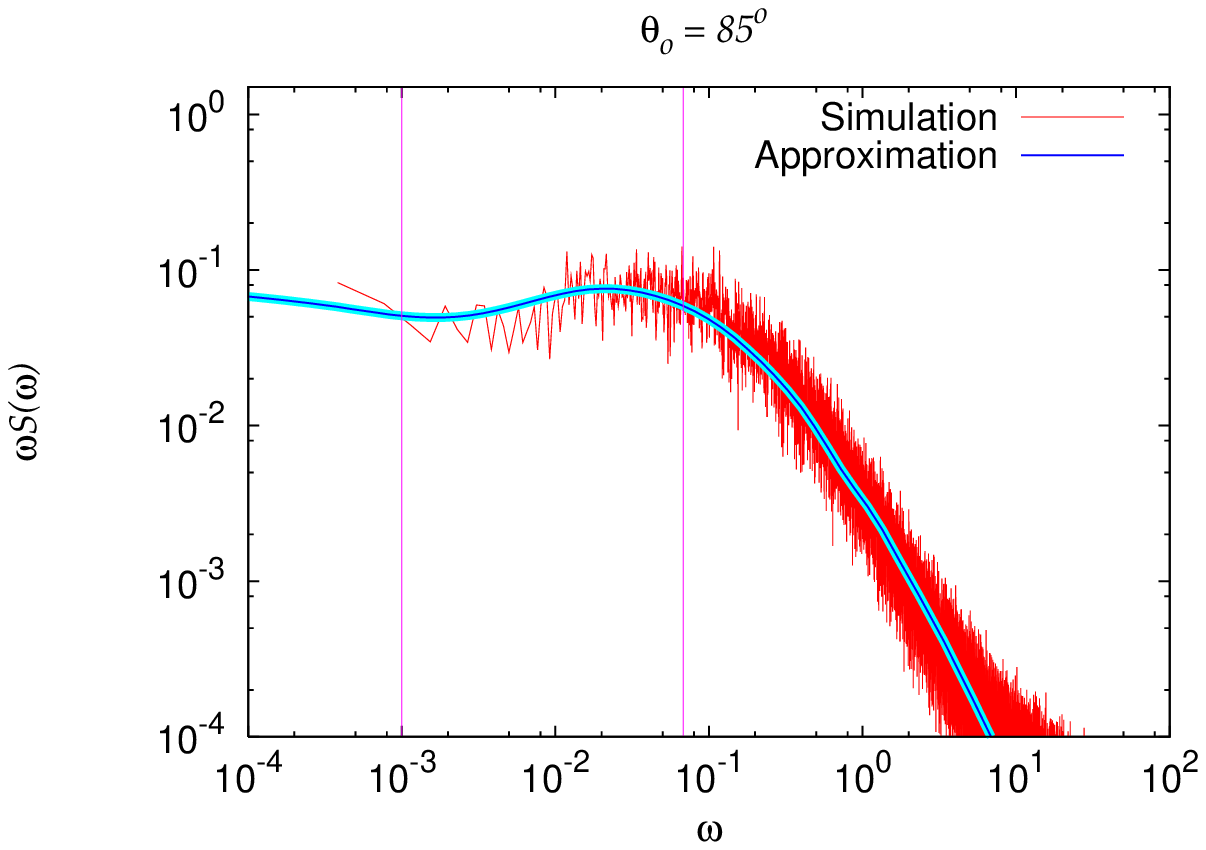}
\caption{Upper panels: Power spectra from the Poisson-driven spot model,
calculated for  a thin accretion disc extending between radii $r=6M$ and
$r=100M$ (in geometrical units), for three inclinations $\theta_{\rm
o}$.  The purpose of this plot is to demonstrate a general agreement
between the PSD calculated from the model light curve and from the
analytical formula. The wiggly (red) curve is a result of direct
numerical simulation, including the relativistic effects. 
The smooth (blue) curve is the analytical result,
derived from Eq.~(\ref{PoissPSD}) by specifying the probability density
function $\zeta(\tau)\propto1/\tau$. The vertical (magenta) lines
represent the Keplerian orbital frequencies, $\Omega(r)$, at the inner
and the outer edges of the disc. Lower panels: Power spectra from the
Hawkes process with the exponential infectivity ($\alpha=7$,
$\nu=0.8$). The analytical curve was calculated by using formula
(\ref{HawkPSD}) and assuming  the same probability density function
$\zeta(\tau)\propto1/\tau$. In all panels we set $\tau_{\rm min}=300$,
$\tau_{\rm max}=5000$.}
\label{SpektObr}
\end{center}
\end{figure*}

\begin{figure*}[bth]
\begin{center}
\includegraphics[width=0.32\textwidth]{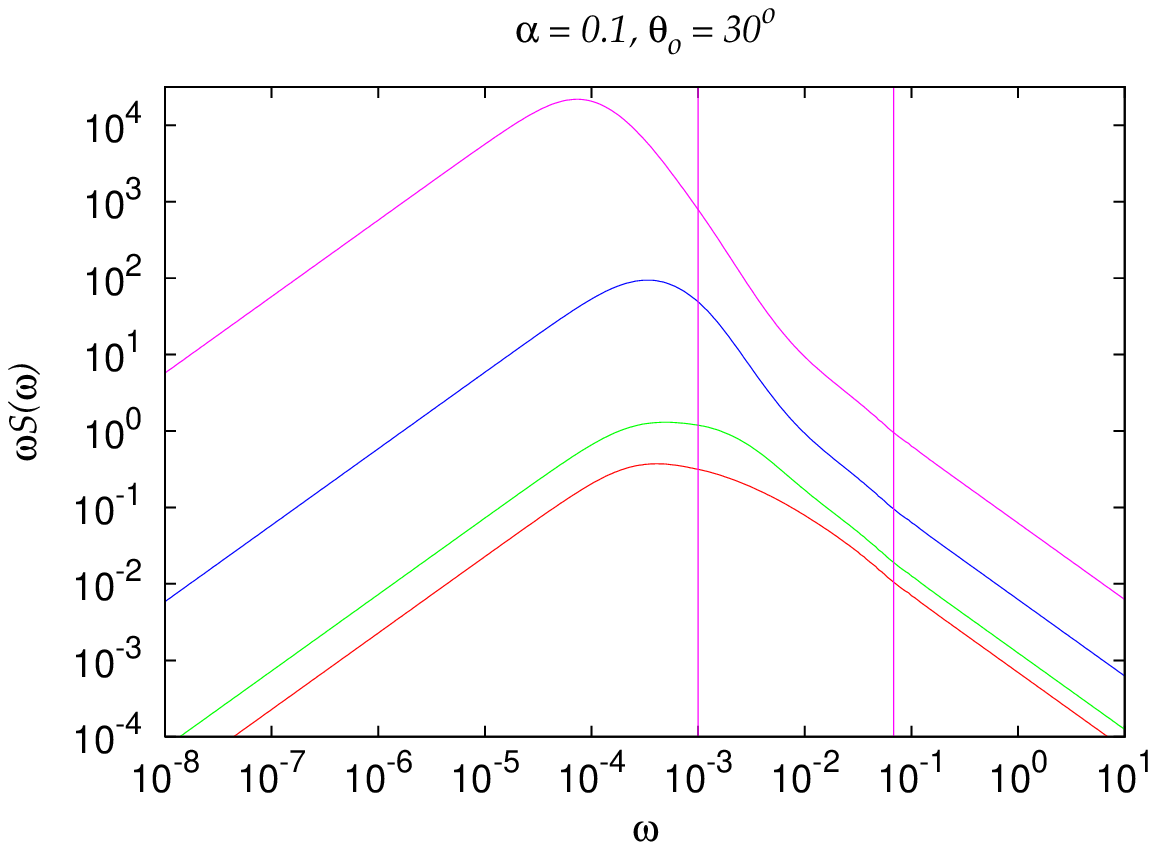}
\includegraphics[width=0.32\textwidth]{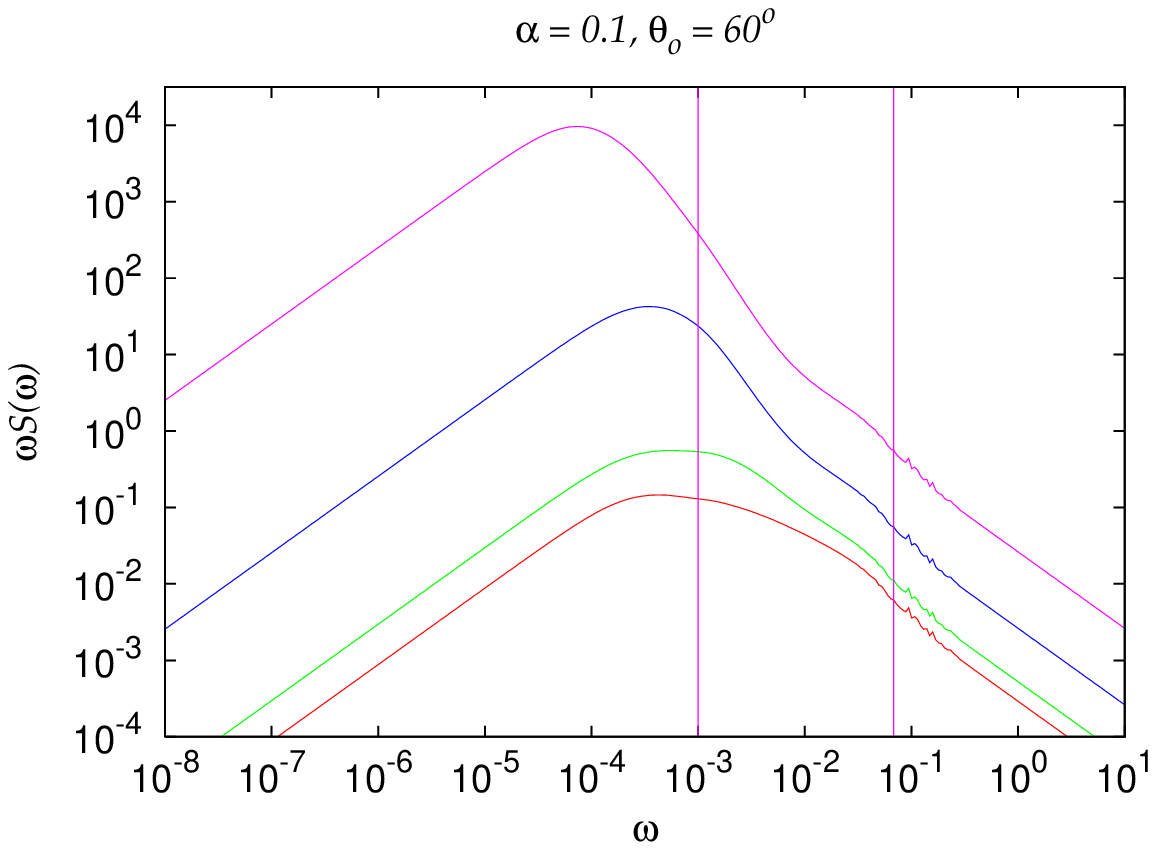}
\includegraphics[width=0.32\textwidth]{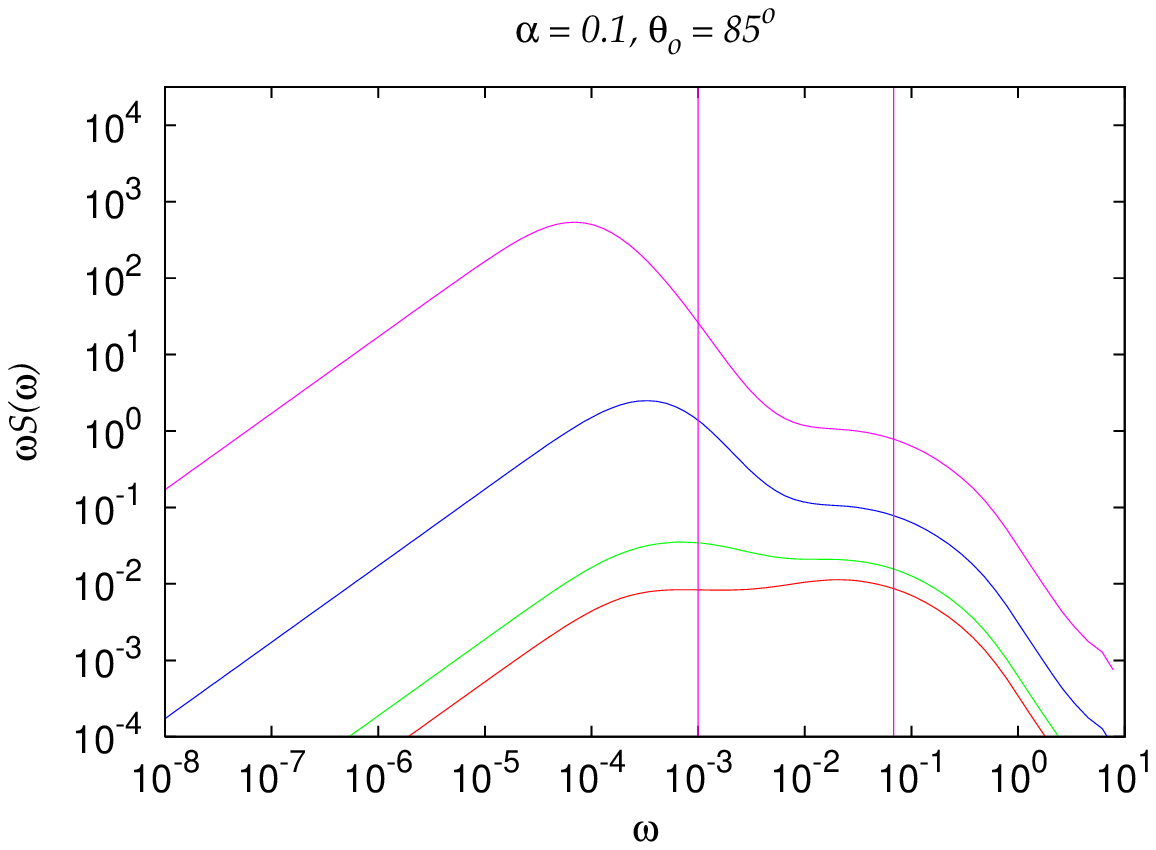}
\includegraphics[width=0.32\textwidth]{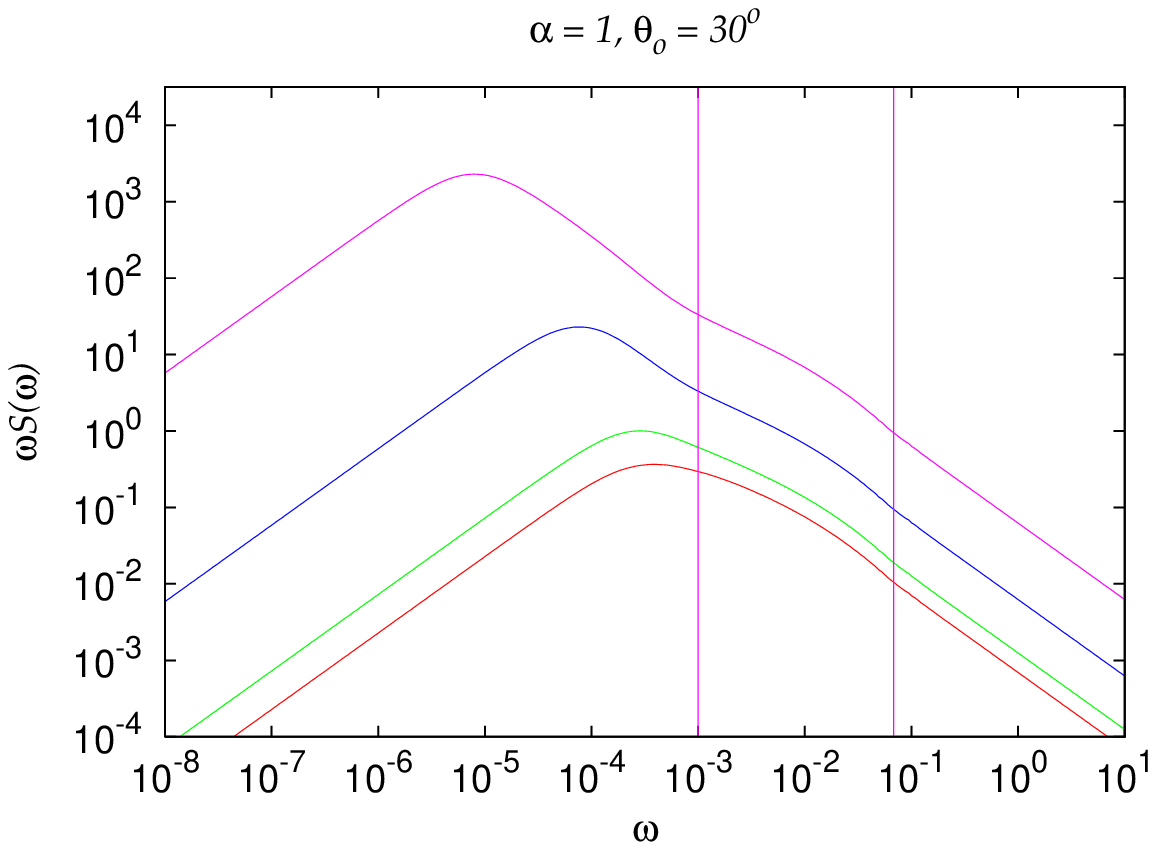}
\includegraphics[width=0.32\textwidth]{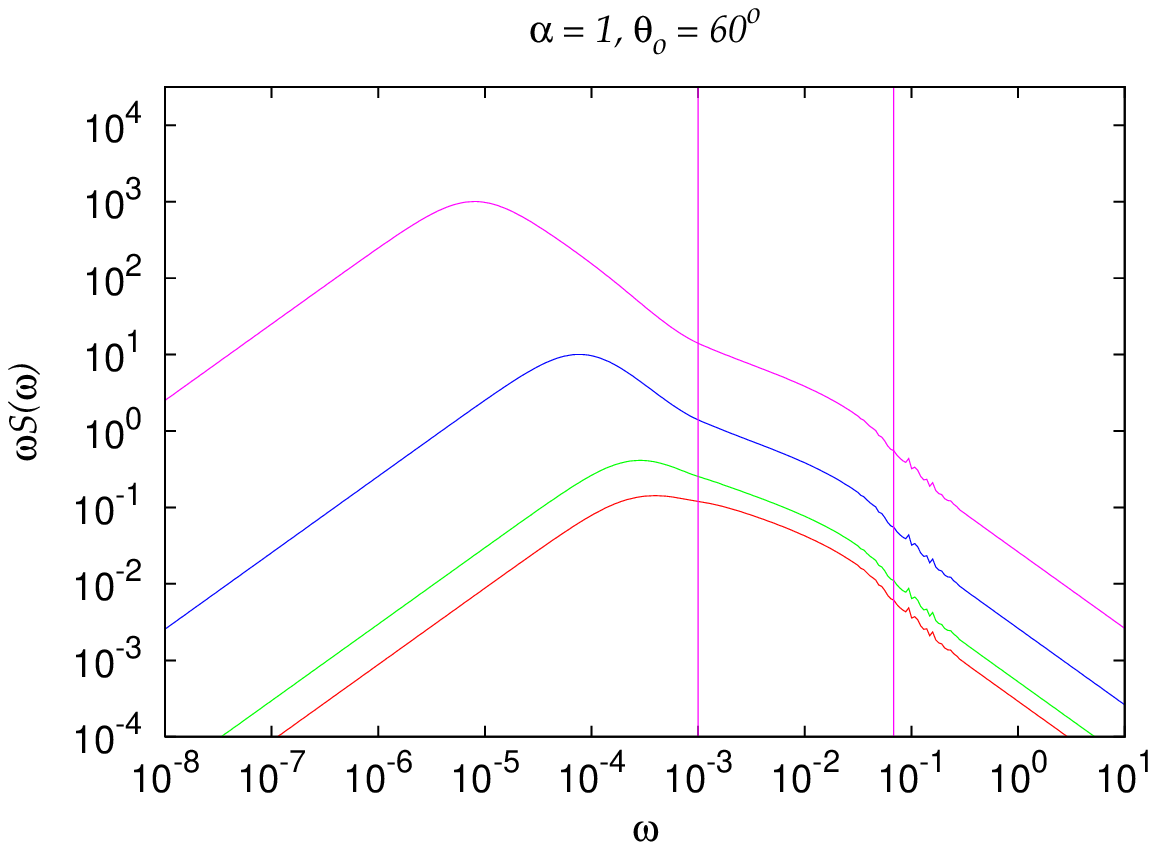}
\includegraphics[width=0.32\textwidth]{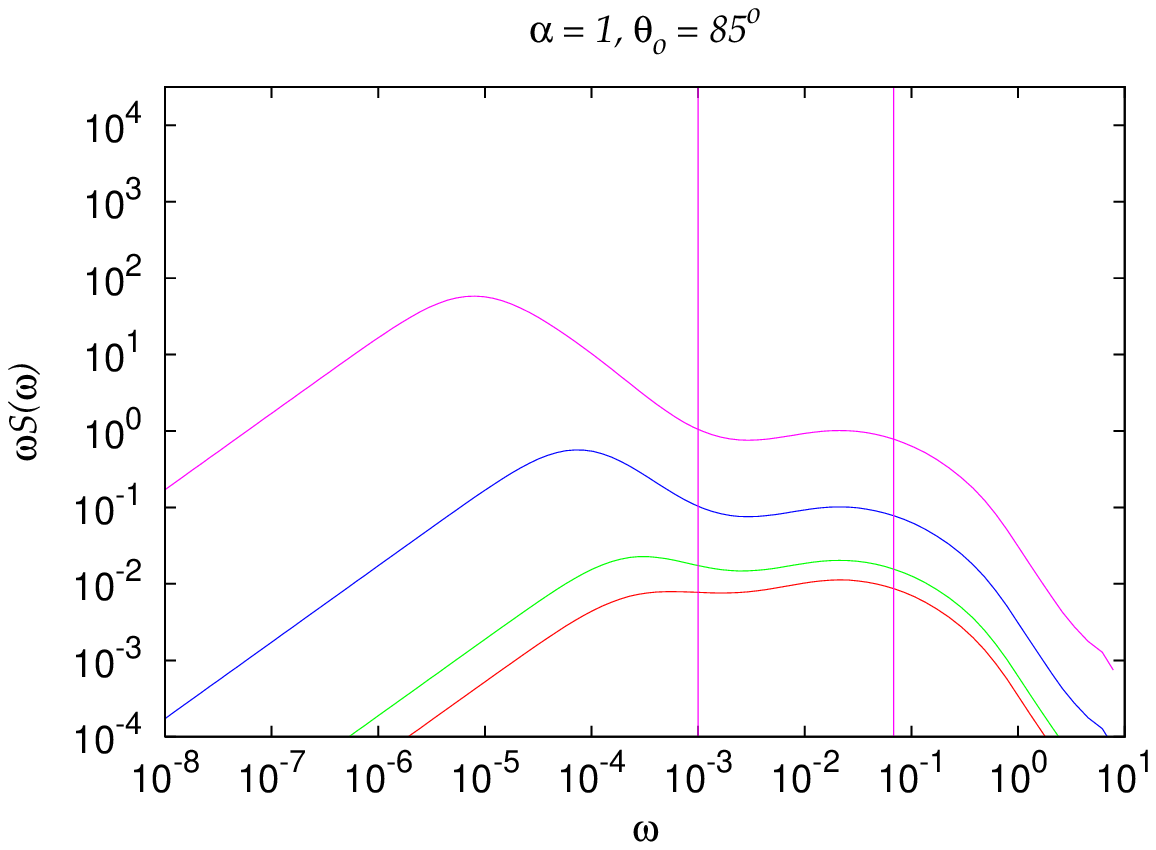}
\includegraphics[width=0.32\textwidth]{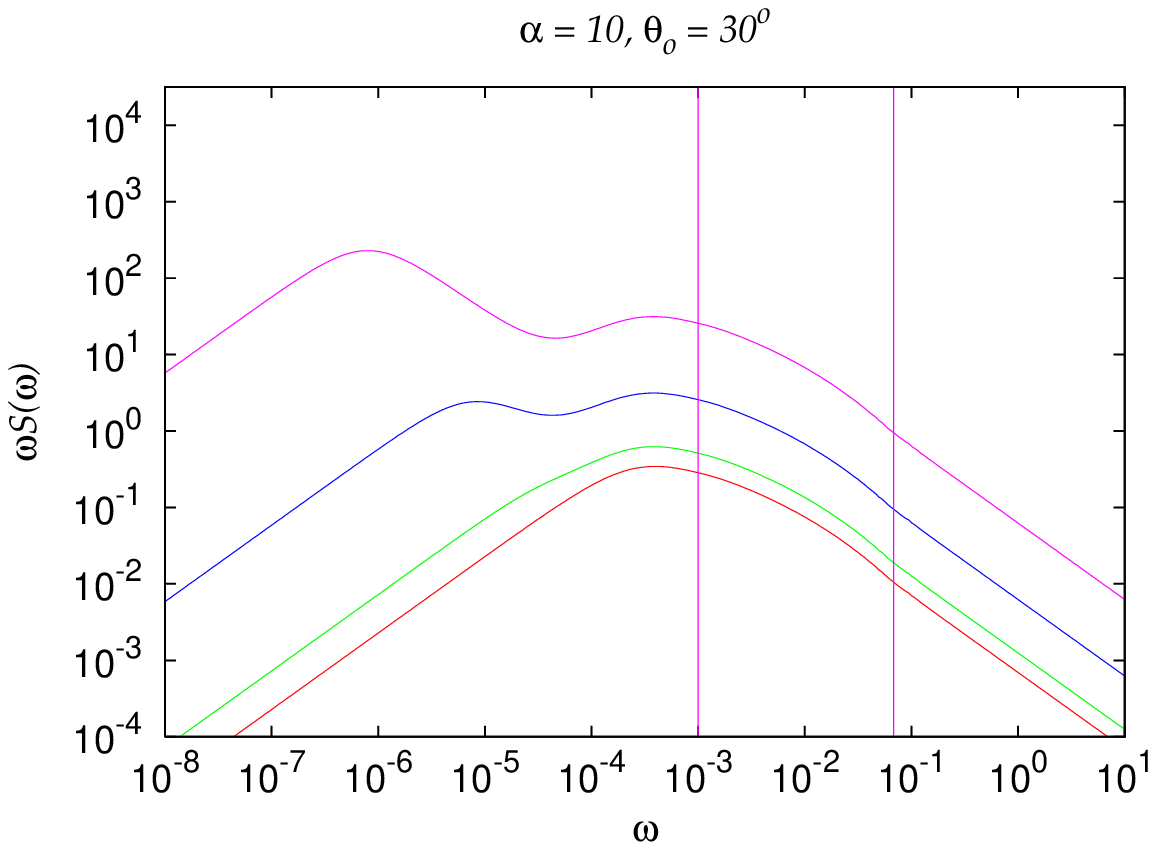}
\includegraphics[width=0.32\textwidth]{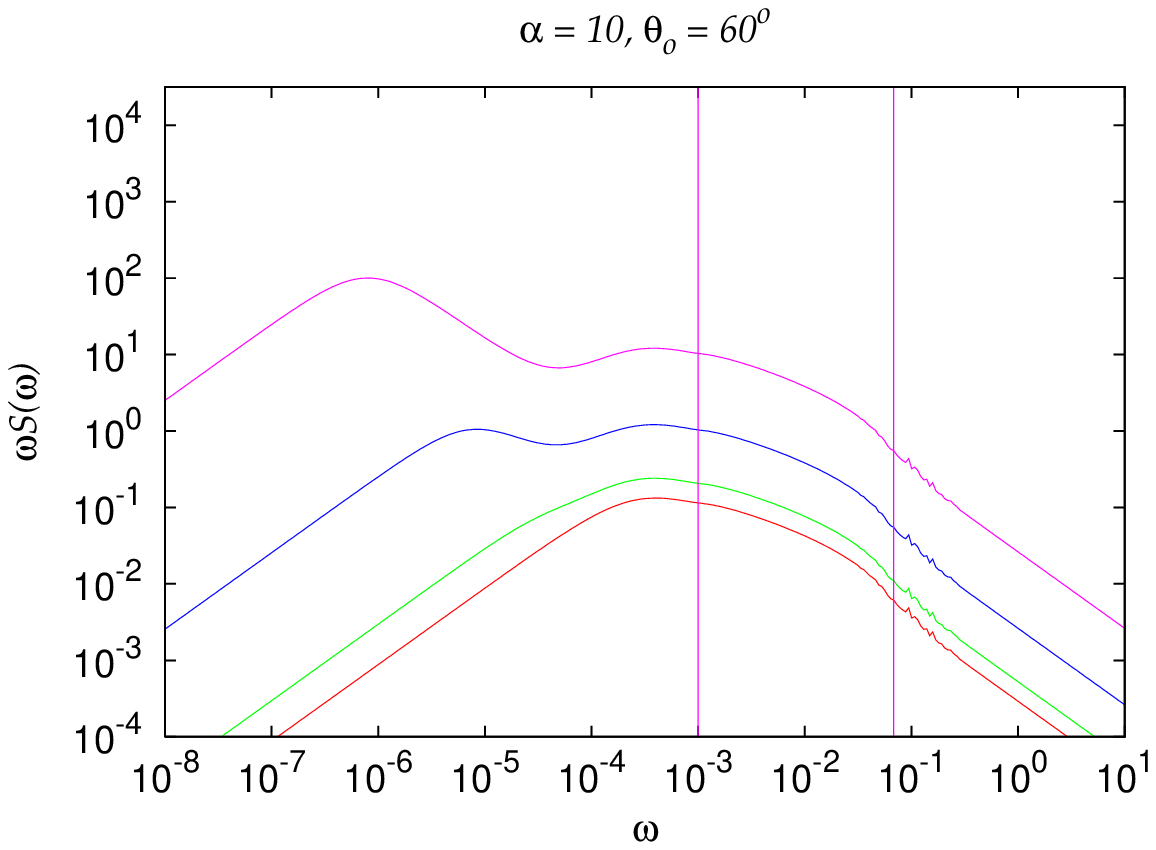}
\includegraphics[width=0.32\textwidth]{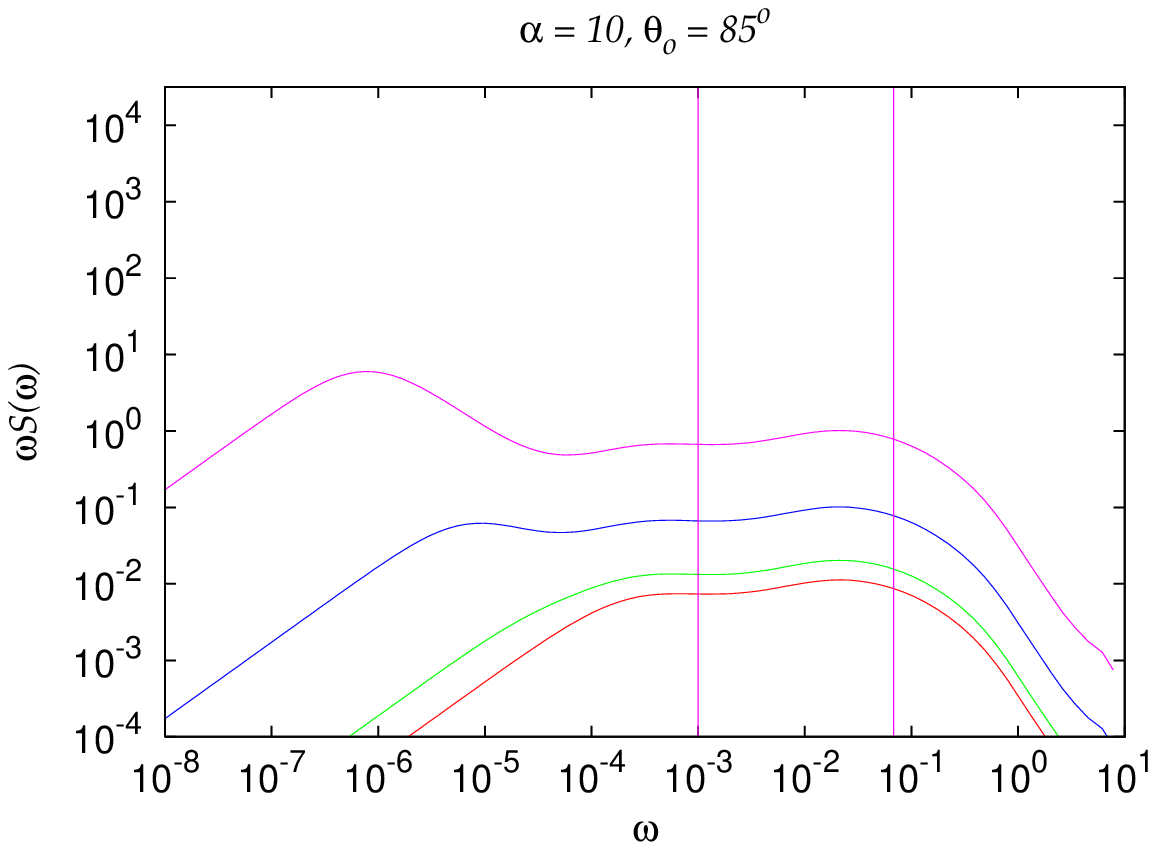}
\end{center}
\caption{The analytical curves of the PSD profiles are plotted, with 
relativistic effects included. In order to allow comparison with
previous figures, we selected appropriate combinations of the model
parameters: the infectivity $\alpha$, the mean number of secondary spots
$\nu$, and the spot distribution are the same as in Fig.~\ref{haw1}. The
inclination $\theta_{\rm o}$ and the spot distribution as in
Fig.~\ref{SpektObr}. The vertical lines again indicate the range of
orbital frequencies corresponding to the assumed range of radii,
$6M<r<100M$, where the spots are distributed. The analytical form is
quick to evaluate, hence it is convenient to obtain the PSD form for
variety of different situations.}
\label{SpektObr2}
\end{figure*}

\begin{figure*}[bth]
\begin{center}
\includegraphics[width=0.32\textwidth]{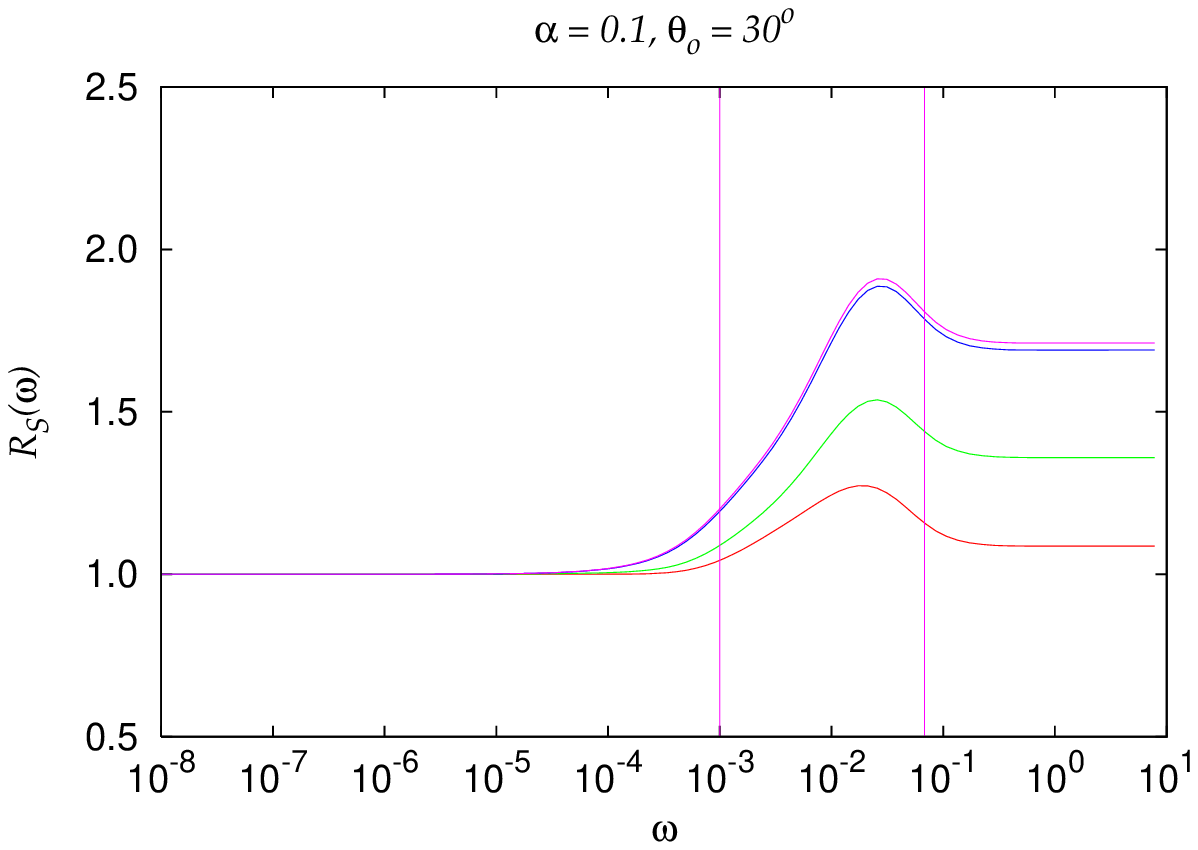}
\includegraphics[width=0.32\textwidth]{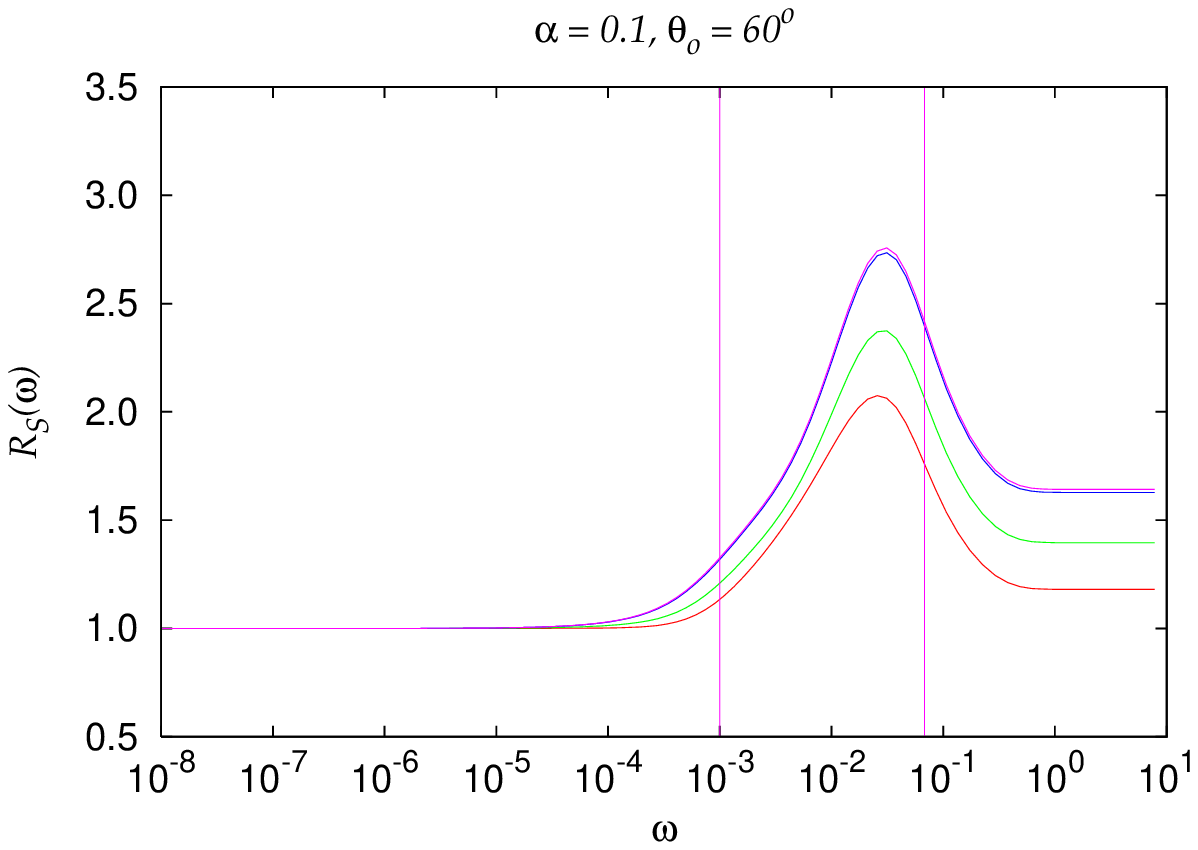}
\includegraphics[width=0.32\textwidth]{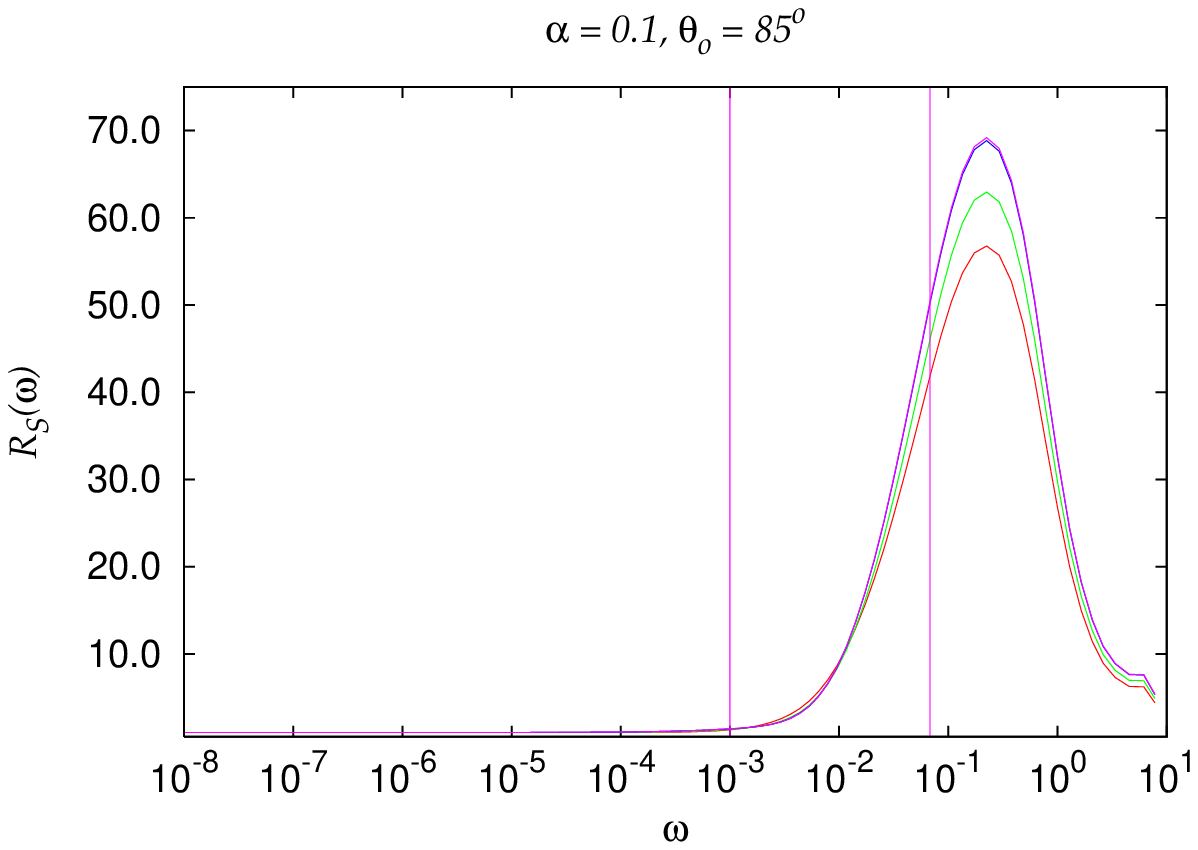}
\includegraphics[width=0.32\textwidth]{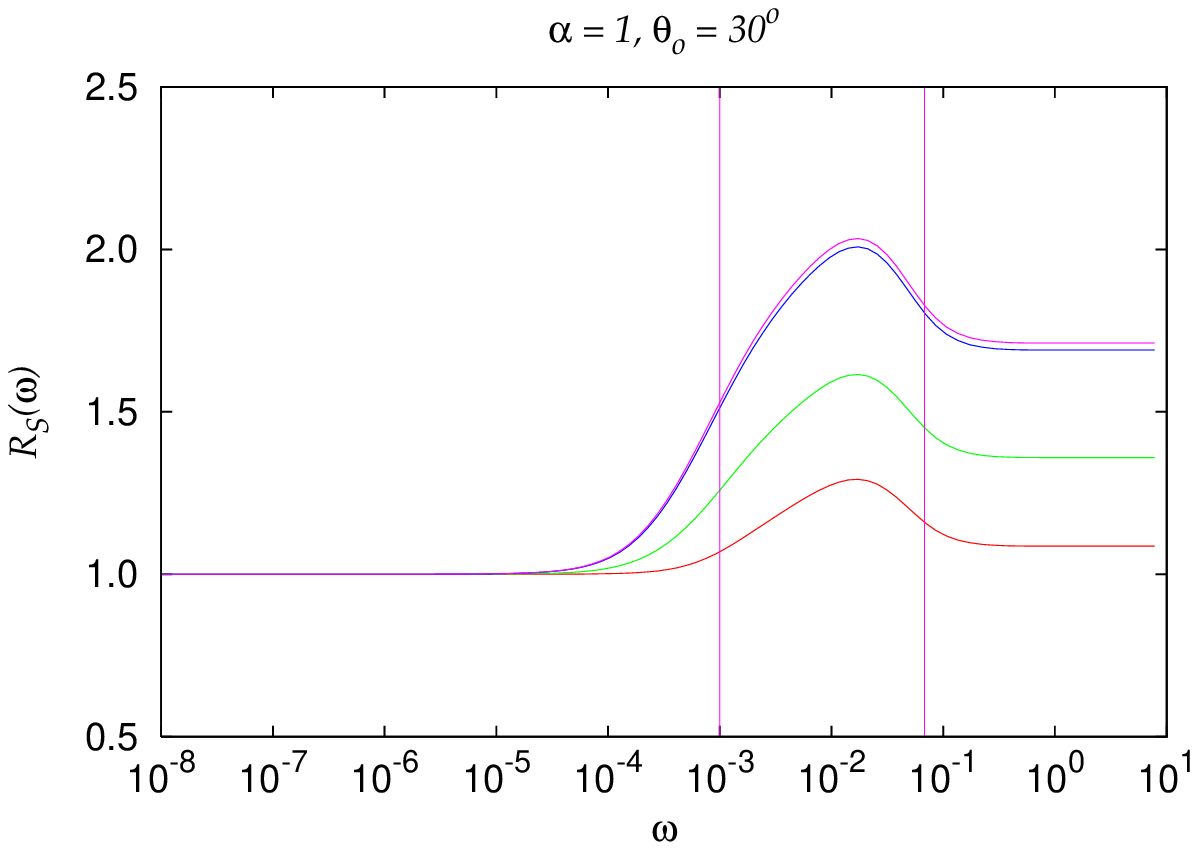}
\includegraphics[width=0.32\textwidth]{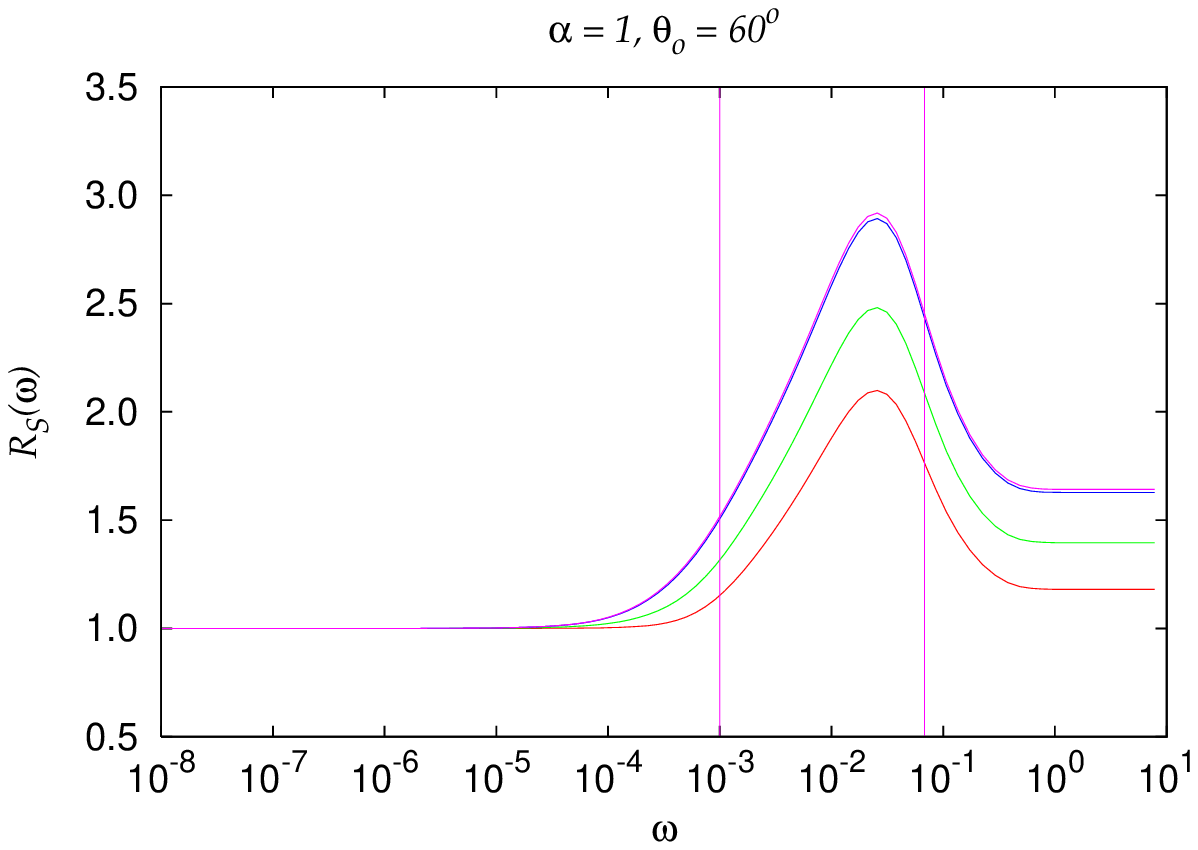}
\includegraphics[width=0.32\textwidth]{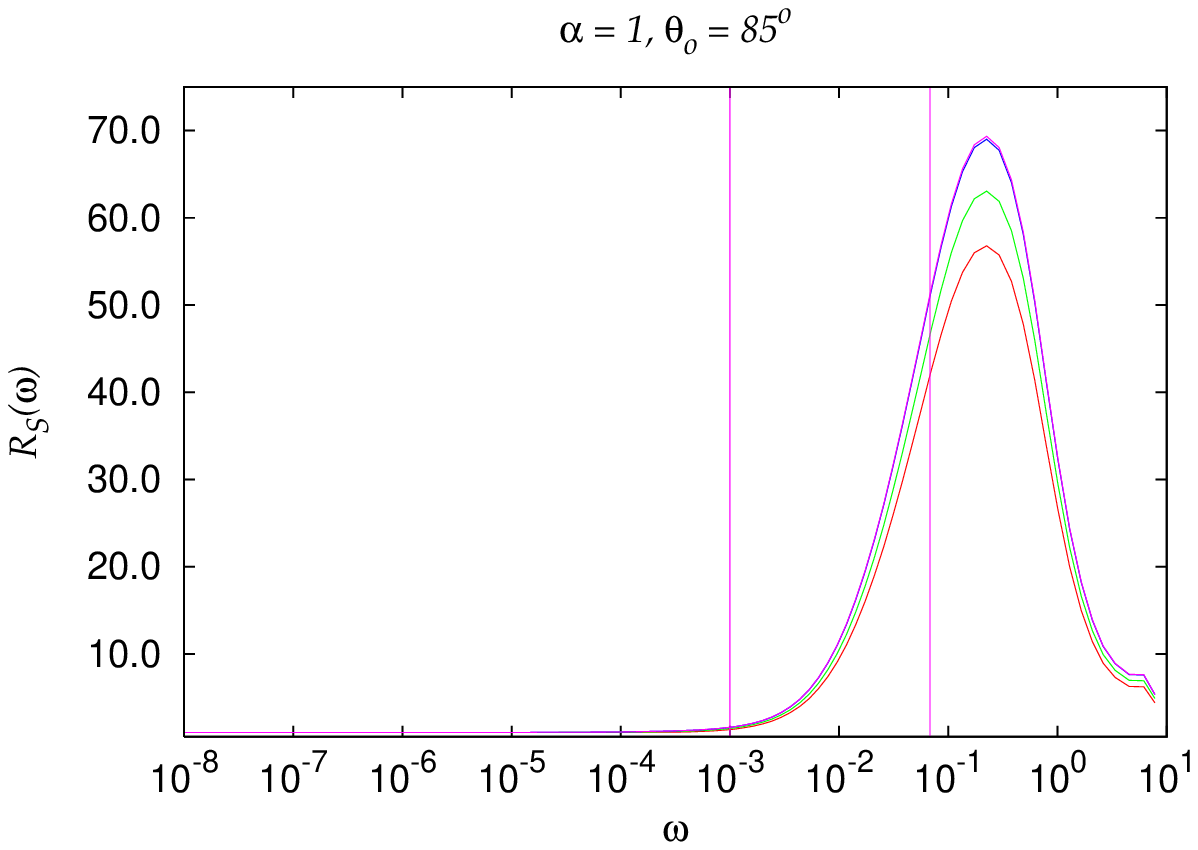}
\includegraphics[width=0.32\textwidth]{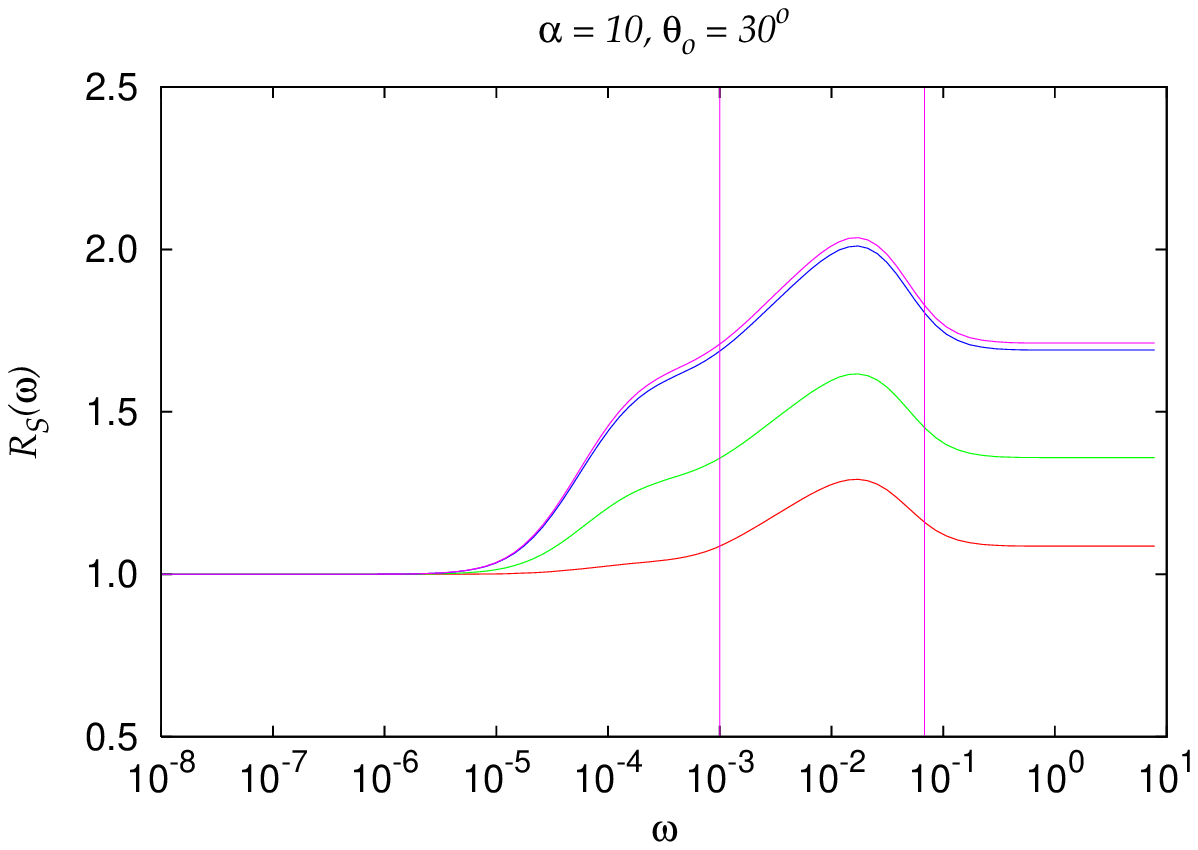}
\includegraphics[width=0.32\textwidth]{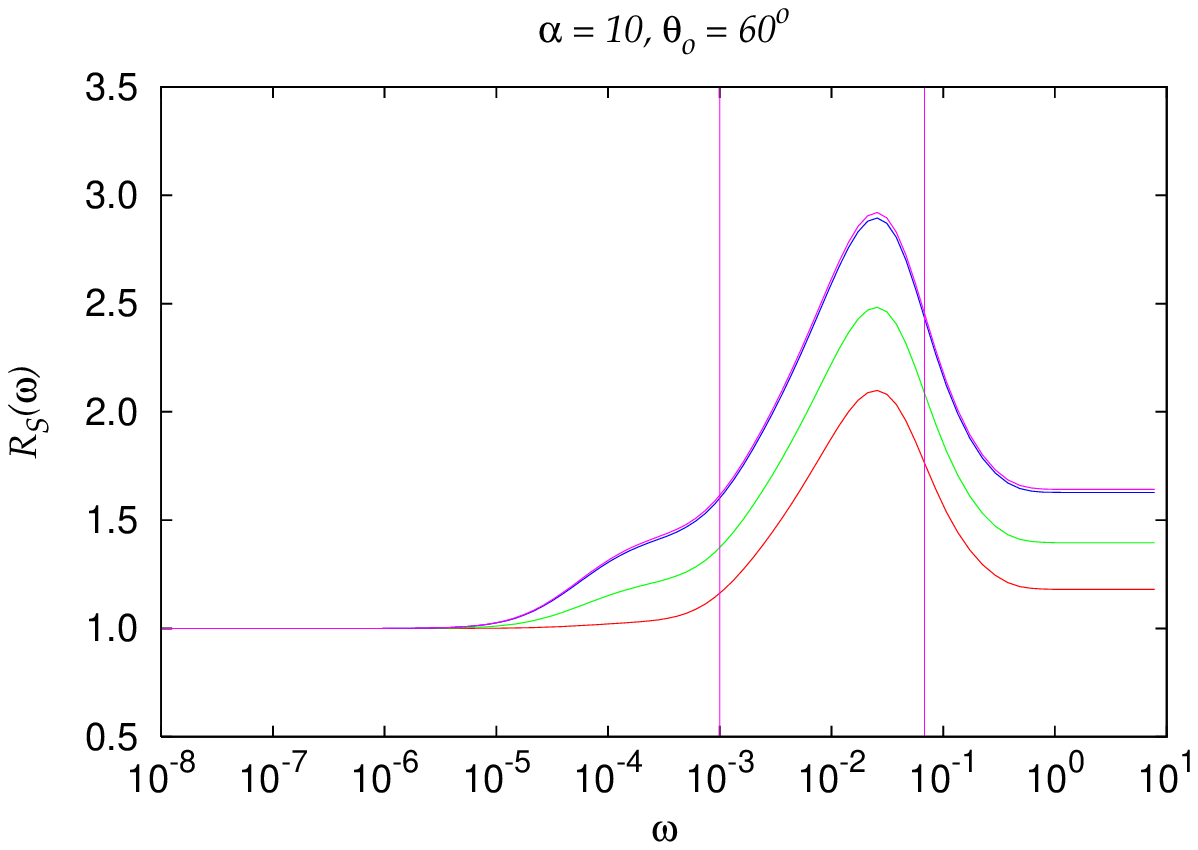}
\includegraphics[width=0.32\textwidth]{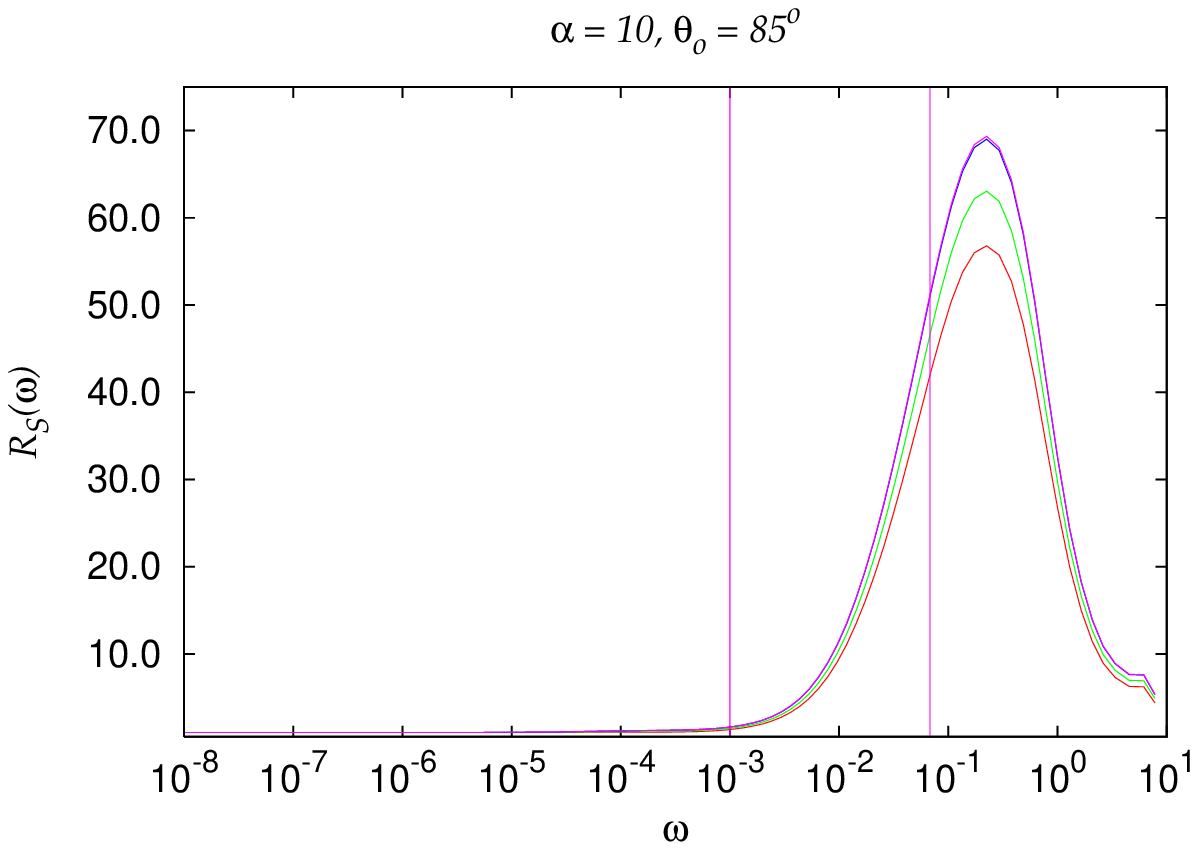}
\end{center}
\caption{The ratio 
$R_{\rm{}S}(\omega)$ of PSD computed with and without relativistic effects.
The difference is most visible around intermediate frequencies;
Keplerian orbital frequencies at the outer/inner edges of the spot radial
distribution are again indicated by vertical lines. The arrangement of 
panels and the assumed parameter values are the same
as in the previous Fig.~\ref{SpektObr2}. In particular, the vertical
lines indicate the minimum and the maximum orbital frequencies corresponding
to the outer and the inner edges of the spot distribution.}
\label{ratios}
\end{figure*}

\subsection{The pulse avalanche model}
\label{sec:avalanche}
The pulse avalanche model was discussed in the context of various
astronomical objects whose light curves exhibit signs of stochastic
behaviour. They are, namely, gamma-ray bursting sources
\citep{1996ApJ...469L.109S}. As a framework to describe the timing
characteristics of accreting black holes, the model was
studied by \citet{1999MNRAS.306L..31P}. Details of the process are
different in those two papers and our description is closer to the
latter one.

The basic properties of the pulse avalanche model can be summarised as 
follows. (i)~The observed signal consists of pulses of the form
$I(t,\tau)=I_0\,g(t,\tau)\,\theta(t)$, where $\tau$ is their
characteristic duration. (ii)~Each pulse gives birth to $b$ baby pulses;
the number of baby pulses varies, obeying the Poisson  distribution
$P(b)=\nu^b/(b!\,e^\nu)$ with the mean value $\nu$. (iii)~The baby
pulses are delayed with respect to the parent ones by $\Delta t$, which
is a random value with exponential distribution, $P(\Delta
t)=(\alpha\tau_0)^{-1}e^{-\Delta t/(\alpha\tau_0)}$. (iv)~Some pulses
occur spontaneously, according to the Poisson process operating at the
mean rate $\lambda$. Finally, (v)~temporal constants, $\tau$, for all
pulses are mutually independent and drawn from the same distribution
function, $\zeta(\tau)$. Such a process is clearly of the form
(\ref{proces}).

The underlying point process is a cluster process operating on the set
$\mathbb{R}\times \langle\tau_{\rm min},\,\tau_{\rm max}\rangle$
with properties similar to the Hawkes process. (In fact, we will see below that
the Hawkes process can be considered as a special case of the pulse avalanche 
process.) The center process is the Poissonian one, and its intensity is
$\lambda_{\rm c}(t,\tau)=\lambda\zeta(\tau)$. The clusters are driven by the
Poissonian branching process \citep{1965cox} with the parameter measure
\begin{equation}
\mu(t,\tau\,|\,t_0,\tau_0)=\frac{\nu\,\zeta(\tau)}{\alpha\tau_0}\,
\exp\Big\{-\frac{t-t_0}{\alpha\tau_0}\Big\}\;\theta(t-t_0).
\label{PAMmira}
\end{equation}
One can prove, by direct integration, that
\begin{eqnarray}
{\rm E}\big[\Delta t\,|\,t_0,\tau_0\big]\!&=&\!\frac{1}{\nu}\int\limits_{-\infty}^\infty\!\int\limits_{\tau_{\rm min}}^{\tau_{\rm max}}\!
(t-t_0)\,\mu(t,\tau\,|\,t_0,\tau_0)\,{\rm d}\tau\,{\rm d}t
=\alpha\tau_0,\nonumber\\
{\rm E}\big[b\,|\,t_0,\tau_0\big]\!&=&\!\int\limits_{-\infty}^\infty\int\limits_{\tau_{\rm min}}^{\tau_{\rm max}}
\mu(t,\tau\,|\,t_0,\tau_0)\,{\rm d}\tau\,{\rm d}t=\nu.
\end{eqnarray}
In analogy to Eqs.~(\ref{HawkClusInt}) and (\ref{HawQ}) one can derive a set
of linear integral equations for the first and second order moments of
the cluster process. The chosen form (\ref{PAMmira}) of the kernel
$\mu(t,\tau\,|\,t_0,\tau_0)$ greatly simplifies the task, as it can be
again transformed into a convolutory form and solved in the Fourier
domain.

The resulting functions $\tilde{m}_{[1]}$ and $\tilde{m}_{[2]}$ are 
complicated and their inverse Fourier transforms can be directly found
only in very special cases, e.g.\ by assuming $\zeta(\tau)=\delta(\tau-a)$.
In that case the pulse avalanche model is transformed into the Hawkes process
with the exponential infectivity measure. 

Fortunately, we do not need the explicit form of the moments to find the
final formula for the PSD. This result is provided by
Eq.~(\ref{HawkPSD}) with
\begin{equation}
S_{\!1}(\omega)=4\pi^2\!\! \int\limits_{\tau_{\rm min}}^{\tau_{\rm max}}
\int\limits_{\tau_{\rm min}}^{\tau_{\rm max}}\!\!
S_{\rm\! P}(\omega,\tau,\tau')\,Q_{{_\mathcal{K}}}(\omega,\tau)\,Q_{{_\mathcal{K}}'}^{*}(\omega,\tau')\;{\rm d}\tau\;{\rm d}\tau'
\label{paPSD}
\end{equation}
and
\begin{eqnarray}
S_{\rm\! P}(\omega,\tau,\tau')&=&\frac{\lambda}{1-\nu}\,\zeta(\tau)\,\zeta(\tau')
\big[f_1(\omega)\,|f_2(\omega)|^2\nonumber\\
&&+ \frac{\nu}{1-i\alpha\tau\omega}\,f_2(\omega)+\frac{\nu}{1+i\alpha\tau'\omega}\,f_2^*(\omega)\big],
\end{eqnarray}
\begin{eqnarray}
f_1(\omega)&=&\int_{\tau_{\rm min}}^{\tau_{\rm max}}\frac{\nu^2\,\zeta(y)}{1+\alpha^2\omega^2y^2}\,{\rm d}y,\\
f_2(\omega)&=&\Big[1-\int_{\tau_{\rm min}}^{\tau_{\rm max}}\frac{\nu\zeta(y)}{1+i\alpha\omega y}\,{\rm d}y\Big]^{-1}.
\end{eqnarray}

We show several graphs of the resulting model PSD curves in
Fig.~\ref{haw4}. The overall form of the PSD resembles the previous
examples derived for the Hawkes process, although they are not the same. 
The two models -- the pulse avalanche process and the Hawkes process --
are identical in their high-frequency and low-frequency limits, but they
differ from each other in the middle range of frequency. In order to
demonstrate the difference in a clear way we plot the ratio of the
models in the right column. Fig.~\ref{haw4} assumes
$\zeta(\tau)=1/(\tau_{\rm max}-\tau_{\rm min})$. We also studied the
case of $\zeta(\tau)=[\tau\ln(\tau_{\rm max}/\tau_{\rm min})]^{-1}$ and
we found that the general form of the PSD profiles exhibits very similar
trends. We see only quite minor differences in the limiting slopes
of the PSD and in the overall PSD shape in the middle range of
frequencies. Therefore, we do not show these plots here (see
\citeauthor{2008Pechacek} \citeyear{2008Pechacek}).

\subsection{PSD with relativistic effects}
Relativistic effects influence the photon energy in the course of light
propagation through the curved spacetime, towards a distant observer. In
this paper we consider only the energy-integrated light curves, but even
those are affected because the energy shifts modify the observed flux
level. Furthermore, light rays are bent as they pass near the black
hole, causing the light-focusing effect. In consequence the observed
light curves differ from their intrinsic profiles produced at the point
of emission, and this further complicates the decomposition of the PSD
spectrum into elementary components. Therefore, in general the power
spectrum cannot be rigorously expressed as a combination of Lorentzians.
However, in most circumstances the relativistic effects are not very
prominent -- only a small fraction of rays passing very close to the
black hole horizon and those crossing the  caustics are affected. One
expects that they cannot be ignored if the accretion disc extends down
to the innermost stable orbit or if some non-negligible emission arises
below that orbit. Also, high-inclination objects are affected more
because in those cases the disc is seen edge-on and the intrinsic
fluctuations of the emission are considerably amplified.

Exemplary power spectra, including the relativistic effects are modelled
in Fig.~\ref{SpektObr} where the numerical simulation is compared with
the analytical result. The upper panels assume that orbiting spots are
generated by Poissonian process; the lower ones show the PSD derived
from Eq.~(\ref{HawkPSD}). The assumed radial distribution of the spots
was $\rho(r)=\rho_0\,(1-\sqrt{6/r})\,r^{-2}$, where $\rho_0$ is
normalisation constant. It turns out that the relativistic effects 
influence the final PSD especially at high frequencies
and high inclinations.

We remind the reader that frequencies in these plots are given in 
geometrical units (in physical units frequencies scale inversely with
the mass of the black hole). We notice that a high-frequency part of the
spectrum decays as $\omega S(\omega)\propto\omega^{-1}$, whereas the
break occurs towards lower frequencies. These plots
provide us with graphical comparisons between the analytical form and
the corresponding  results of numerical simulations. We note that
the adopted approximation of relativistic effects
\citep{2005A&A...441..855P} holds for moderate inclination
($\lesssim70$~deg). It loses accuracy when the view angle becomes almost
edge-on, although the main trend of the PSD remains unchanged. 

The main advantage of the analytical method is, obviously, in the
possibility of obtaining a general form of the PSD, including the
relativistic effects. We are ale to search systematically through the
vast parameter space of different models for which the model PSD can be
explored across a wide range of frequencies. We take the advantage of
this approach and plot variety of profiles in Fig.~\ref{SpektObr2}. Here,
we omit the numerically-simulated curves of the previous figure, so the
entire graph is constructed very efficiently. We assumed that spots
are generated by the Hawkes mechanism.

Basic trends of the PSD shape are readily recognised. In particular,
the curves have either one or two maxima, prominence and position of
which changes with parameters. Notice, for example, the bottom right
panel in which the PSD is almost flat over several decades of frequency,
well below the typical orbital frequency in the disc (vertical lines
denote the Keplerian  frequencies at the edges of the assumed spot
distribution). The flattish part of the spectrum can be extended further,
to lower frequencies, by enlarging the $\nu$ parameter towards unity,
i.e., by protracting the sequence of avalanches. Next, for large
inclinations we notice that relativistic effects produce a
prominent bump. This feature occurs near the orbital frequency of the
inner disc. Relativistic effects are the main cause of differences
between this figure and Fig.~\ref{haw1}, in which those effects were
neglected.

As already mentioned, general relativity affects mainly the
high-frequency domain of the PSD, around the orbital frequency of the
inner disc, where it adds power to the observed PSD. In physical units
the relevant frequency generated at radius $r$ is  $T_{\rm
orb}^{-1}\approx10^{-4}\,(r/10R_{\rm g})\,(M/10^7M_{\odot})$~[Hz]. On
the other hand, it does not influence the middle part of the spectrum,
i.e.\ at frequency $\lesssim T_{\rm orb}^{-1}(r_{\rm out})$,
neither it changes the asymptotical form at far ends of the frequency
range (where the PSD decays as power law). It has been argued that the
additional signal is actually not present in the data of MCG--6-30-15
\citep{2005MNRAS.359..308Z}, although the light curve from the long 
observation should reveal some excess. However, the situation here is
more complex because of the avalanches contributing power to lower 
frequencies. In fact, our model predicts rather weak enhancement near
the inner edge orbital frequency -- unless the inclination is almost
edge-on (which is unlikely). More power is therefore typically expected
at moderate frequencies, lower than the inner-edge orbital frequency.

In order to see at which frequencies the relativistic effects are 
most important, we construct the ratio $R_{\rm{}S}(\omega)$ of 
the PSD calculated with (denoted by superscript ``gr'') and without 
(``cl'') these effects taken into account:
\begin{equation}
R_{\rm{}S}(\omega)=N_0\;\frac{S_{\rm{}P}^{\rm{}gr}(\omega)}{S_{\rm{}P}^{\rm{}cl}(\omega)}\,,
\end{equation}
where the normalisation factor is $N_0\equiv
{S_{\rm{}P}^{\rm{}cl}(0)}/{S_{\rm{}P}^{\rm{}gr}(0)}$. (Doppler and
lensing effects are neglected in the calculation of 
${S_{\rm{}P}^{\rm{}cl}(\omega)}$.) Figure~\ref{ratios} shows that the PSD
graphs are affected only by a  constant shift at their low-frequency as
well as at the high-frequency limits, i.e., the ratio $R_{\rm{}S}(\omega)$
reaches a constant value at both ends of the frequency range. 
This is quite easy to understand as at the ends
ends of the frequency domain the relativistic PSD acquires the same slope
as the non-relativistic one. Relativity shapes the PSD especially within
the range of orbital frequencies of the assumed spot distribution.

\section{Conclusions}
\label{conclusions}
We adopted the viewpoint that the
variability pattern is determined by the interplay among the bulk
orbital motion, relativistic effects, and the intrinsic changes of the
inner accretion disc. We concentrated our attention solely on the PSD
characteristics. The spots have a certain kind of memory in our model.

We gave several examples in which the PSD changes the slope and certain
break frequencies. The frequency of the break depends on the interplay
of model properties, i.e., the intrinsic form of the spot light curves,
which determine the individual contributions to the total signal
together with the avalanche mechanism. {\em The location of spots on the
disc and the inclination of the source define the importance of
relativistic effects.}

In some cases, a double break occurs and the overall PSD profile is 
then approximated by a broken power law. This is a promising feature in
view of applications to real sources with accreting black holes. The
broken power-law profile either resembles a combination of the
Lorentzians or, in some cases, an intermediate power law develops and
connects the two peaks across the middle frequencies. The change of the PSD
slope is clearly visible and well-defined in some cases, though under
most circumstances it appears rather blurry. The low-frequency limit of
the PSD slope is a constant; the-high frequency behaviour depends mainly
on the shape of the spot emission profile, including the general
relativity effects. In our calculations the emissivity was decaying
exponentially and the slope of the PSD was equal to $-2$ at high
frequencies. {\em In between those two limits the intrinsic PSD is
influenced by both the individual light curve profile and the underlying
process.}

It is interesting to notice that the doubly-broken power law occurs only
for certain assumptions about the intrinsic light curves of the
individual spots or avalanches -- their onset and the decay; in other
cases the break frequencies are not well defined, or the broken
power-law PSD is not preferred at all. {\em We stress that if two
Lorentzian seem to dominate the PSD (i.e., two peaks show up), it still
does not necessarily mean that two single oscillation mechanisms operate 
simultaneously. Instead, it may well be the manifestation of the
avalanche mechanism.}

We employed a general statistical approach to the variability of a black
hole accretion disc with orbiting spots that continuously arise and
decay. The origin and evolution of spots were described by Poissonian
and Hawkes' processes, the latter one representing a category of
avalanche models. We derived analytical formulae for the PSD, Eqs.\
(\ref{PoissPSD}) and (\ref{HawkPSD}), and we discussed their limitations
and accuracy. {\it The main advantage of the analytical form is the
insight into the properties and the fast evaluation that captures the
main trends of the PSD shape.}

It is worth noting that the PSD does not maintain all information about
the light curves that can be studied by Fourier methods
\citep{1992ApJ...391..518V}. Extensions have been discussed and compared
with real data
\citep{1993ApJ...402..432K,1997MNRAS.288...12K,1999ApJ...510..874N,2008arXiv0802.0391V},
but this would go beyond the scope of our present work.

Our approach allows us to investigate the influence of the assumed
mechanism, which describes the creation of parent spots and of
subsequent cascades of daughter spots. In particular, we can discuss the
PSD slope at different frequency ranges and locate the break frequencies
depending on the model parameters. The relationship between the
mathematical nature of the process and the PSD of the resulting signal
is an interesting consequence of this investigation, as it provides a
way to grasp and rigorously constrain the physical models of the source.
Therefore we believe that the method that we described is very helpful
for identifying the underlying mechanisms that shape the PSD in black
hole accreting sources.

\begin{acknowledgements} 
We thank Dr.\ L.~\v{S}ubr for helpful comments. We appreciate the continued
support from research grants of the Academy of Sciences (ref.\
300030510) and of the Czech Science Foundation (ref.\ 202/06/0041). Part
of this work was supported via the ESA Plan for European Cooperating
States (98040). VK is a member of the Center for Theoretical
Astrophysics in Prague, LC06014.
\end{acknowledgements}

\appendix
\section{Spots as a random process}
\label{appa}
\subsection{Assumptions, definitions and preliminaries}
In this Appendix, we briefly introduce the formal mathematical approach and
notation used throughout the paper. We employ the concept of random
values on probability space, $\left(\Omega,\Sigma,P\right)$
\citep{1950kolmogorov,1965cox}, where $\Omega$ is the sample space
(i.e., the set of all possible outcomes of an experiment), $\Sigma$
denotes the $\sigma$-algebra on $\Omega$, and $P$ is non-negative,
$\sigma$-additive measure satisfying conditions $P(\emptyset)=0$,
$P(\Omega)=1$. It is usually assumed on physical grounds that real 
signals satisfy all mathematical prerequisites.

A real random value is a map $X$ from $\Omega$ to real numbers,
$X:\Omega\rightarrow\mathbb{R}$.
The distribution function $F(x)$ and the probability density function 
$f(x)$ are then defined by
\begin{equation}
F(x)=P\left(\{\varpi\in\Omega\,|\,X(\varpi)<x\}\right),\quad
f(x)=\frac{{\rm d}}{{\rm d}x} F(x).
\end{equation}
The mean value operator E acts as
\begin{equation}
{\rm E}\big[g(X)\big]=\int\limits_{\mathbb{R}}g(x)\,F({\rm d}x)
=\int\limits_{\mathbb{R}}g(x)\,f(x)\,{\rm d}x.
\end{equation} 

Random values are characterised by their moments, defined as ${\cal
M}_k\equiv{\rm E}[X^k]$. The above-given relations can be  generalised
to higher dimensions by introducing a $k$-dimensional vectorial random
value, $\mbox{\boldmath $X$}=(X_1,\dots,X_k)$, the distribution function
$F(\mbox{\boldmath $x$})$, and the probability distribution
\begin{equation}
f(\mbox{\boldmath $x$})=\frac{\partial}{\partial x_1}\dots\frac{\partial}{\partial x_k}F(\mbox{\boldmath $x$}).
\end{equation}
From the common distribution, $F(\mbox{\boldmath $x$})$, marginal distributions
$F_i(x_i)$ can be
derived, as well as the mean value ${\rm E}[h]$ of quantity $h(\mbox{\boldmath $X$})$:
\begin{equation}
F_i(x_i)=\!\!\int\limits_{\mathbb{R}^{k-1}}\!\!\!F\Big[\prod\limits_{j\not=i}{\rm d}x_j\Big],\quad
{\rm E}[h]=\!\int\limits_{\mathbb{R}^k}\!h(\mbox{\boldmath $x$})\,
F\Big[\prod\limits_{j=1}^{k}\!{\rm d}x_j\Big].
\label{meanvec}
\end{equation}

Most important for applications are the first and the second moments,
which receive their own designation:
$\mu_i={\rm E}[X_i]$, and $R_{ij}={\rm E}[X_i\,X_j]$,
respectively. The meaning of $R_{ij}$ is the correlation matrix. 
Finally, it is useful to introduce the covariance matrix,
\begin{equation}
C_{ij}={\rm E}\big[X_i\,X_j\big]-{\rm E}\big[X_i\big]\;{\rm E}\big[X_j\big]
=R_{ij}-\mu_i\,\mu_j.
\end{equation}
It can be proven that statistically independent random values are 
always uncorrelated, in which case $C_{ij}$ is a diagonal matrix. 

\subsection{Random processes}
Random processes are a mathematical description of
measurements of a physical quantity which is either
disturbed by noise or subject to non-deterministic evolution by itself.
A random process is a collection of random values $\{X(t):t\in \mathbb{T}\}$ with 
respect to $(\Omega,\Sigma, P)$, where $\mathbb{T}$ is 
a time set (usually a subset of $\mathbb{R}$ or a
set of integers). For every finite set of points
$\{t_1,\dots,t_n\}\subset \mathbb{T}$, one can find the distribution function
$F_{t_1,\dots,t_n}(x_1,\dots,x_n)$ corresponding to the random
vector $\mbox{\boldmath $X$}_n=(X(t_1),\dots,X(t_n))$.

The random process can be interpreted as a function of two variables,
$X\equiv X(t,\varpi)$. For a fixed value of some $\varpi^\prime\in\Omega$, 
$X(t,\varpi^\prime)$ is a function of time, called the trajectory or
the realisation of the random process. For a fixed value of
$t^\prime\in \mathbb{T}$, the function  $X(t^\prime,\varpi)$ is a real
random value. A process is
called stationary if $F_{t_1,\dots,t_{n}}(x_1,\dots,\
x_{n})=F_{t_1+r,\dots,t_{n}+r}(x_1,\dots,x_{n})$ for every
$r\in \mathbb{T}$. This implies that all moments of $X(t)$ are independent of 
time,
\begin{equation}
{\rm E}\big[X^k(t)\big]={\rm E}\big[X^k(t+r)\big]={\cal M}_k,
\end{equation}
for all $r$ and $t$. A weaker form of this condition is often used: a
random process is called a weak-sense stationary if
\begin{equation}
{\rm E}\big[X(t)\big]={\rm E}\big[X(t+r)\big],\quad
{\rm E}\big[X^2(t)\big]={\rm E}\big[X^2(t+r)\big].
\end{equation}

A stationary random process does not change its nature with time.
However, in general there is no way to calculate the statistical
characteristics of such a process knowing only a single realisation. For
this purpose, a stronger assumption has to be made: a random process is
called an ergodic one if for all fixed $r\in \mathbb{T}$ and a real
function $h(x)$
\begin{equation}
{\rm E}\big[h\left(X(r)\right)\big]=
\lim\limits_{T\rightarrow\infty}\frac{1}{T}\int\limits_0^T h\left(X(t)\right)
{\rm d}t.
\end{equation}
It is commonly assumed that real processes satisfy this condition.

Stationary processes can be characterised by autocorrelation $R(t)$
and autocovariance $C(t)$ functions,
\begin{equation}
R(t)={\rm E}\big[X(r)\,X(r+t)\big],\quad
C(t)=R(t)-\mu^2,
\end{equation}
which, according to stationarity, are independent of $r$. By setting
$r=-t$ we find that $R(t)=R(-t)$.

Another way of characterising processes is by their spectral properties,
which are also our main interest in this paper.
Power spectral function of a stationary stochastic process $X(t)$ is
\begin{equation}
S(\omega)=\lim\limits_{T\rightarrow\infty}\frac{1}{2T}\;
{\rm E}\left[\left|\mathcal{F}_T[X(t)](\omega)\right|^2\right],
\label{powersp}
\end{equation}
where
\begin{equation} 
\mathcal{F}_T\big[X(t)\big]=\int\limits^T_{-T}X(t)\,e^{-i\omega t}\,{\rm d}t
\label{cast_four} 
\end{equation}
is the incomplete Fourier transform. According to the 
Wiener--Khinchin theorem the power spectral function can be 
calculated from the autocovariance:
$S(\omega)=\mathcal{F}\left[C(t)\right](\omega)$.

\subsection{Random point processes}
The concept of point processes was originally developed to describe 
random configurations of points in space \citep{1965cox,Daley:2003}. One 
way to characterise such random configurations in some
topological space $\mathcal{X}\subset\mathbb{R}^n$ is by means of
the counting measure, $N(A)$. For every 
$A \subset \mathcal{X}$, the counting measure gives the number of 
points lying in $A$. 

Similar to random processes, a random point process can be characterised 
by its mean value and moments. The first-order moment is called 
the intensity measure,
\begin{equation}
M_1(A)={\rm E}\big[N(A)\big].
\end{equation}
For every $A\subset\mathcal{X}$, $M_1(A)$ is the mean number of points 
in $A$. The second-order moment measure is then
\begin{equation}
M_2(A\times B)={\rm E}\big[N(A)N(B)\big].
\label{M2}
\end{equation}

Let $\{x_i\}$ be one possible configuration of points, i.e.\ the support of
some $N(.)$. For the functions $h(x)$ and $g(x,y)$ on $\mathcal{X}$
and $\mathcal{X}^{2}$, respectively, it follows \citep{Campbell:1909,Daley:2003}
\begin{eqnarray}
{\rm E}\Big[\! \sum\limits_{i} h(x_i)\Big]
&=&\int\limits_{\mathcal{X}}h(x)\,M_1({\rm d}x),\label{Camb1}\\
{\rm E}\Big[\! \sum\limits_{i,j} g(x_i,\,x_j)\Big]
&=&\int\limits_{\mathcal{X}^{2}}g(x,y)\,M_2({\rm d}x\times{\rm dy}).
\label{stred2}
\end{eqnarray}

The concept of point process can be further generalised by adding {\em
marks}, $\kappa_i$, from the mark set $\mathcal{K}$ to each coordinate
$x_i$ from $\{x_i\}$. Marks carry additional information (for example,
they can be employed to describe the radial distribution of
spots and some relations among spots located at different radii; see 
Sect.~\ref{marks}). The resulting  point process on the set
$\mathcal{X}\times\mathcal{K}$ is called the marked point process if
for every $A\subset\mathcal{X}$ it fulfils the condition  $N_{\rm
g}(A)\equiv N(A\times\mathcal{K})<\infty$.

The random measure $N_{\rm g}(A)$ represents the {\em ground process} of
the marked process $N$. If the dynamics of the process is governed
only by the ground process and the marks are mutually independent and
random values with the distribution functions
$G({\rm d}\kappa)$, then the measure of the marked process fulfils 
\begin{eqnarray}
M_1({\rm d}x\!\times\!{\rm d}\kappa)\!&=&\!M_{{\rm g}1}({\rm d}x)\;G({\rm d}\kappa),\label{markpp1}\\
M_2({\rm d}x_1\!\times\!{\rm d}\kappa_1\!\times\!{\rm d}x_2\!\times\!{\rm d}\kappa_2)
\!&=&\!M_{{\rm g}2}({\rm d}x_1\!\times\!{\rm d}x_2)\,G({\rm d}\kappa_1)\,G({\rm d}\kappa_2)\label{markpp2}.
\end{eqnarray}

Finally, it is useful to introduce the factorial measures, which will be 
used later to simplify various expressions. They satisfy definitional relations
\begin{eqnarray}
M_{[1]}(A)&\equiv& M_1(A),\\
M_{[2]}(A\times B)&\equiv& M_2(A\times B)-M_1(A\cap B).
\label{kvadmir}
\end{eqnarray}

\bibliographystyle{aa} 
\bibliography{9720} 

\end{document}